\documentclass[times,twocolumn,final]{elsarticle}

\usepackage{preprint}
\usepackage{framed,multirow}

\usepackage{amsmath}
\usepackage{amssymb}
\usepackage{latexsym}

\usepackage{xcolor}

\usepackage[switch,pagewise]{lineno} 
\usepackage{subcaption}
\usepackage[colorlinks=true]{hyperref}

\begin{document}
\newcommand{\norm}[1]{\left\lVert#1\right\rVert}

\verso{Accepted for publication}
\begin{frontmatter}
\title{Analysis of a reduced-order model for the simulation of elastic geometric zigzag-spring meta-materials}
\author[1]{Kurt {Leimer}\corref{cor1}}
\cortext[cor1]{Corresponding author.}
\emailauthor{kurt.leimer@tuwien.ac.at}{Kurt Leimer}

\author[1,2]{Przemyslaw {Musialski}}
\address[1]{Technische Universit\"at Wien (TU Wien)}
\address[2]{New Jersey Institute of Technology (NJIT)}

\accepted{6 October 2021}

\begin{abstract}

We analyze the performance of a reduced-order simulation of geometric meta-materials based on zigzag patterns using a simplified representation. 
As geometric meta-materials we denote planar cellular structures which can be fabricated in 2d and bent elastically such that they approximate doubly-curved 2-manifold surfaces in 3d space. They obtain their elasticity attributes mainly from the geometry of their cellular elements and their connections.  {In this paper we focus on cells build from so-called zigzag springs.} 
The physical properties of the base material (i.e., the physical substance) influence the behavior as well, but we essentially factor them out by keeping them constant. 

The simulation of such complex geometric structures comes with a high computational cost, thus we propose an approach to reduce it by abstracting the  {zigzag cells} by a simpler model and by learning the properties of their elastic deformation behavior. 
In particular, we analyze the influence of the sampling of the full parameter space and the expressiveness of the reduced model compared to the full model. Based on these observations, we draw conclusions on how to simulate such complex meso-structures with simpler models.

\end{abstract}


\end{frontmatter}

	\section{Introduction}\label{sec:intro}
	
\begin{figure*}
    \begin{subfigure}{0.19\textwidth}
    	\includegraphics[width=\textwidth]{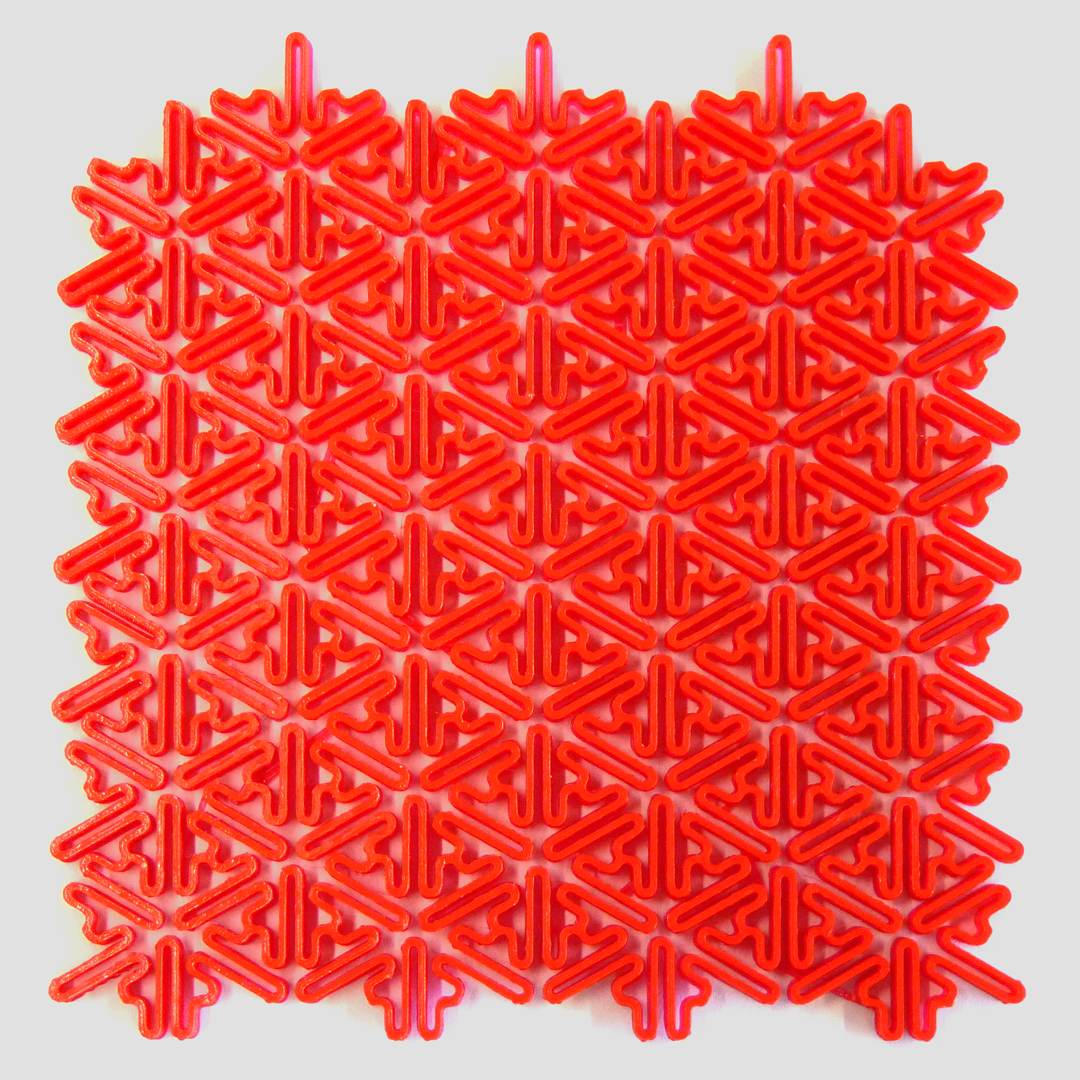}
    	\includegraphics[width=\textwidth]{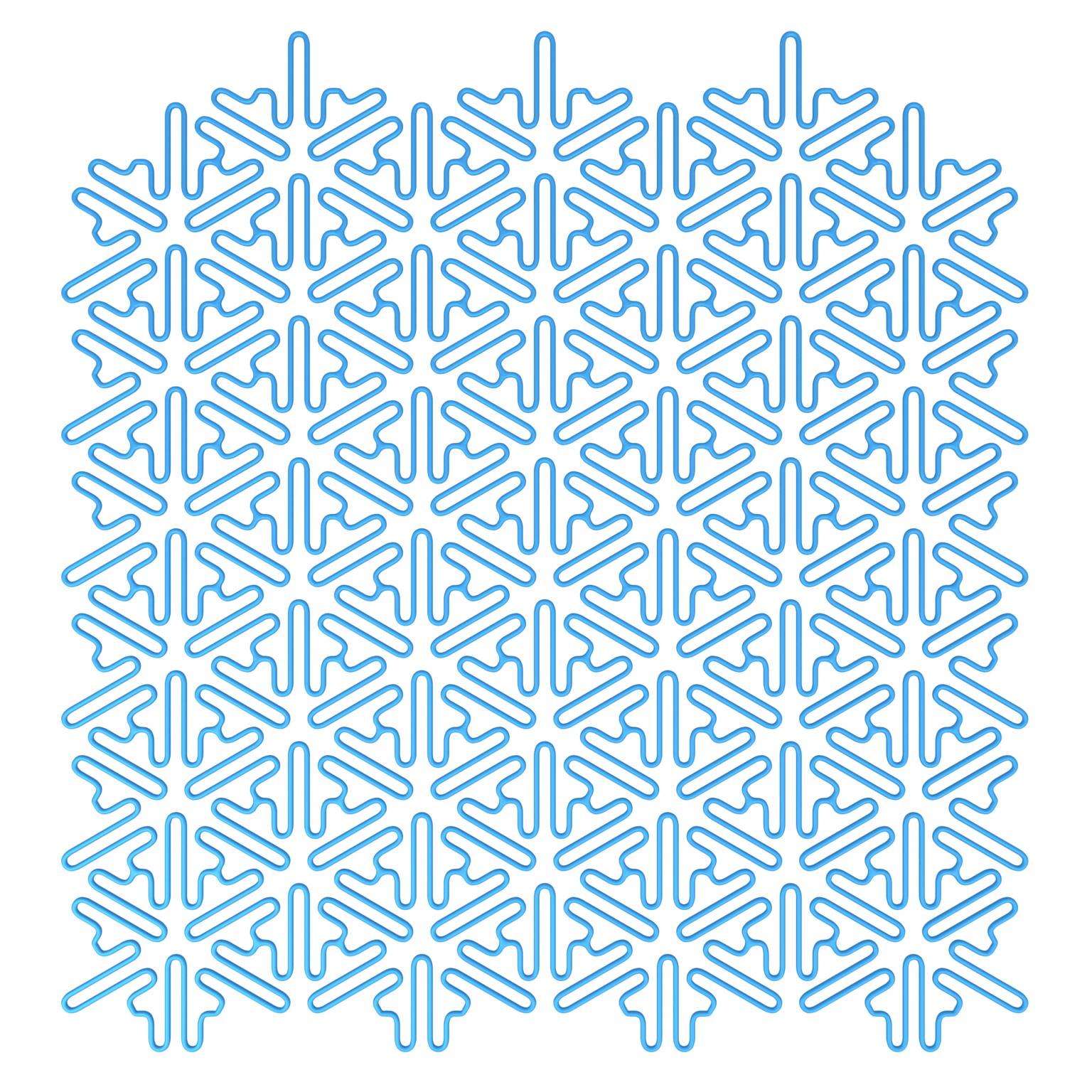}
    	\caption{Flat Grid}
    \end{subfigure}
    \hfill
    \begin{subfigure}{0.19\textwidth}
    	\includegraphics[width=\textwidth]{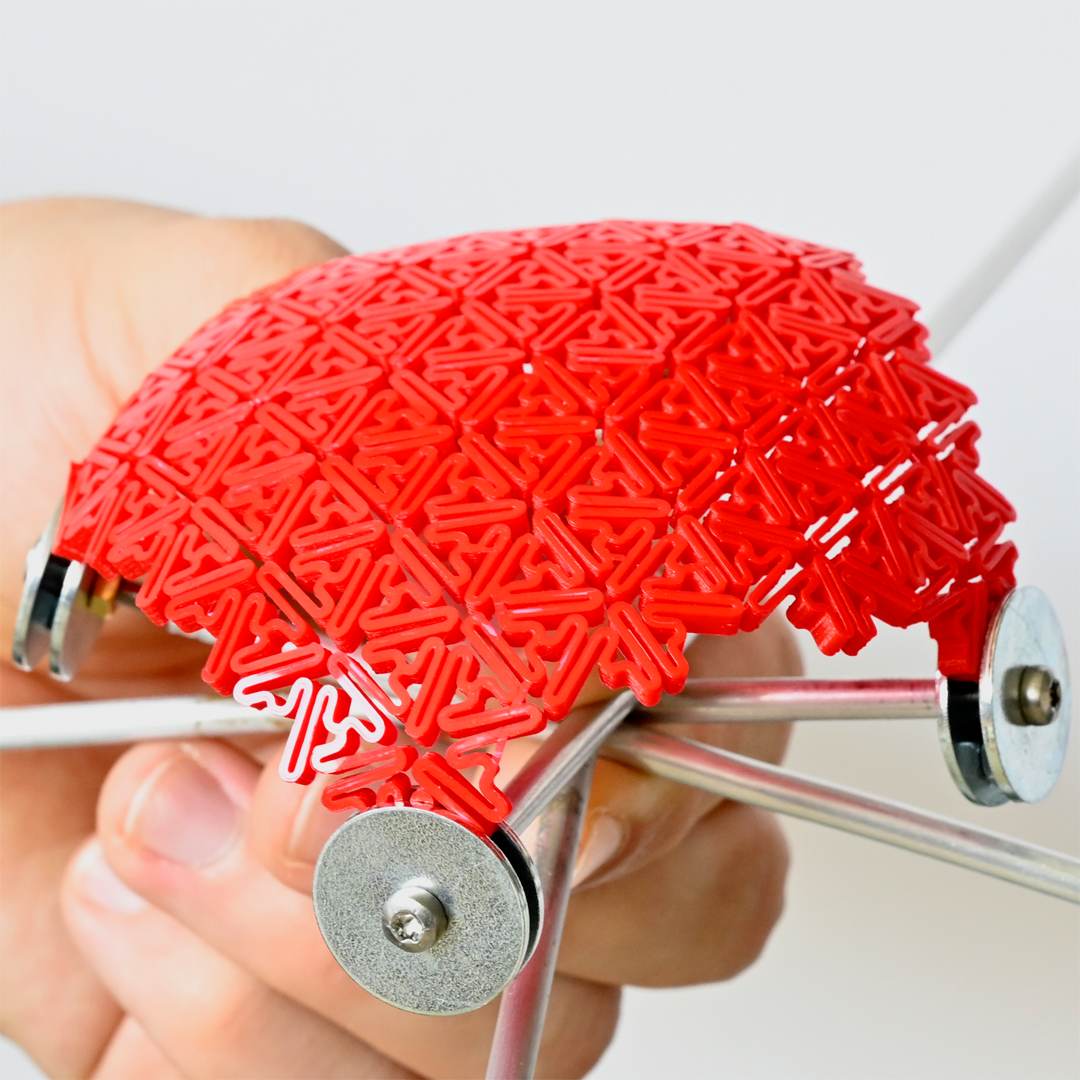}
    	\includegraphics[width=\textwidth]{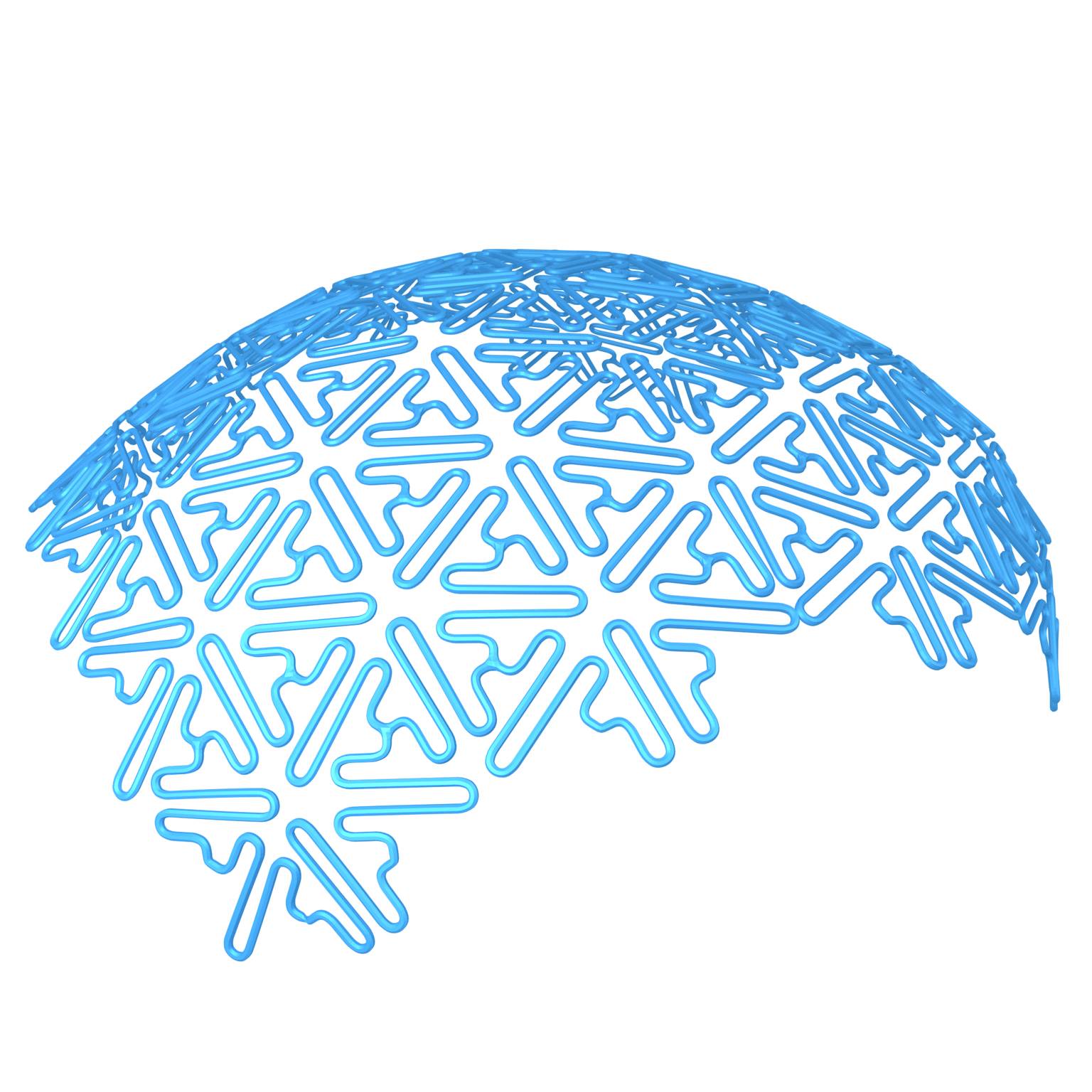}
    	\caption{Dome}
    \end{subfigure}
    \hfill
    \begin{subfigure}{0.19\textwidth}
    	\includegraphics[width=\textwidth]{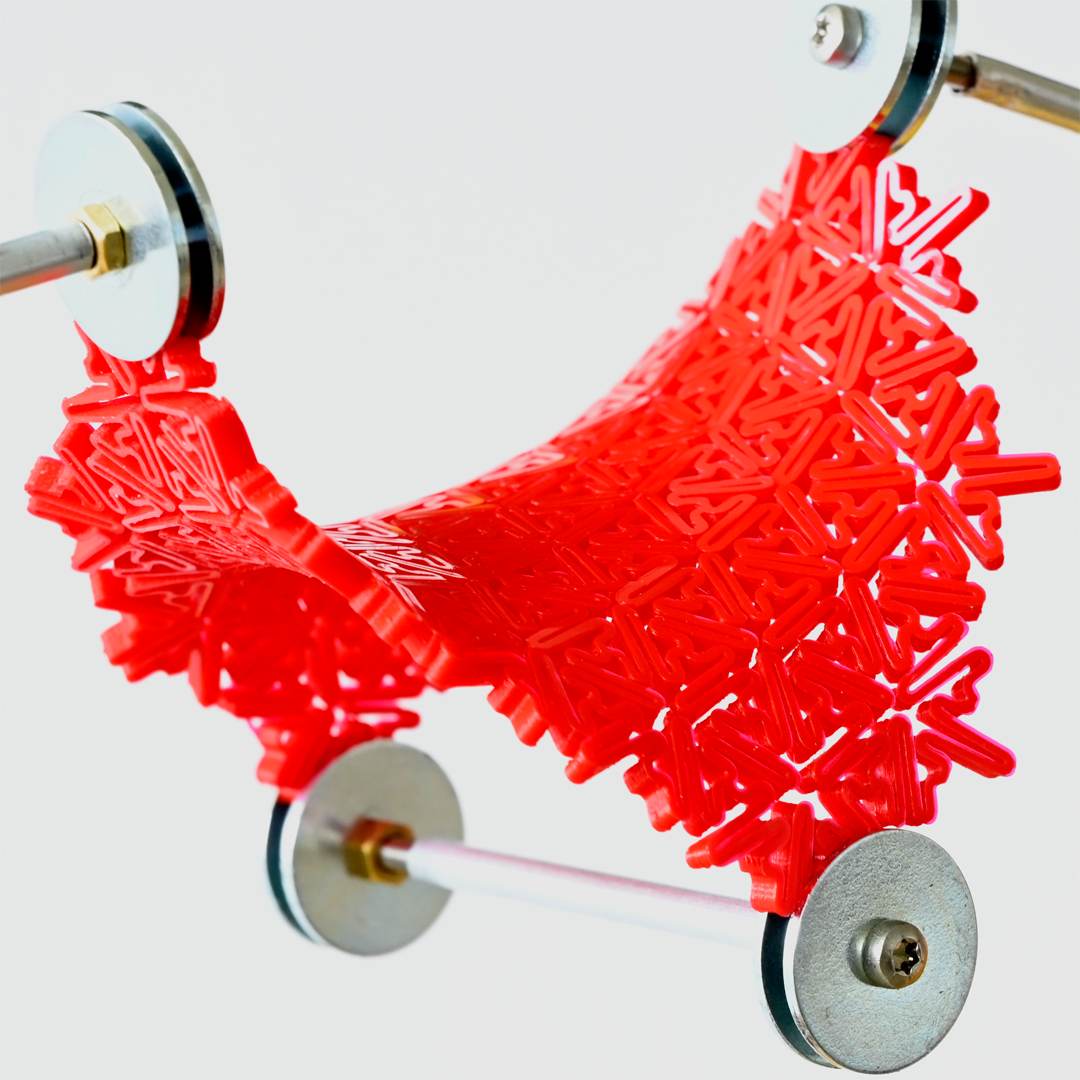}
    	\includegraphics[width=\textwidth]{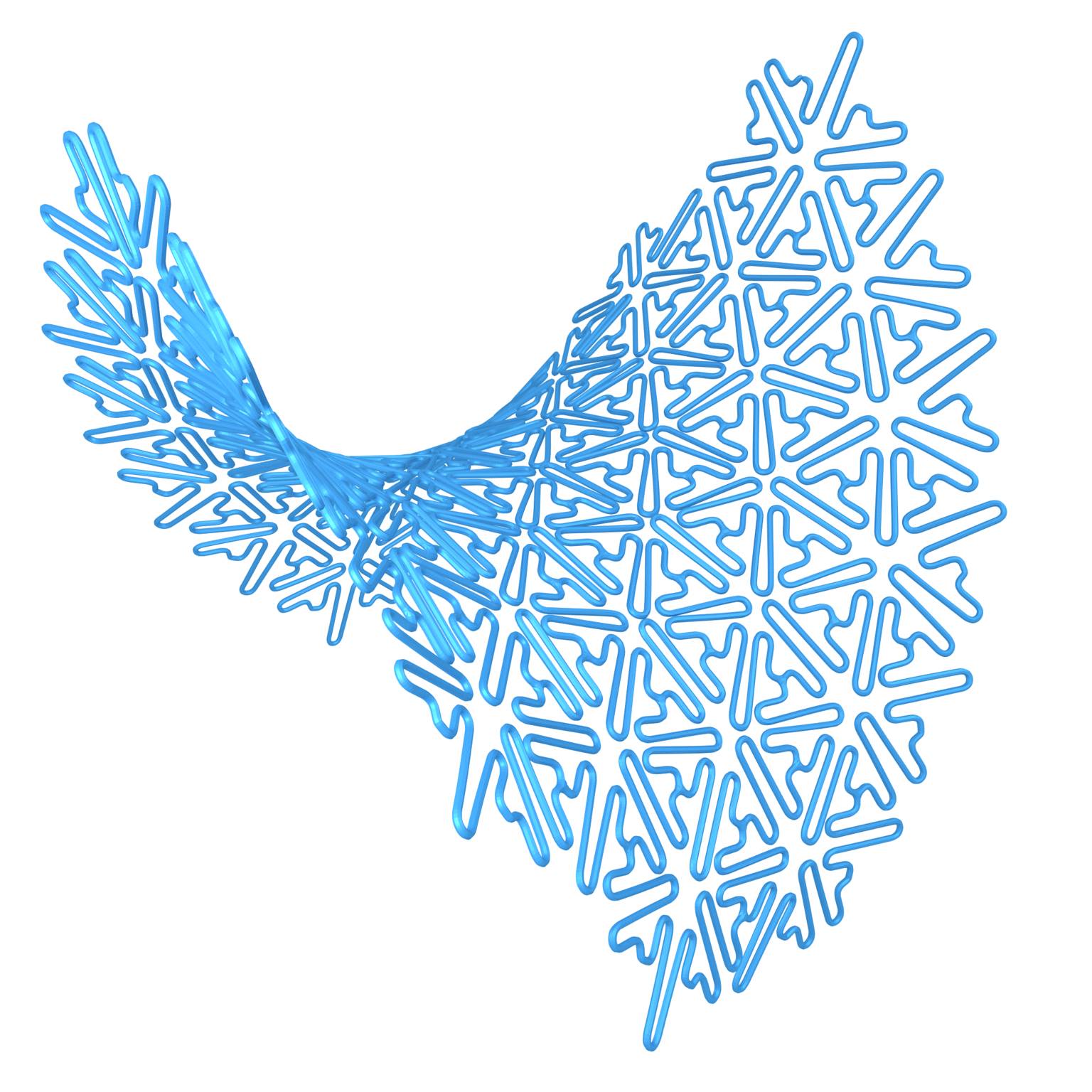}
    	\caption{Saddle}
    \end{subfigure}
    \hfill
    \begin{subfigure}{0.19\textwidth}
    	\includegraphics[width=\textwidth]{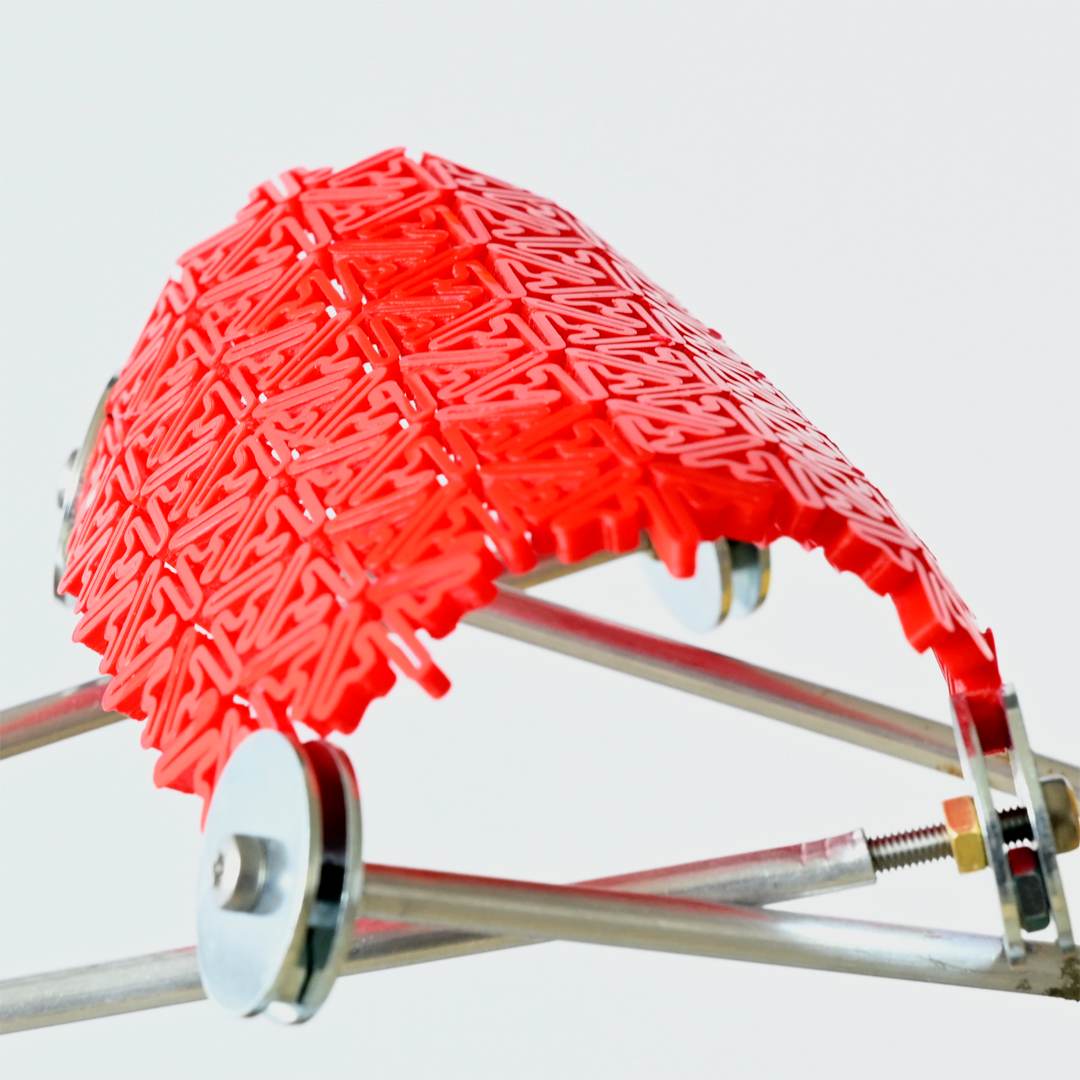}
    	\includegraphics[width=\textwidth]{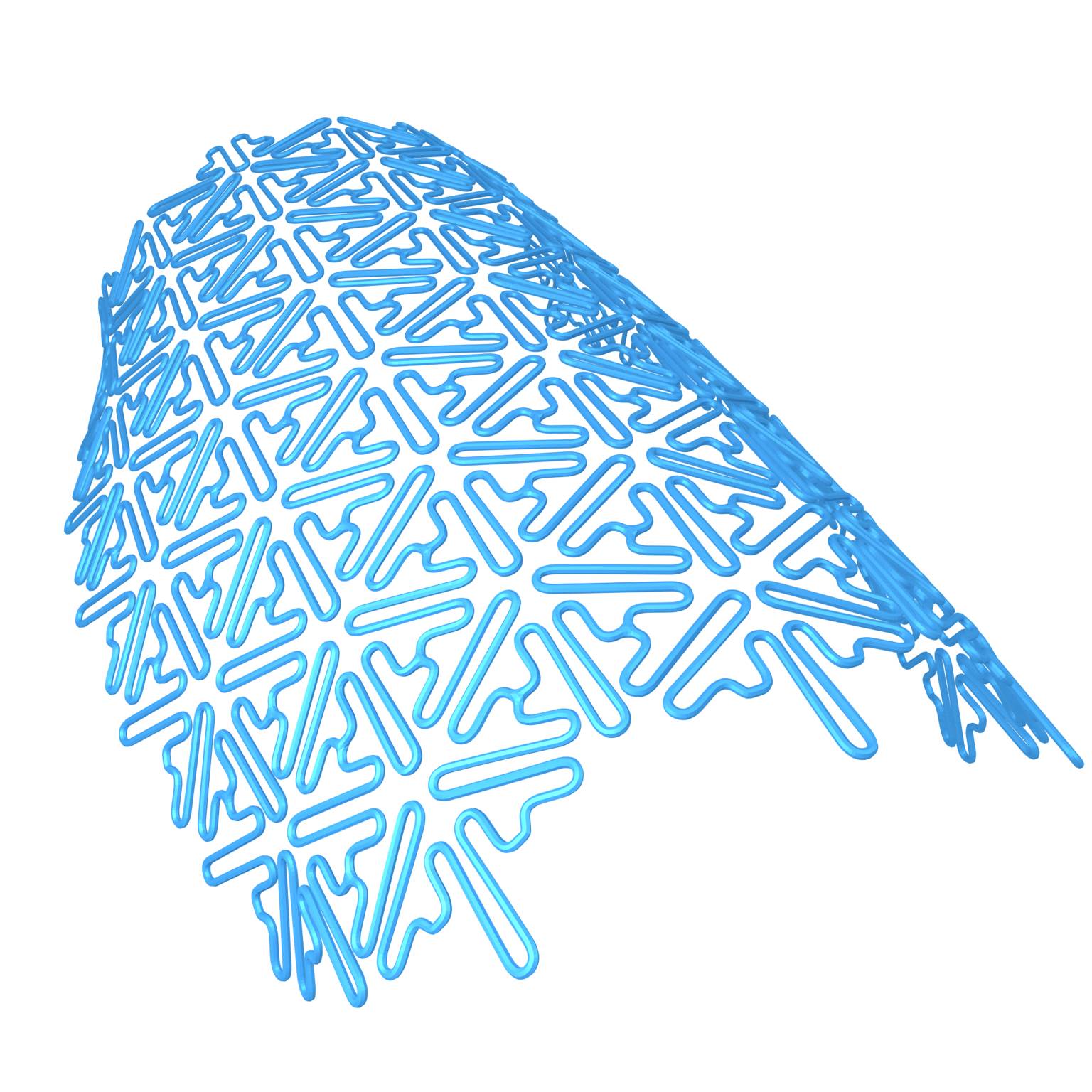}
    	\caption{Tunnel}
    \end{subfigure}
    \hfill
    \begin{subfigure}{0.19\textwidth}
    	\includegraphics[width=\textwidth]{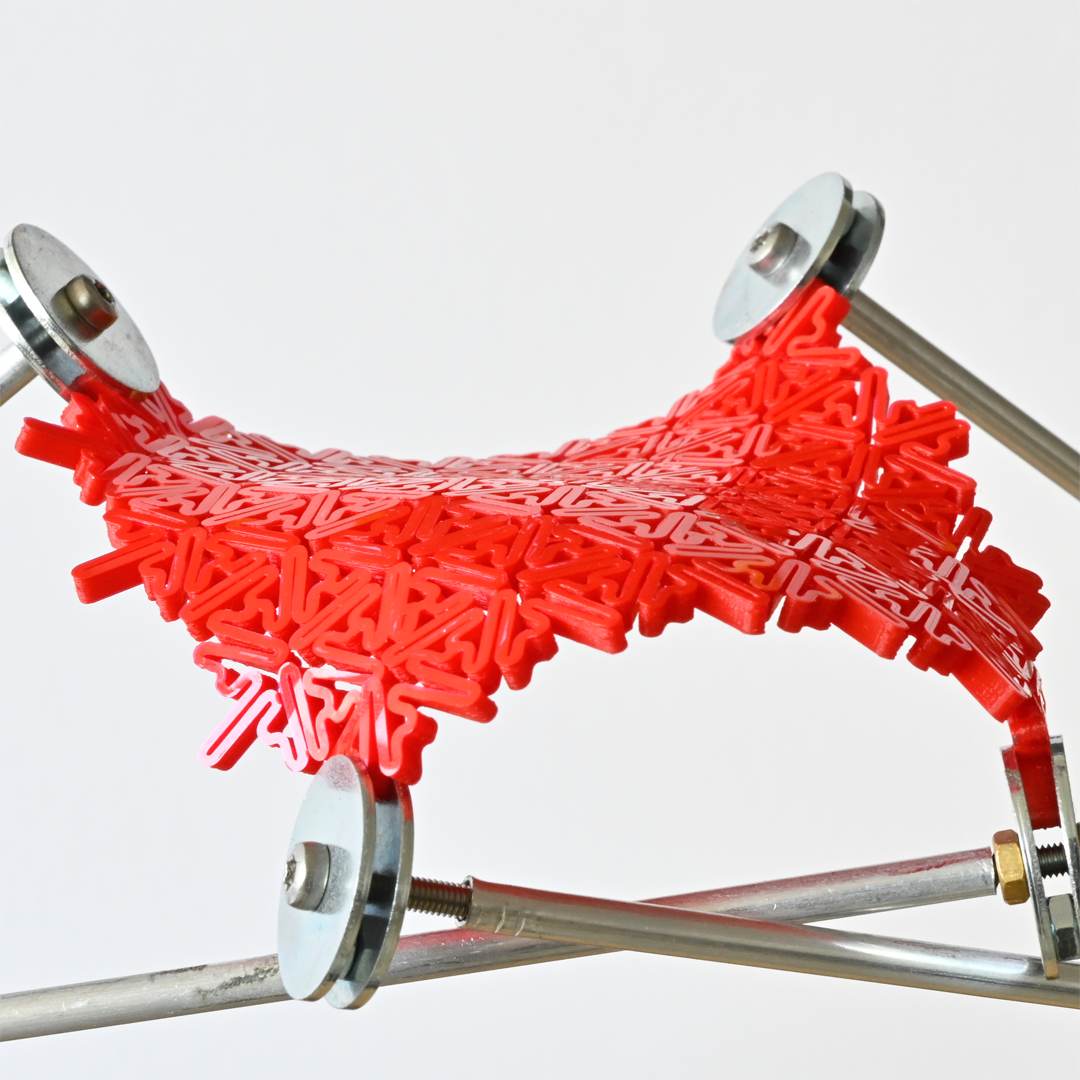}
    	\includegraphics[width=\textwidth]{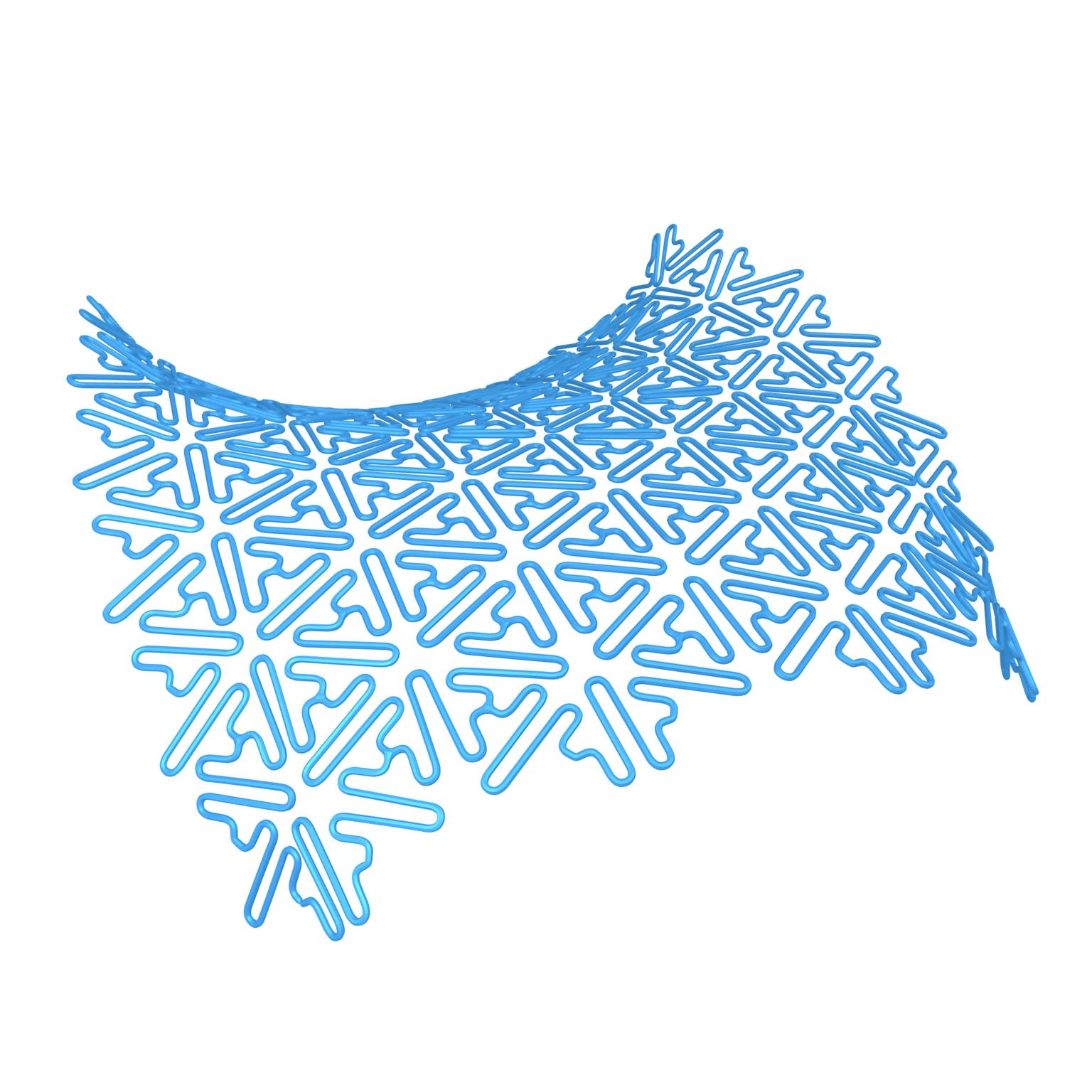}
    	\caption{Manta}
    \end{subfigure}
    \caption{Patterns made from rigid material can be made flexible by introducing meso-structures. Top: 3d-printed pattern deformed into various shapes. Bottom: Simulation of pattern with similar structure.}
    \label{fig:grids}
\end{figure*}

In the field of computational design of meta-materials, complex patterns of meso-structures have seen increased interest because of the ability to locally control their flexibility through adjustment of the meso-structure parameters. 
Such structures come with a number of advantages like, despite the simplicity of their fabrication, their ability to nestle to sophisticated free-form surfaces. 
We can think of such materials as meshes composed of nodes that are connected by edges of complex geometry. Additionally, while these meshes are planar in their rest state, they do have a certain thickness (cf. Fig.~\ref{fig:grids}).

The assumption is that mechanical properties of such structures can be homogenized~\cite{Panetta2015}, i.e., the way how the structure stretches, bends, shears, or breaks, depends continuously on the interplay of individual connections, their thickness and size, and their angular deflections about nodes. If such a structure is exposed to internal and external loads, the forces spread across the elements and stress individual cells depending on the number of their connections and how the edges transmit the deformation. 

This behavior is very valuable since it allows the structure to naturally nestle to complex shapes, in contrast to traditional flexible materials, like cardboard, fabrics, or sheets of metal. 

The deformation behavior of such materials can be controlled by incorporating meso-structures with the desired mechanical properties \cite{Martinez2019} that may also form elaborate aesthetically pleasing patterns, which can be used to approximate 3D surfaces well.

However, using such complex patterns for inverse shape design leads to two problems: First, their distribution on the surface is a non-trivial geometric task. Second, there is the problem of the performance of their simulation. 

The pattern can be of very complex geometry and of high resolution, hence its simulation is of significant computational cost. To speed up the optimization process, simplifications need to be made. 
	
In this work,  {we focus on structures composed of zigzag spring patterns, and} we address the problem by simplifying the structure of the pattern by encoding its mechanical properties into the material parameters used in the physical simulation and further performing a homogenization of the meta-material on the coarse level.

Recently, Leimer and Musialski \cite{Leimer2020a,Leimer2020b} proposed a structural simplification for such  {zigzag spring}  meso-structures to reduce the computational burden while keeping the input-output behavior. 
This paper is an extension of that work and provides a more rigorous approach to the sampling of simulated examples used for the material parameter optimization and systematic analysis of their influence.
The contributions of this paper are as follows:
\begin{enumerate}
    \item We introduce a pattern generator for zigzag springs that is capable of creating flexible hexagonal or rectangular patterns whose flexibility can be locally controlled.
    \item We propose an optimization scheme for encoding the deformation behavior of  {a zigzag pattern} into the material parameters of a simple pattern, thus greatly reducing the computational cost of physical simulation.

    \item This optimization scheme requires a number of example deformations of the full pattern in order to find a good fit for the reduced model parameters. We analyze how the choice of example deformations influences the accuracy of the optimization and propose a small set of examples that performs well in terms of both accuracy and computational cost.
    \item We apply our reduced model to the task of shape design by formulating an optimization problem that aims to find the ideal boundary constraints for a flexible  {zigzag pattern} such that the pattern deforms to approximate a given target shape. The advantage of the reduced model is demonstrated by comparing the computational cost of the optimization with that of the full pattern.
\end{enumerate}

	\section{Related Work}\label{sec:related}
	
\begin{figure*}[t]
	\begin{centering}
	\begin{subfigure}[b]{0.27\textwidth}
		\includegraphics[width=\textwidth]{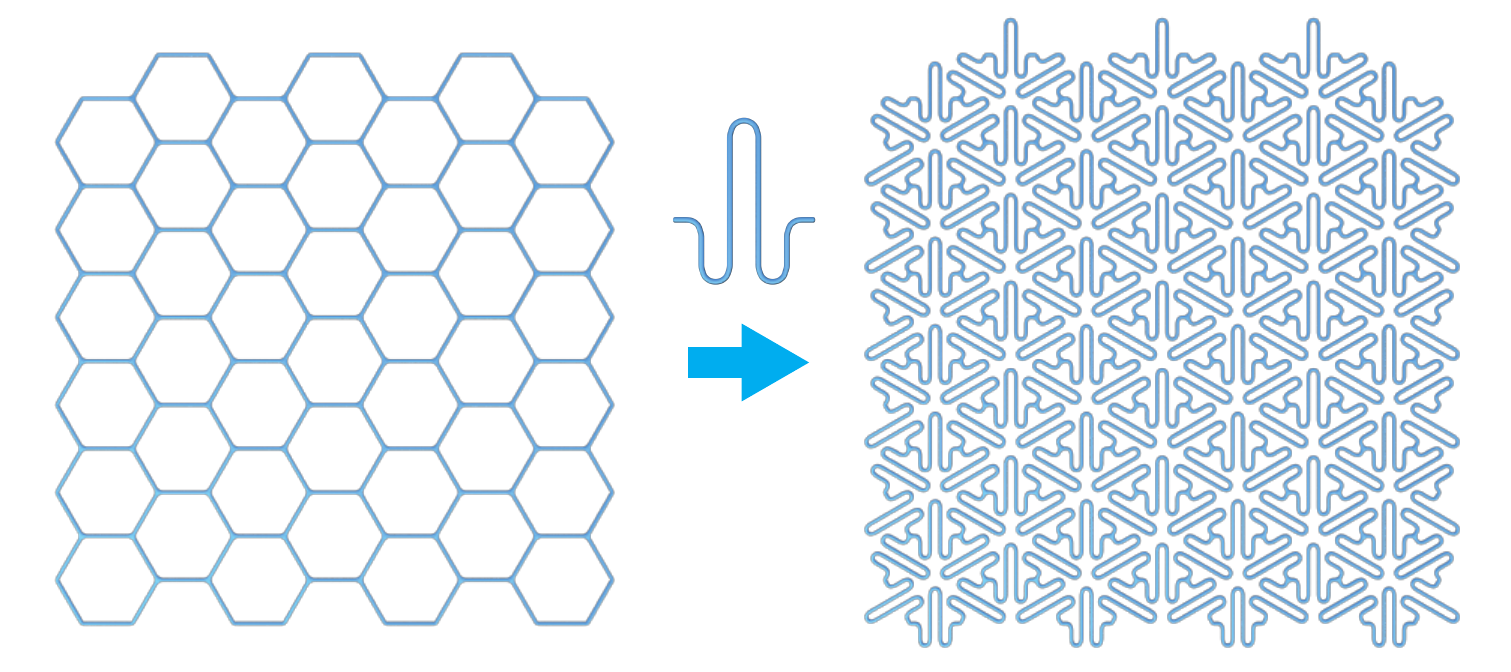}
		\caption{Pattern generation}
	\end{subfigure}
	\hfill
	\begin{subfigure}[b]{0.34\textwidth}
		\includegraphics[width=\textwidth]{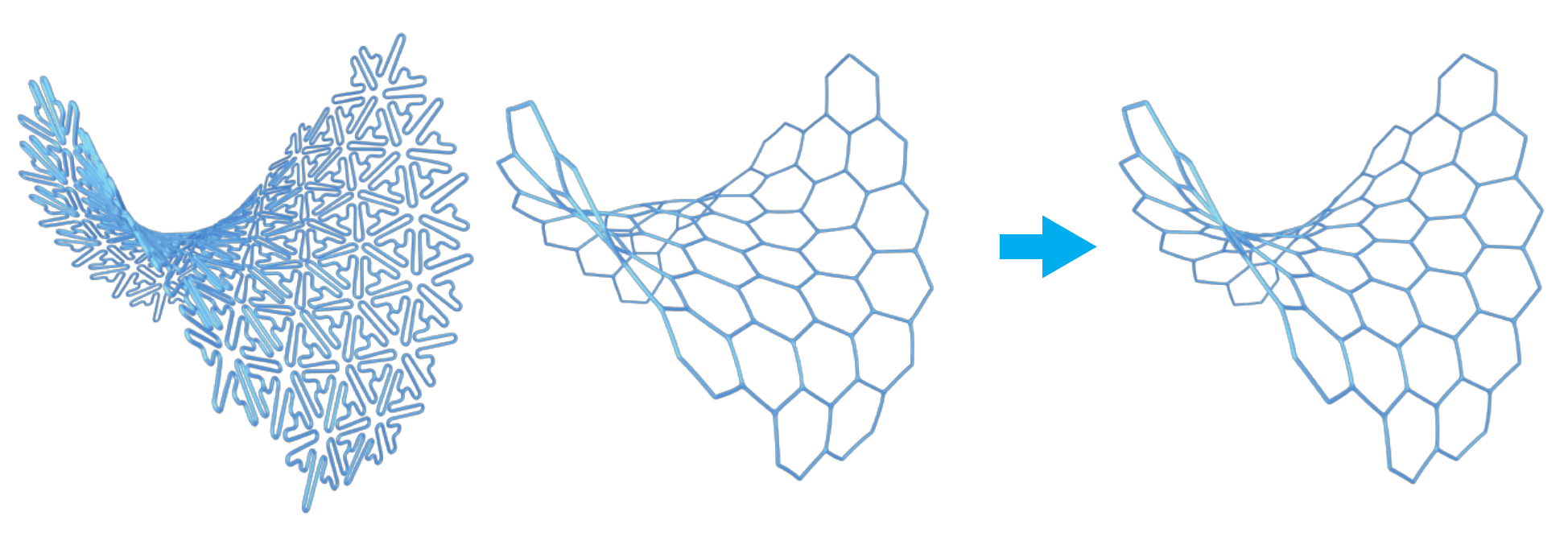}
		\caption{Pattern simplification}
	\end{subfigure}
	\hfill
	\begin{subfigure}[b]{0.33\textwidth}
		\includegraphics[width=\textwidth]{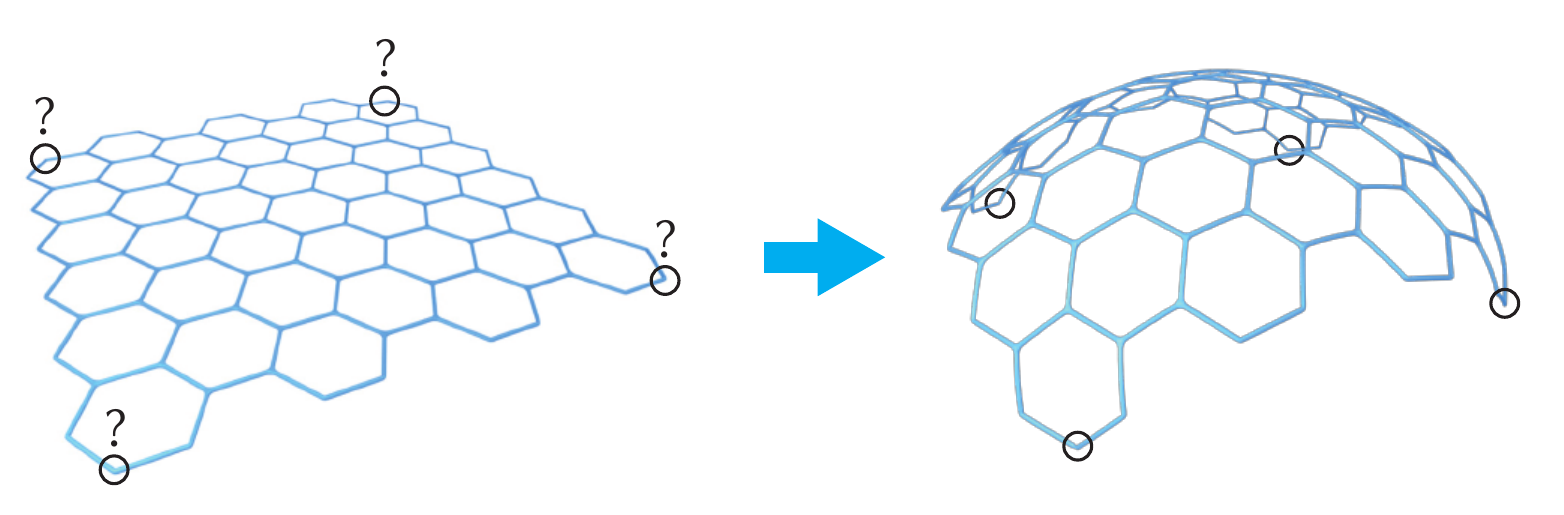}
		\caption{Shape optimization}
	\end{subfigure}
	\caption{Overview of our proposed method. (a) We first generate a  {flexible zigzag pattern} that can be deformed elastically due to its structure. (b) The pattern is simplified by encoding its deformation behavior into the material parameters of a reduced-order pattern. (c) The simplified pattern can then be used for the task of shape optimization.}
	\label{fig:overview}
	\end{centering}
\end{figure*}

The popularity of modern additive manufacturing techniques has led to many advancements in the field of computational design of meta-materials. Increased attention has been given to the possibility of building 3D objects out of volumetric structures composed of regular \cite{Panetta2015} as well as stochastically sampled \cite{Martinez2017} grids. Another branch are flat flexible material sheets \cite{malomo} which can be used to form 3D surfaces. Additionally, their advantage is that they can be easily adapted to various shapes and easily produced, transported, and assembled on site.

The materials we aim at can be classified as cellular structures, i.e., they consist of more or less regular cellular elements. 
The research of such structures has been approached across various disciplines due to their very interesting properties. 
They have been shown to be mechanically extremely efficient since they are very robust and can absorb impacts from large impulse forces on the one hand, but are of extremely low density and thus lightweight and flexible on the other \cite{Wadley2003}. 

Especially the case of so-called honeycomb structures~\cite{Pottmann2015}, which primarily consist of hexagonal elements, has been studied extensively in the fields of mechanical engineering and material sciences, where computational models for their mechanics, i.e., the structure's response to forces, have been introduced \cite{Gibson1982,Masters1996}. In addition, the influence of the wall thickness variations on the elastic properties \cite{Li2005},  and the effects of hierarchy \cite{Taylor2011} have been studied.

Arrays of cellular structures allow the definition of computational design problems that aim at the automatic computation of structural or mechanical designs that suit some desired goals. They have been studied in computational industrial design as form-finding problems, as well as in statics and mechanical engineering as structural optimization problems \cite{Haftka1992}. 

Due to the advances in the digital fabrication field, such problems increasingly arise in various personal fabrication design tasks \cite{Garg2014,dumas2015}. 
Of particular relevance to our work is the design of structures made out of rod-like elements \cite{Perez2015, Zehnder2016, vekhter, malomo, Panetta2019, pillwein2020, Pillwein2021a}, which is made possible through efficient methods of simulating the deformation behavior of such elements \cite{bergou2008,bergou2010}.

This leads to programmable elastic structures which are based on both bending and tensile energy, for instance, by using prestressed latex membranes to actuate planar structures into free-form shapes \cite{Guseinov2017} and evolve to double-curved surfaces over time \cite{Guseinov2020}. 
A combination of elastic rods and membranes leads to Kirchhoff-Plateau surfaces that allow easy planar fabrication and deployment \cite{Perez2017a}. Another combination of precomputed flexible meso-cells in planar configurations leads to surface approximations which can deform to certain desired shapes if appropriate boundary conditions are applied \cite{malomo}. 

\begin{figure*}[t]
		\centering
		\includegraphics[width=0.99\textwidth]{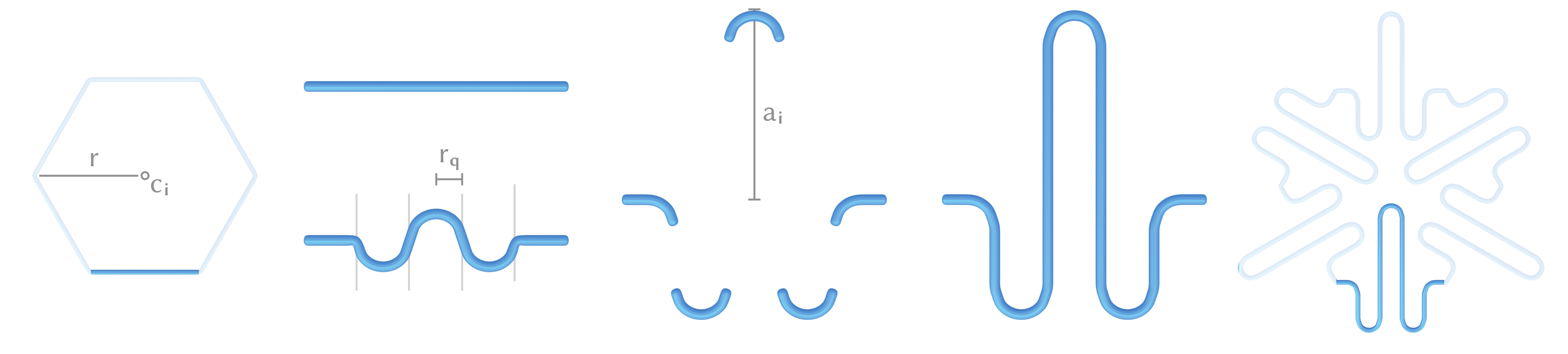}
		\caption{Transforming a hexagon edge into a zigzag spring. Given an edge of a hexagon cell with center $c_i$ and circumradius $r$, the segments of the edge are replaced by semicircles with radius $r_q$, which are then offset in $y$-direction by amplitude $a_i$.}
		\label{fig:create_zigzag}
\end{figure*}

Physical simulation of materials has a long tradition in computer graphics \cite{Terzopoulos1987}, however, they have been usually studied for the sake of virtual materials in animation or surgery simulation \cite{Nealen2006a}. 
More recently, due to the emergence of the 3d-printing paradigm, computer graphics researchers turned towards real material simulation methods. For instance, 
Bickel et al. developed in a series of articles a computational system for data-driven design of non-linear elastic materials, e.g., soft tissues \cite{Bickel2009,Bickel2010}. Their approach provides also goal-based material design which partially inspired our approach.

Other scientific and engineering fields also research the mechanics of cellular meta-materials~\cite{zheng2014,ion2016,schaedler2016} due to their strong structural properties and light-weightiness~\cite{zheng2014}. Unstructured cell distributions, like foams~\cite{ashby2006}, and nano-materials~\cite{meza2014} belong to this category in the broadest sense as well. An additional advantage of such structures is that they allow programming specific properties, like negative Poisson ratio~\cite{clausen2015}, or adaptable mechanical structures~\cite{jenett2017}.

	\section{Meta-Material Design Overview}\label{sec:overview}
	
	We focus on flat sheets of elastic meta-materials which can be fabricated in planar configurations. Such flat sheets can be bent into spatial surfaces of a desired shape if programmed accordingly. They obtain their shape and properties also due to the stress which is constantly stored in the structure if spanned in a precomputed configuration. 
	
	The first step in the design of such materials is to choose which type of pattern and meso-structure to use.  {For the pattern, we decide on a regular tiling of hexagons which allow for a regular tessellation of the Euclidean plane. For the meso-structure, we use a type of spring which we will refer to as \textit{zigzag springs}}. 
	
	 {We decide on this type of spring for two main reasons: First, since they wind in a space-filling manner, they cover the cells of the regular tiling quite uniformly with material, such that if the shape is deformed, no large gaps between the cells occur. This is in contrast to spiral-like patterns~\cite{malomo}, which are more flexible, but at the same time they develop large variation of the coverage of the surface with material. Our goal is to achieve an as homogenous surface coverage as possible. Second, the given pattern exhibits the desired degrees of freedom: it slightly stretches and compresses, allows rather small shear and twist, and its bending deformation is the dominant one. And third, we can easily create the geometry of flexible sheets from polygonal tilings by simply replacing polygon edges with zigzag-springs.
	In addition, due to their nature they can develop both locally positive or  negative Poisson ratios~\cite{Eidini2015} which also goes along with our goal to achieve patches with elliptic and hyperbolic surface characteristics.}

	To create this meso-structure, we transform each hexagon edge by introducing a natural rest curvature in the underlying 2D plane so that instead of a straight line we obtain a winding curve (see Figure \ref{fig:create_zigzag} for an example). The number of windings and their amplitude are the free parameters of this meso-structure. The process of creating these patterns is explained in further detail in Section \ref{sec:patterns}.

	We use a physical simulation based on the Discrete Elastic Rod (DER) formulation \cite{bergou2008,bergou2010} to approximate the deformation behavior of the pattern. For this purpose, appropriate material parameters are necessary to achieve a realistic behavior. Usually, these values are set based on empirical experiments with the corresponding real-world materials. However, simulating the deformation behavior of such complex patterns comes with a high computational cost. To alleviate this problem, we want to look at the pattern at a lower level of detail by simplifying each meso-structure to a simple straight edge made of a meta-material with different material parameters instead.
	
	As an example, pulling at the ends of a zigzag spring causes in-plane bending in certain regions of the meso-structure. However, when looking at the corresponding edge in the simplified pattern, which is the straight line connecting the ends of the zigzag edge, this deformation corresponds to stretching of the edge instead. Thus, to properly imitate this deformation behavior in the simplified pattern, we need to treat the meta-material of the simplified pattern as extensible even though the real-world material of the pattern is not. This is done by choosing appropriate meta-material parameter values. The process of how we optimize these parameters such that the deformation behavior of the simplified pattern matches that of the original pattern is explained in Section \ref{sec:simplification}.
	
	Furthermore, we would like to use the simplified pattern for shape design. Since the deformation behavior of flexible patterns is very complex, it can be difficult to figure out how a pattern must be deformed to approximate a desired target shape. In Section \ref{sec:optimization} we propose an optimization scheme to find the optimal positions and directions of a set of boundary constraints such that the pattern deforms into the desired shape. 

We demonstrate the results of our proposed framework in Section \ref{sec:results}. In particular, we evaluate our approach by how well the optimized material parameters of the simplified pattern capture the deformation behavior of the complex pattern in Section \ref{sub:results_stiffness}. Furthermore, our choice of example shapes from which we learn the material parameters is evaluated in Section \ref{sub:results_training} by analyzing how the quantity and types of shapes used as examples influence the result. In Section \ref{sub:results_shape}, we demonstrate how the simplified pattern can be used for the task of shape design, highlighting the advantage of this approach by comparing its performance to the use of the full pattern for the same task. Finally, we discuss potential future work in Section \ref{sec:future} and we conclude this paper in Section \ref{sec:conclusions}.

	\section{Zigzag-Pattern Generation}\label{sec:patterns}

	In this section, we explain how we generate the complex patterns that we use as examples throughout our work. Our pattern generator is capable of creating regular tilings of either quads or hexagons containing zigzag springs as edges that control the deformation behavior of the pattern.  {By extension, it is also possible to create triangular tilings by triangulation, but the reduced space within the triangles introduces additional space constraints for the design of the pattern, making them less suitable for our purpose.}
	
	For the remainder of this work, we will mostly focus on hexagonal tilings for a number of reasons. First, hexagons are one of only three polygon types that allow for regular tilings, the others being triangles and rectangles \cite{jiang}. Second, the even number of edges allows for a consistent orientation of the zigzag springs---when traversing the edges of a hexagon, the middle peak of the zigzag springs always alternates between pointing inward and outward on consecutive edges. Finally, from the perspective of meta-materials, a hexagonal pattern leads to isotropic material behavior \cite{schumacher}. While rectangles also have the second property, their meta-material behavior only shows a tetragonal symmetry.
	
	Our pattern generation process consists of two main steps, followed by an optional post-processing step. In the first step, we generate a regular tiling of polygons that serves as the coarse structure of the pattern. Then we transform the edges of the tiling to create the final pattern. Finally, we prepare the pattern for the simulation stage by removing potential causes for error and adjusting the resolution of the pattern.
	
	The input parameters for the first step include the coordinates of the origin $c_0$, which is the center of the bottom left element of the tiling, the number of elements in x- and y-direction, as well as the circumradius $r$ of the elements (cf. Fig.~\ref{fig:create_zigzag}). The output consists of vertices $V$, faces $F$, and edges $E$, as well as a vector $s$ containing a sign for each edge that determines the winding direction of the transformed edge. 
	
	For creating a hexagonal tiling, we first compute the center position of each hexagon based on the origin and circumradius. The hexagon centers $c_i$ of the first row are given by
	\begin{equation*}
	    c_i = c_0 + \sum_{j=0}^{i-1} r \begin{bmatrix}  1+\sin(\pi/3) \\ (-1)^{j} \cos(\pi/3) \end{bmatrix} \,.
	\end{equation*}
	The hexagon centers of any other row can be computed by simply offsetting the centers of the previous row by $d = 2r \sin(\pi/3)$ in y-direction, which is the diameter of the hexagon incircle.

	Then the position of each corner vertex is computed by offsetting each hexagon center by $\left( \cos(\frac{i\pi}{3}), \sin(\frac{i\pi}{3}) \right)$ with $i=0,\dots,5$. We create a face matrix $F$ with one row per hexagon, each containing the indices of the corresponding hexagon vertices. Since this process creates six vertices per hexagon, we afterward weld the pattern by removing duplicate vertices and updating their indices in the face matrix. To create the list of signs $s$, we simply traverse each hexagon edge-by-edge while alternating between $1$ and $-1$, only appending the value to the list if the current edge has not been visited before.
	
	In the second step, we transform each hexagon edge into a zigzag spring. This operation takes a vector of amplitudes $a$ as input - the dimension $n$ of the vector determines the number of peaks, while the values $a_i$ with $0 \leq a_i \leq 1$ determine the height of each peak.  {The computations in this step are performed in the local coordinate system of each edge, with the edge being parallel to the x-axis with start point $(0,0)$ and end point $(1,0)$ and the hexagon center being located at $(0.5,1)$. Since the unit vectors $(1,0)$ and $(0,1)$ in this local coordinate system map to vectors with length $r$ (the circumradius of the hexagon) and $\sqrt{0.75}r$ (the radius of the hexagon incircle) in the global coordinate system, the amplitudes $a_i$ are given relative to the incircle radius.}
	
	We partition each edge into $n+2$ segments. Every segment excluding the first and last segment, which are treated separately, is replaced by a semicircle. The radius of the semicircle is given by $r_q = 1 / (2(n+2))$. We use the top or bottom half of a circle depending on the sign of the peak $\hat{s}_i = s_e (-1)^i$, where the sign $s_e$ of edge $e$ determines the overall direction of the zigzag spring, while $(-1)^i$ causes the direction of consecutive peaks to alternate. The semicircle centers are then offset in y-direction by $\hat{s}_i (a_i - r_q)$. In other words, the amplitude $a_i$ of the $i$-th peak determines how far each semicircle is moved from the original edge, with the positive or negative direction depending on the sign of the peak. In the special case that $a_i < r_q$, we replace the semicircle with a semi-ellipse of width $r_q$ and height $a_i$.
	
	For the first segment of the edge, we leave the first half as it is and replace the second half with a quarter circle such that it can connect to the next segment without self-intersections. The last segment is treated similarly but mirrored horizontally. Please refer to Figure \ref{fig:create_zigzag} for a visual depiction.
	
	Finally, we perform one additional postprocessing step as pre\-par\-ation for simulating the deformation behavior of the pattern. We first remove any zero-length edges in the pattern--these can occur in cases where $a_i < r_q$ since the start and end points of consecutive peaks become identical. Then we subdivide any edges that are longer than a user-provided maximum to ensure that the resolution of the pattern is sufficiently high.
	
    \begin{figure}[t]
			\centering
			\includegraphics[width=0.23\textwidth]{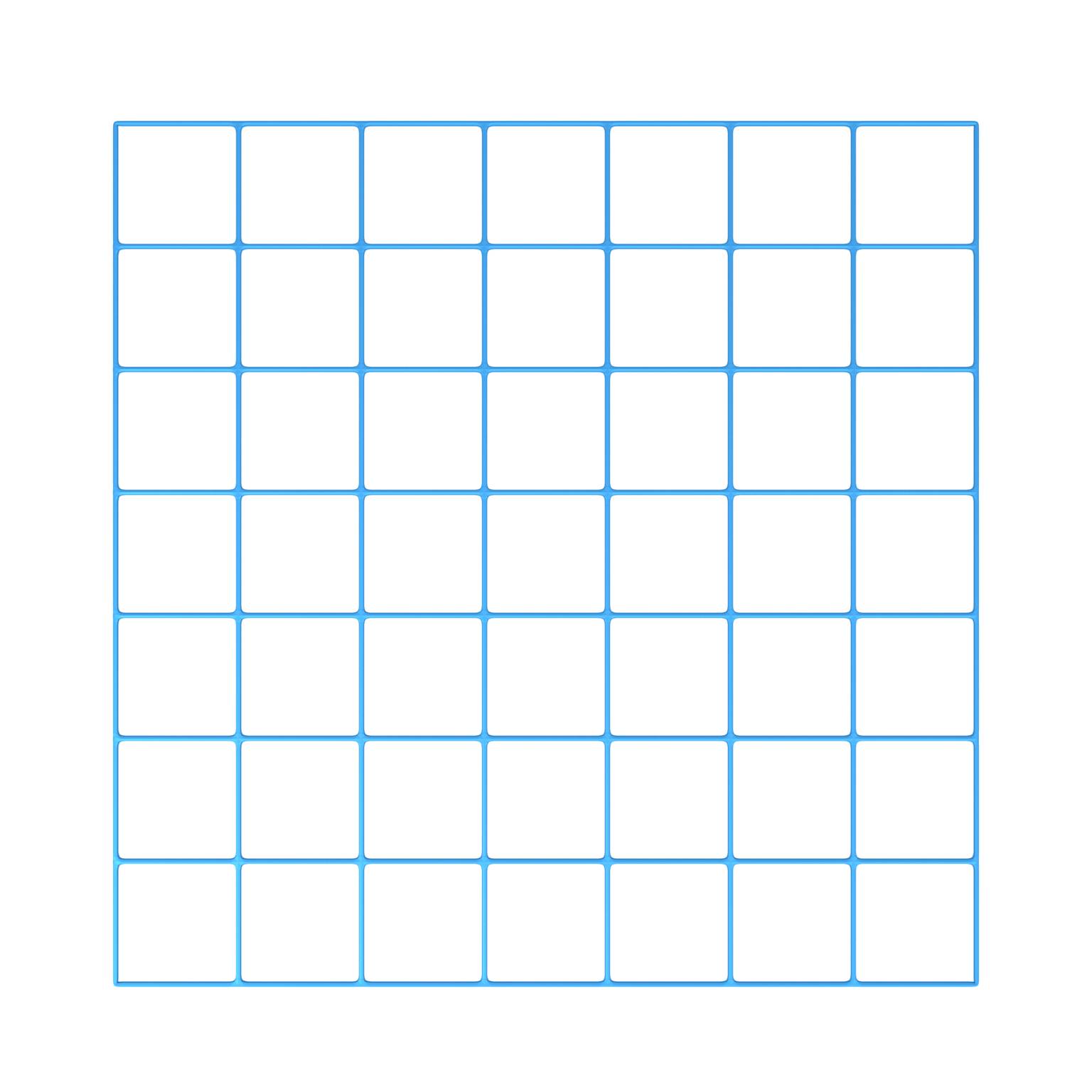}
			\includegraphics[width=0.23\textwidth]{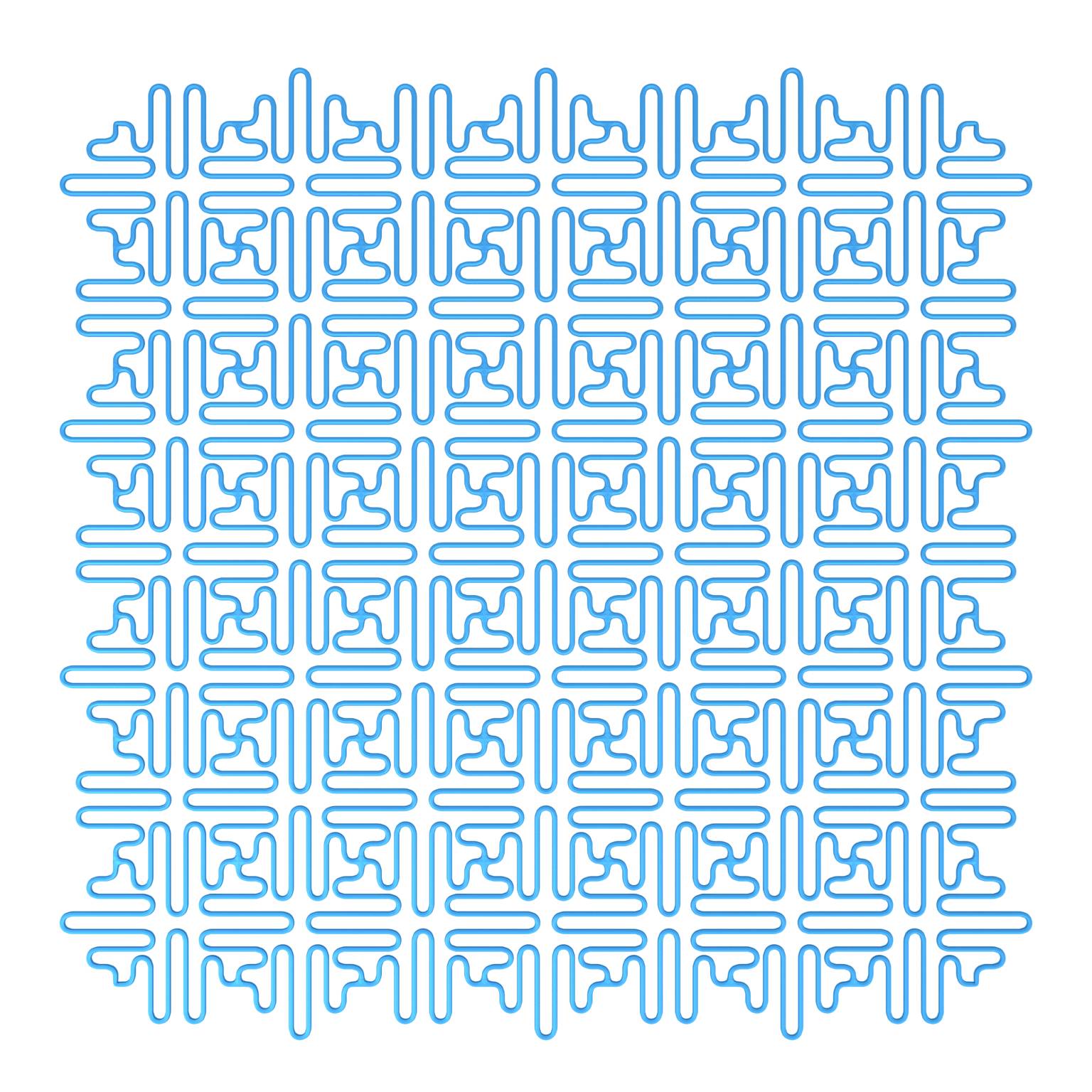}
			\caption{An example of a pattern based on a rectangular tiling. While the orientation of zigzag springs in patterns consisting of hexagons is the same for each cell, they are inverted for each pair of adjacent rectangles.}
			\label{fig:quad}
	\end{figure}
	
	The generation of a rectangular pattern is handled in a similar manner. The main difference is how the list of signs $s$ is created. Just like with the hexagonal tiling, we traverse the pattern edge-by-edge, alternating between positive and negative signs. For the rectangular tiling, however, we additionally make sure that the signs of adjacent rectangles are inverted. An example can be seen in Figure \ref{fig:quad}.

\section{Zigzag-Pattern Simplification}\label{sec:simplification}
	
As mentioned previously, simulating the deformation behavior of such complex meso-patterns has a high computational cost. Our goal is to reduce this cost by simplifying the pattern, encoding the structural properties of the complex pattern into the material parameters of the simplified pattern instead. In this section, we explain this process.
	
Consider two patterns, a complex pattern $T$ and a simple pattern $S$, that are generated from the same tiling. $T$ is generated by replacing the edges of the tiling with zigzag springs (see Section \ref{sec:patterns}), while the simple pattern $S$ is created by subdividing the edges. The deformation behavior of these patterns can then be simulated, which is explained in Section \ref{sub:simulation}. Within the simulation, we treat each edge of the original tiling as a flexible rod that is connected to other rods at the vertices of the original tiling. Therefore, both the complex pattern $T$ and the simple pattern $S$ have the same number of rods and the same number of connections where the ends of these rods meet.  {We denote the position of these connections as $V_T$ and $V_S$, and the material directions at these connections as $N_T$ and $N_S$ for the complex pattern $T$ and simple pattern $S$ respectively. The material directions describe the orientation of the cross-section around the rod centerline in the deformed state. In the rest state of the structure, the material frame at an edge segment is defined by 3 orthogonal vectors $d_1$, $d_2$ and $d_3$, with $d_3$ being the tangent of the rod centerline and $d_1$ and $d_2$ describing the thickness- and width directions of the rod. In the deformed state, the material direction of a rod segment is obtained by rotating the thickness direction vector $d_1$ around the centerline tangent $d_3$, taking into account the twisting deformation of the rod. Since the material directions are given for each rod segment, we obtain the material directions $N_T$ and $N_S$ at the rod connections by averaging the material directions of the incident rod segments.} Having a direct correspondence between the connections of the complex and simple pattern makes it easier to compare the deformation behavior of the two patterns by directly comparing the positions $V_T$ and material directions $N_T$ of the deformed complex pattern with $V_S$ and $N_S$ of the deformed simple pattern.

Since the structure of the simple pattern is very different from the complex pattern, the deformation behavior will also be different. For the simple pattern to imitate the deformation behavior of the complex pattern, it is necessary to find appropriate material-parameters $k_S$ that encode the structural properties of the zigzag edges of the complex pattern into the meta-material properties of the straight edges of the simple pattern. We do this by formulating an optimization problem which is explained in Section \ref{sub:material_parameters}.

	\begin{figure}[t]
			\centering
			\includegraphics[width=0.48\textwidth]{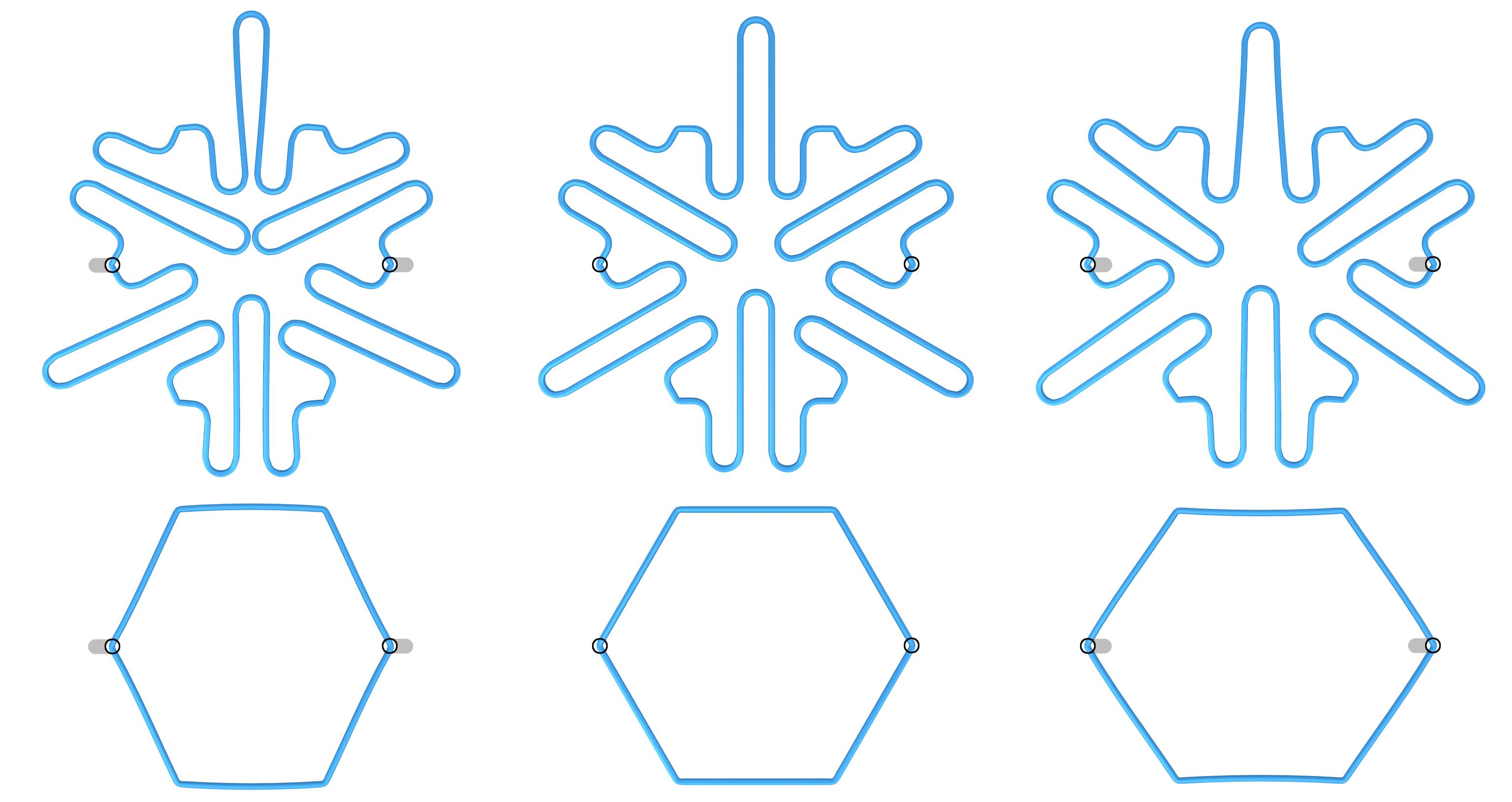}
			\caption{Deformation behavior when the anchor constraints (circles) are moved toward or away from the center. The corner vertices of the simple hexagons (bottom) match their corresponding vertices of the complex hexagons (top).}
			\label{fig:hexes}
			
	\end{figure}

\subsection{Simulation} \label{sub:simulation}
Our physical simulation is based on the Discrete Elastic Rods (DER) formulation \cite{bergou2008,bergou2010}, treating each hexagon edge as a rod. Additionally, the corner vertices of the hexagons where multiple rods come together are treated as rigid-body connections \cite{Perez2015} whose orientation is defined by additional degrees of freedom \cite{Zehnder2016}. This formulation allows the computation of the elastic stretching, bending, and twisting energies of each rod under a set of constraints, based on the geometry of the pattern, and to find the equilibrium state of the structure by minimizing the elastic energy. Building upon the framework of Vekhter et al.~\cite{vekhter}, we formulate an optimization problem to find this equilibrium state of the pattern by minimizing the energy functional
\begin{equation*}
    E_{rod} = E_r + E_a,
\end{equation*}
where $E_r$ is the internal energy of the rods consisting of stretching, bending and twisting energies and $E_a$ denotes the energy of the anchor constraints. For a detailed explanation of the internal energy, we refer the reader to the work of Bergou et al.~\cite{bergou2010}. A single anchor constraint is defined by a vector of size $9$. The first $3$ entries consist of a rod index, segment index $i$ and barycentric coordinate $\beta$ that specify the point $p = (1-\beta) p_{i} + \beta p_{i+1}$ on the corresponding rod. The other entries define the target position $p_a$ and material vector direction $m_a$, leading to the definition of the anchor constraint energy
\begin{equation}\label{eq:anchor}
E_a =  \norm{ p - {p}_a }^2 + {  \phi \left( m, m_a \right) }^2 ,
\end{equation}
with the function
\begin{equation*}
    \phi(m,m_a) = 2 \operatorname{atan2} \left( (m \times m_a)^T (m \times m_a),\norm{m} \norm{m_a} + m^T m_a \right)
\end{equation*}
returning the angle between the material vector $m$ of segment $i$ and the target direction $m_a$.

\subsection{Material Parameter Optimization} \label{sub:material_parameters}

Aside from the geometry of the pattern, the deformation behavior is determined by the respective material parameters $k_T$ and $k_S$. This includes the stiffness for stretching, bending, and twisting of the rods, as well as the thickness and width of the rod cross-section. The material parameters $k_T$ for the complex pattern are set based on the real-world material used for fabricating the pattern. Our goal is then to find the optimal material parameters $k_S$ for the simple pattern, such that the deformation behavior of the simple pattern matches that of the complex one. 

For this optimization process, we first simulate the deformation of both patterns under a set of anchor constraints $A$. We can then define the difference between the two deformed patterns as the energy
\begin{equation*}
    E_{simpl}(A, k_s) = \left\Vert V_T(A) - V_S(A, k_s) \right\Vert ^2 + \left\Vert 1 - \langle N_T(A), N_S(A, k_s) \rangle \right\Vert ^2 ,
\end{equation*}
which measures the distance between the positions of the rod's connections $V_T$ and $V_S$, as well as the difference in material directions $N_T$ and $N_S$. 

Our goal is to find material parameters $k_S$ that minimize this energy. However, a single deformed training shape might not contain enough information about the deformation behavior of the pattern. For example, an in-plane deformation of the pattern as demonstrated in Figure \ref{fig:hexes} does not yield any information about the out-of-plane bending and twisting behavior of the rods. To make sure the space of possible deformations is sufficiently covered, we therefore try to minimize the objective function over multiple sets of anchor constraints $A_i$, which we will refer to as the training set. The optimization problem we want to solve is therefore given as
\begin{equation} \label{eq:simpl}
    \min_{k_S} \sum_i E_{simpl}(A_i, k_S) \,.
\end{equation}

 {In order to evaluate the objective function and its gradients in each iteration of the optimization process, we need to simulate the deformation of the simple pattern using the updated stiffness parameters $k_S$ in each iteration. The optimization problem described in Section \ref{sub:simulation} is therefore solved once per iteration to obtain the deformed state of the pattern.}

The choice of training set $A_i$ from which we learn the material parameters is an important one. If the size of the set is too small or the resulting deformations are not varied enough, some aspects of the deformation behavior might not be properly captured by the optimized material parameters. On the other hand, a training set that is too large will unnecessarily increase the computational cost of the optimization for no benefit. We therefore decide to use 4 sets of anchor points that deform the pattern into shapes with different surface properties---one shape each exhibiting mainly positive, negative, or zero Gaussian curvature, as well as one shape with varying Gaussian curvature (as depicted in Figure \ref{fig:grids}). We evaluate our choice of the training set in Section \ref{sub:results_training} by comparing the accuracy and computational cost to the use of larger training sets.

For the optimization process, we make use of the Sequential Quadratic Programming (SQP) algorithm~\cite{nocedal2006numerical} to minimize the objective function. Since we only want to optimize a small set of parameters, we compute the gradient of the objective function numerically.

\section{Shape Design and Optimization}\label{sec:optimization}

Once we have found the appropriate material parameter values for the simplified pattern, we can use them for applications like shape design. As an example, we will consider a pattern $P$ that can be deformed by constraining the position and material direction at specified points on the pattern using a set of anchors $A$. 

As explained in Section \ref{sub:simulation}, anchors $A$ are given as a matrix with one row per anchor and $9$ columns. For each anchor, the first $3$ columns contain the rod index, segment index, and barycentric coordinate $\beta$ that specify the point $p$ on the pattern that is constrained. These values remain fixed during the optimization process. The remaining entries are the target position and material direction. 

Our goal is now to find the optimal anchor positions and material directions such that pattern $P$ in the deformed state matches the deformed target pattern $Q$. 

\subsection{Optimization Problem}

Let $p$ and $q$ be the respective vertex positions and $m$ and $n$ be the respective material frame vectors for patterns $P$ and $Q$. Similar to the pattern simplification problem, we formulate the optimization problem as
\begin{equation}\label{eq:shape_optim}
    \min_{A} E_{shape}(A)\,,
\end{equation}
where 
\begin{equation}\label{eq:shape_energy}
    E_{shape}(A) = \left\Vert q - p(A) \right\Vert ^2 + \left\Vert 1 - \langle n, m(A) \rangle \right\Vert ^2 \,.
\end{equation}
To solve this optimization problem, we first simulate the deformation of the pattern $P$ using an initial guess for the anchors $A_0$. With this deformed pattern $P_0$ as an initial state, we again use the SQP-algorithm to solve the optimization problem. Within each iteration $i$, the deformation of the pattern is simulated and the objective function and its gradient are evaluated to compute the anchors $A_{i+1}$ for the next iteration. Furthermore, since we need to optimize $6$ variables per anchor, we compute the analytical gradient of the objective function, which is explained in the next section. 

\subsection{Gradient} \label{sub:gradient}

Let $x$ be the vector of all state variables of the DER simulation (in our case the vertex positions $p$ and material frame vectors $m$) and let $u$ be the vector of design variables (in our case the target positions and material directions of anchors $A$). Then the analytical gradient of the objective function is given by
\begin{equation} \label{eq:shapegrad}
	\frac{\partial E_{shape}}{\partial u} = \frac{\partial E_{shape}}{\partial x} \frac{\partial x}{\partial u}
\end{equation} 
The derivative $\partial x / \partial u$ is unknown. However, we can compute this derivative by means of sensitivity analysis in a similar manner to \cite{malomo} and \cite{Panetta2019}. We want any change of the design variables to preserve the equilibrium state of the deformed pattern, meaning that the derivative of the forces $f$ acting on the pattern should vanish. Thus we have
\begin{equation*}
	\delta f = \frac{\partial^2 E_{rod}}{\partial x \partial x} \frac{\partial x}{\partial u} \delta u + \frac{\partial^2 E_{rod}}{\partial x \partial u} \delta u = 0,
\end{equation*} 
with $E_{rod}$ being the elastic energy of the pattern in the deformed state. We can now solve for $\partial x / \partial u$ directly with
\begin{equation}\label{eq:direct}
	\frac{\partial x}{\partial u} = - \left( \frac{\partial^2 E_{rod}}{\partial x \partial x} \right)^{-1} \frac{\partial^2 E_{rod}}{\partial x \partial u},
\end{equation}
or use the more efficient adjoint method by inserting Equation \eqref{eq:direct} into Equation \eqref{eq:shapegrad} to get
\begin{equation*}
	\frac{\partial E_{shape}}{\partial u} = - \frac{\partial E_{shape}}{\partial x} \left( \frac{\partial^2 E_{rod}}{\partial x \partial x} \right)^{-1} \frac{\partial^2 E_{rod}}{\partial x \partial u} = \lambda^{T} \frac{\partial^2 E_{rod}}{\partial x \partial u},
\end{equation*} 
and solving for $\lambda$:
\begin{equation*}
	\lambda = -  \left( \left( \frac{\partial^2 E_{rod}}{\partial x \partial x} \right)^{-1} \right)^T \left( \frac{\partial E_{shape}}{\partial x} \right) ^{T}
\end{equation*}

Based on the DER formulation, we compute the bending and twisting energies at each vertex, the stretching energies at each edge, as well as the position- and direction constraint energies for each anchor. For ease of implementation, we only compute the square root of these terms and store them in the vector $r$, such that the total elastic energy of the system is then given by $E_{rod} = r^T r$. The first and second derivatives of the elastic energy are therefore given by
\begin{equation*}
    \frac{\partial E_{rod}}{\partial x} = 2  J_x^T r,
\end{equation*} 
\begin{equation*}
    \frac{\partial^2 E_{rod}}{\partial x \partial x} = 2 \left( J_x^T J_x + r^T H_{xx} \right) ,
\end{equation*} 
\begin{equation*}
    \frac{\partial^2 E_{rod}}{\partial x \partial u} = 2 \left( J_x^T J_u + r^T H_{xu} \right) ,
\end{equation*} 
with $J$ being the Jacobian matrix 
\begin{equation*}
    J_x =  \frac{\partial r}{\partial x},
\end{equation*} 
and $H$ being the 3-dimensional Hessian tensor 
\begin{equation*}
H_{xx} = \frac{\partial^2r}{\partial x \partial x} \,.
\end{equation*}
For a detailed explanation of the derivatives of the stretching, bending, and twisting energies we refer the reader to the work of Panetta et al.~\cite{Panetta2019}, particularly the supplementary material. For the derivatives of the anchor constraint terms, we refer the interested reader to Appendix \ref{app:derivatives}.

\section{Results}\label{sec:results}

In this section, we perform a number of experiments to evaluate our proposed framework. We first demonstrate our pattern simplification method in Section \ref{sub:results_stiffness} and analyze the importance of the choice of training shapes in Section \ref{sub:results_training}. We then demonstrate the advantage of pattern simplification by using the simplified pattern for the task of shape optimization and comparing its computational cost to the cost of optimization with  {the original zigzag pattern} in Section \ref{sub:results_shape}.
    
\subsection{Simplification Results} \label{sub:results_stiffness}

\begin{figure*}
\begin{centering}
    \begin{subfigure}{0.23\textwidth}
		\includegraphics[width=\textwidth]{./figures_compressed/blue_dome_hi}
		\includegraphics[width=\textwidth]{./figures_compressed/blue_saddle_hi}
		\includegraphics[width=\textwidth]{./figures_compressed/blue_tunnel_hi}
		\includegraphics[width=\textwidth]{./figures_compressed/blue_manta_hi}
		\caption{Training shape}
	\end{subfigure}
	\begin{subfigure}{0.23\textwidth}
		\includegraphics[width=\textwidth]{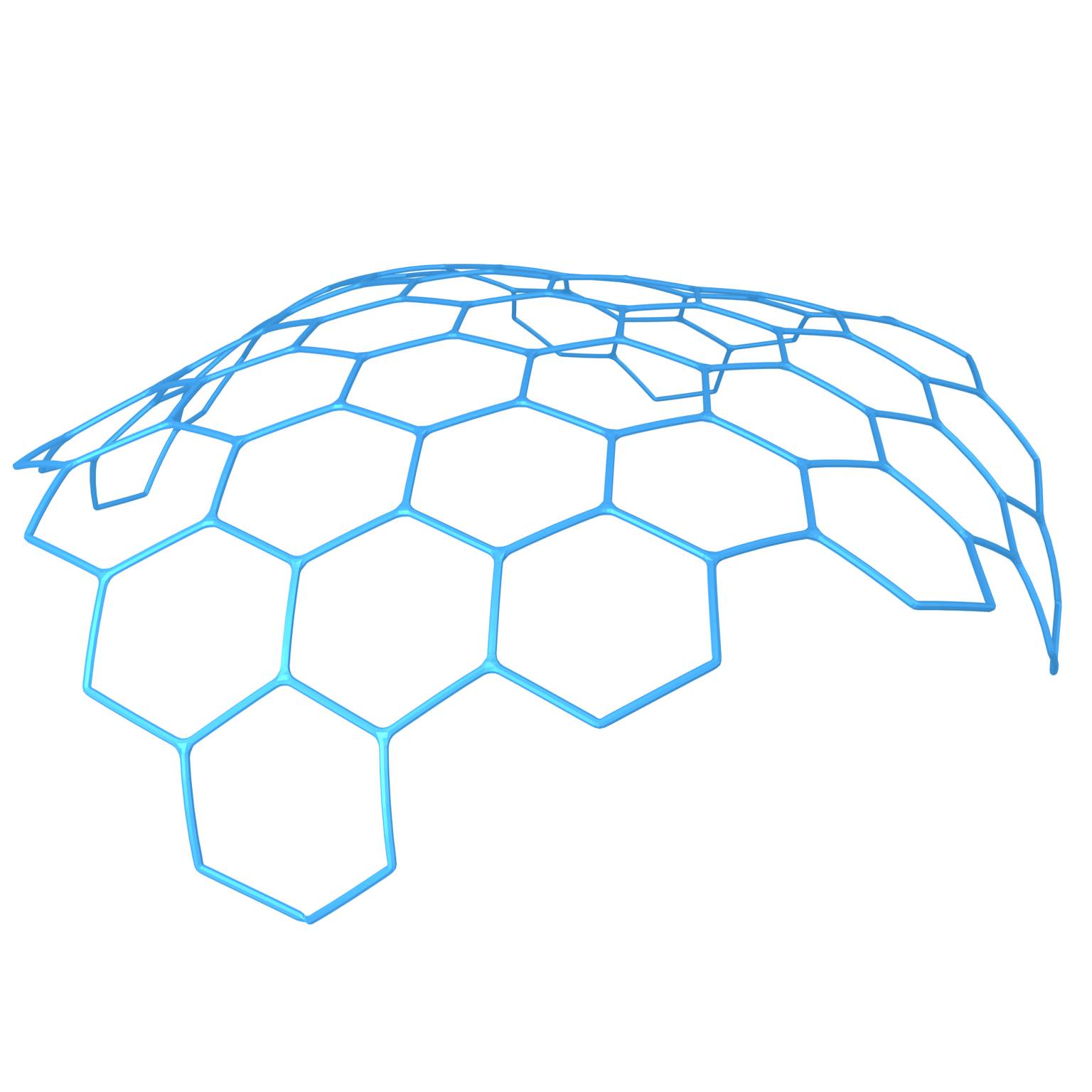}
		\includegraphics[width=\textwidth]{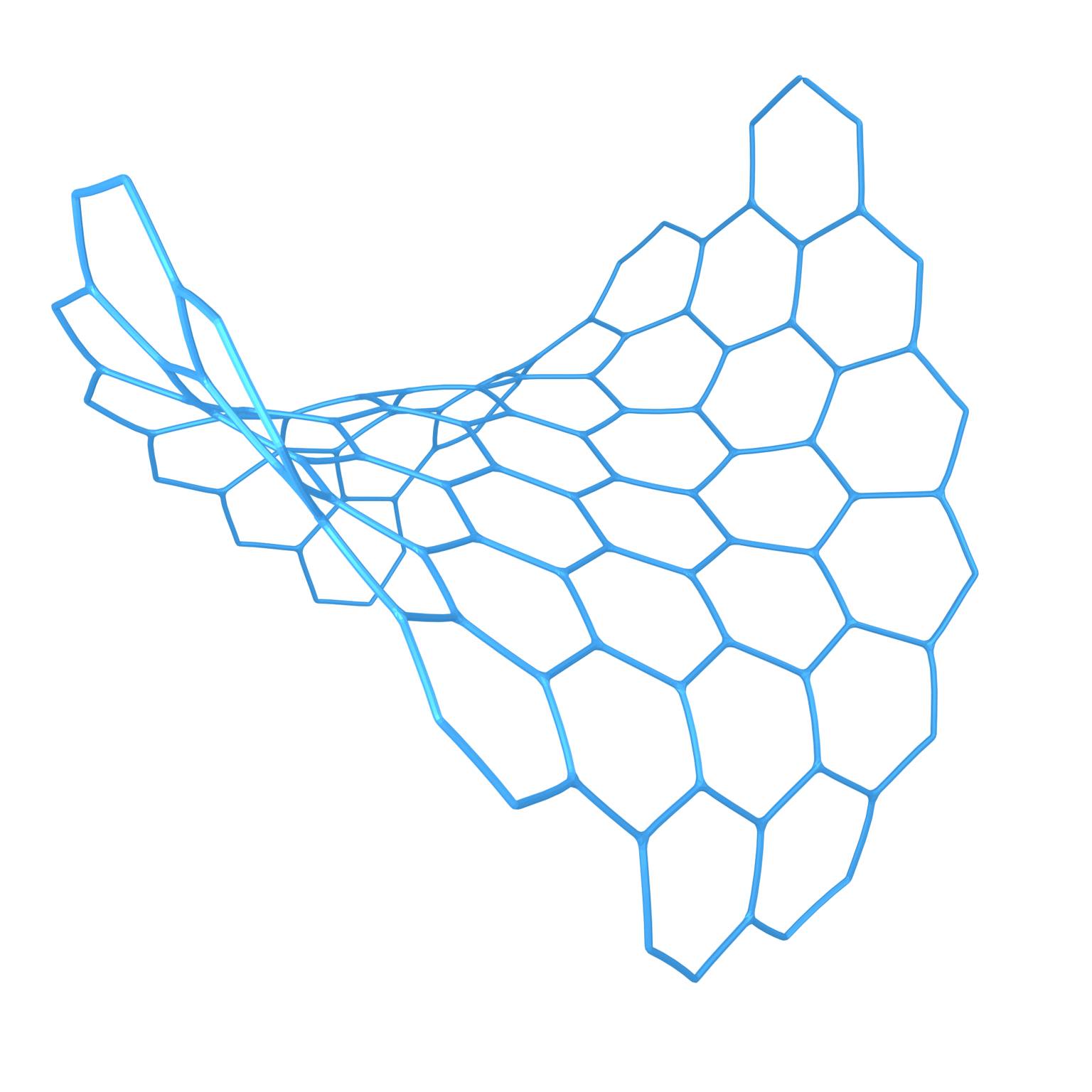}
		\includegraphics[width=\textwidth]{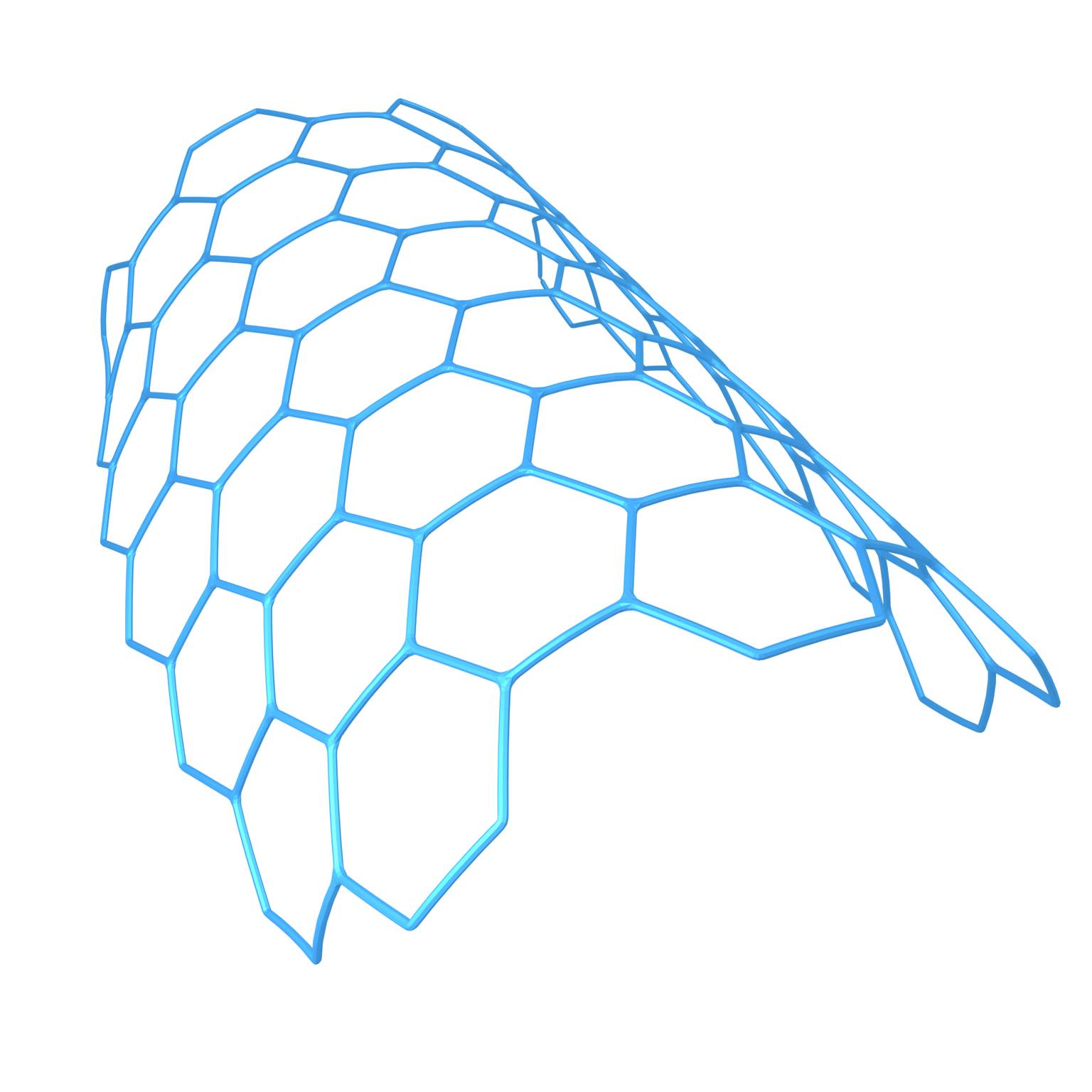}
		\includegraphics[width=\textwidth]{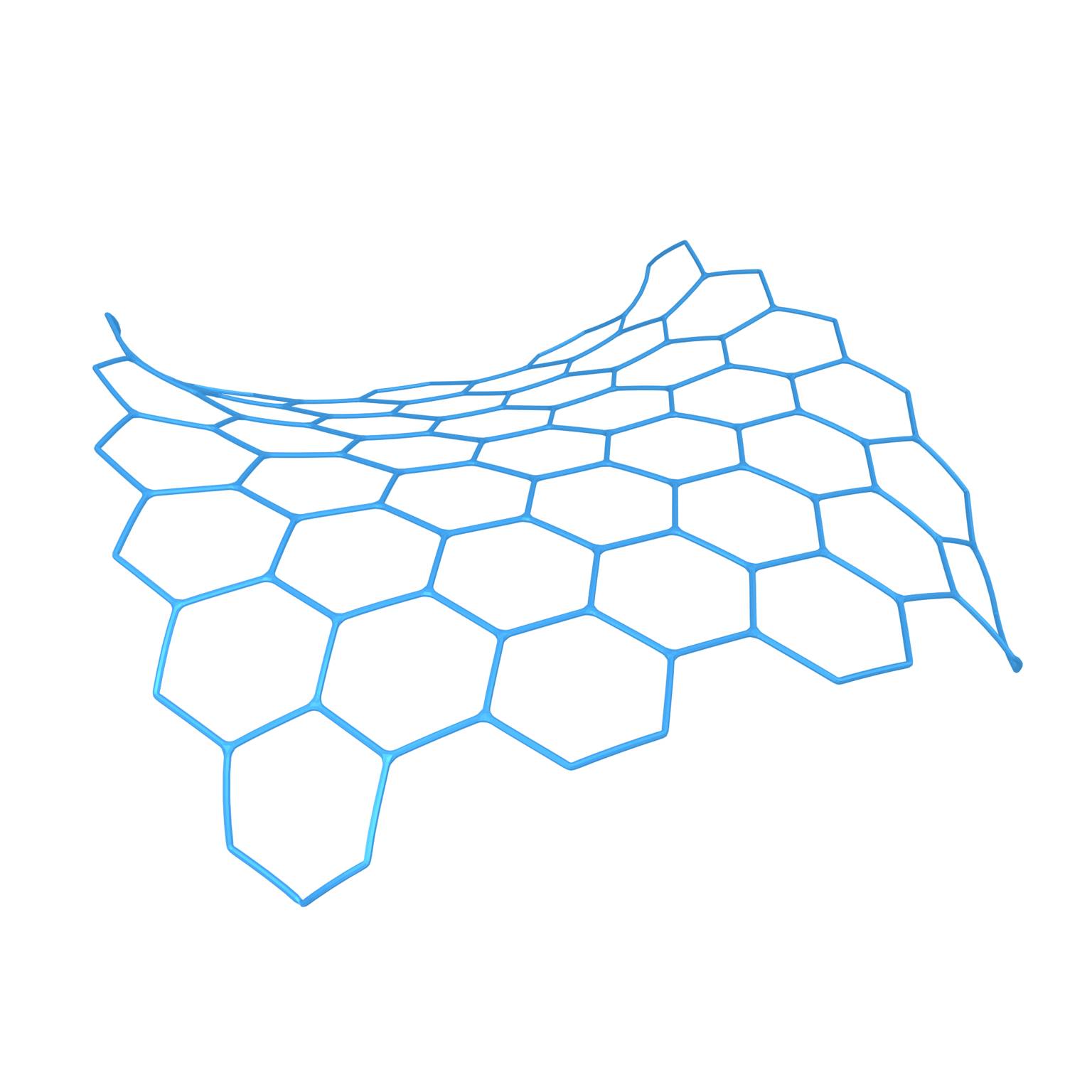}
		\caption{Initial stiffness}
	\end{subfigure}
	\begin{subfigure}{0.23\textwidth}
		\includegraphics[width=\textwidth]{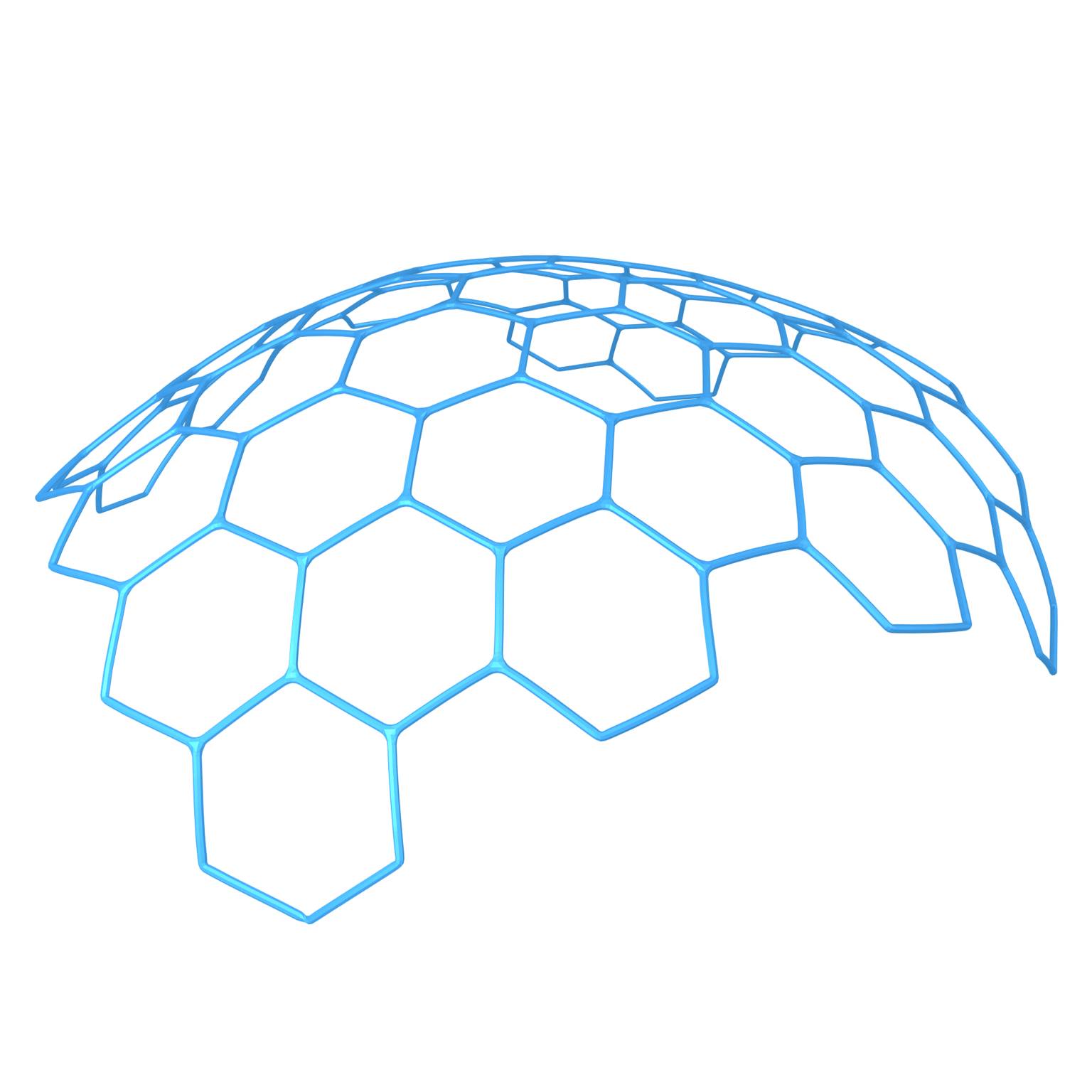}
		\includegraphics[width=\textwidth]{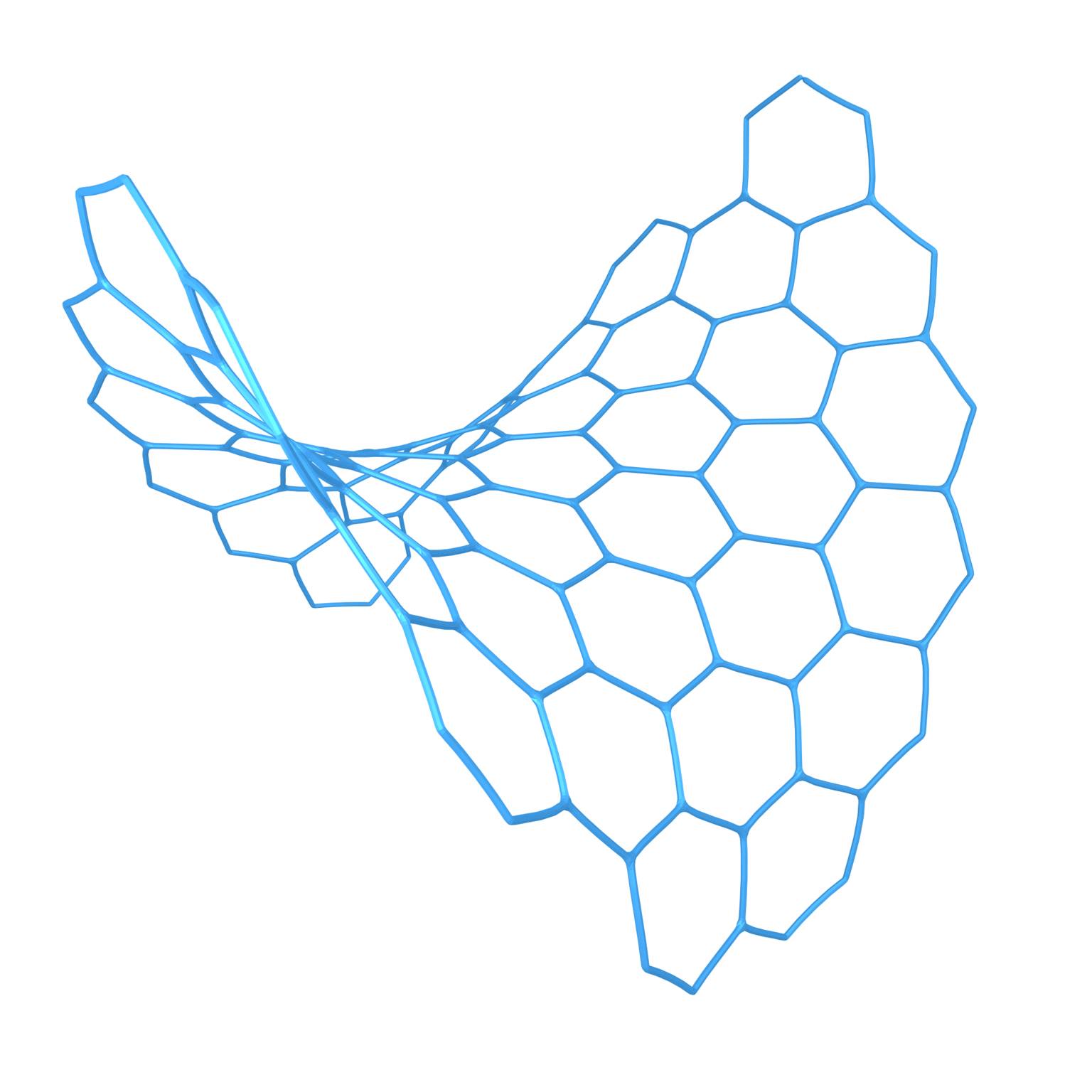}
		\includegraphics[width=\textwidth]{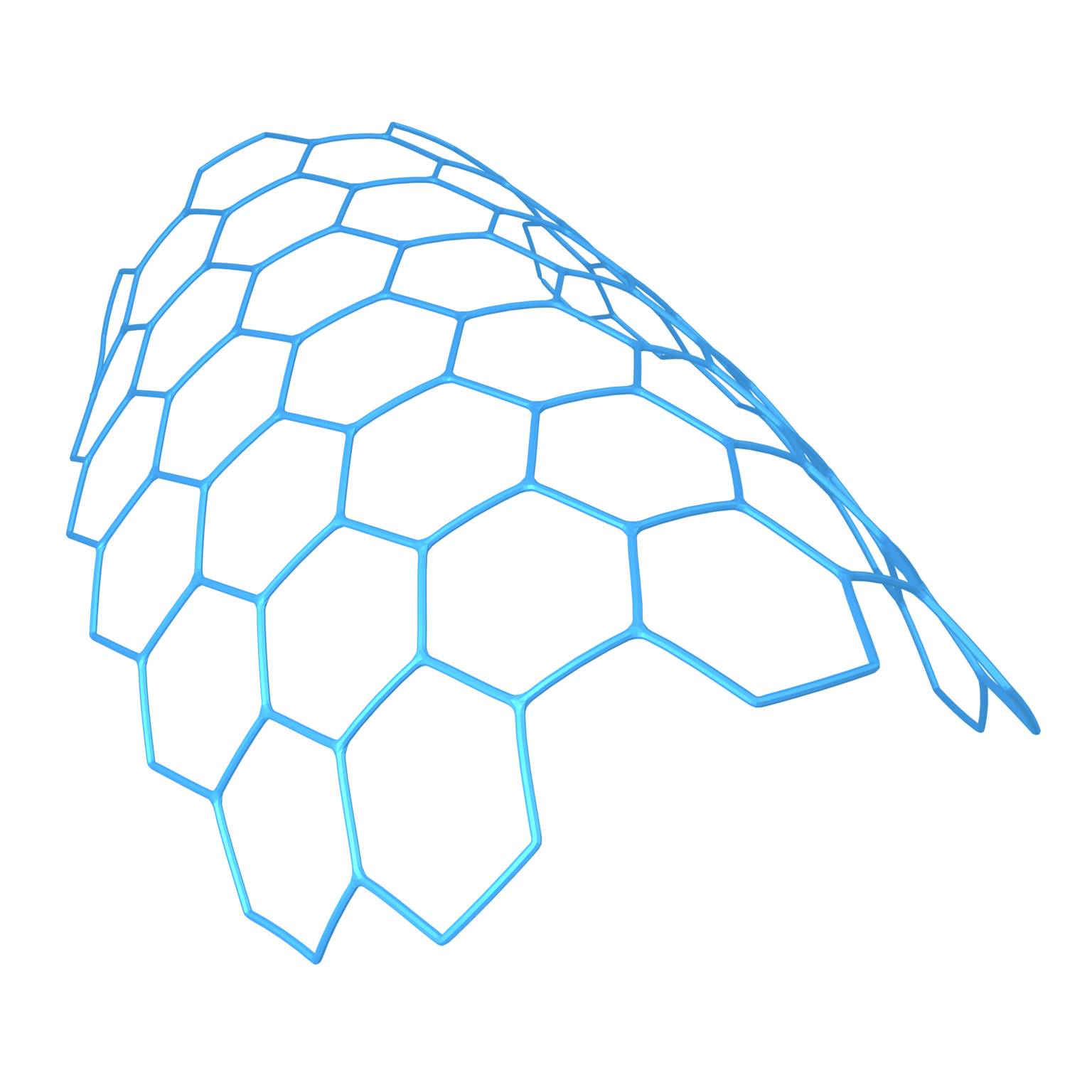}
		\includegraphics[width=\textwidth]{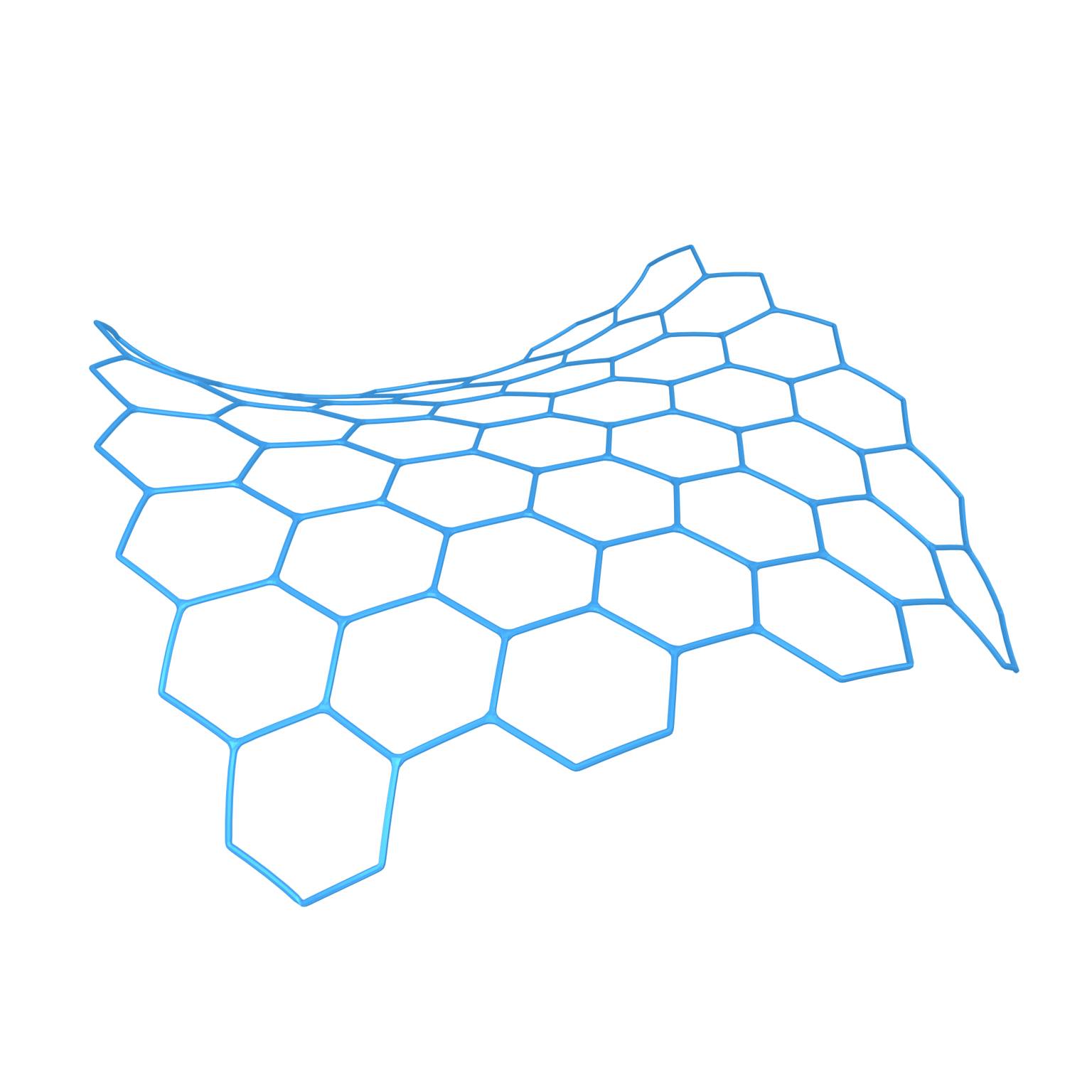}
		\caption{Intermediate result}
	\end{subfigure}
	\begin{subfigure}{0.23\textwidth}
		\includegraphics[width=\textwidth]{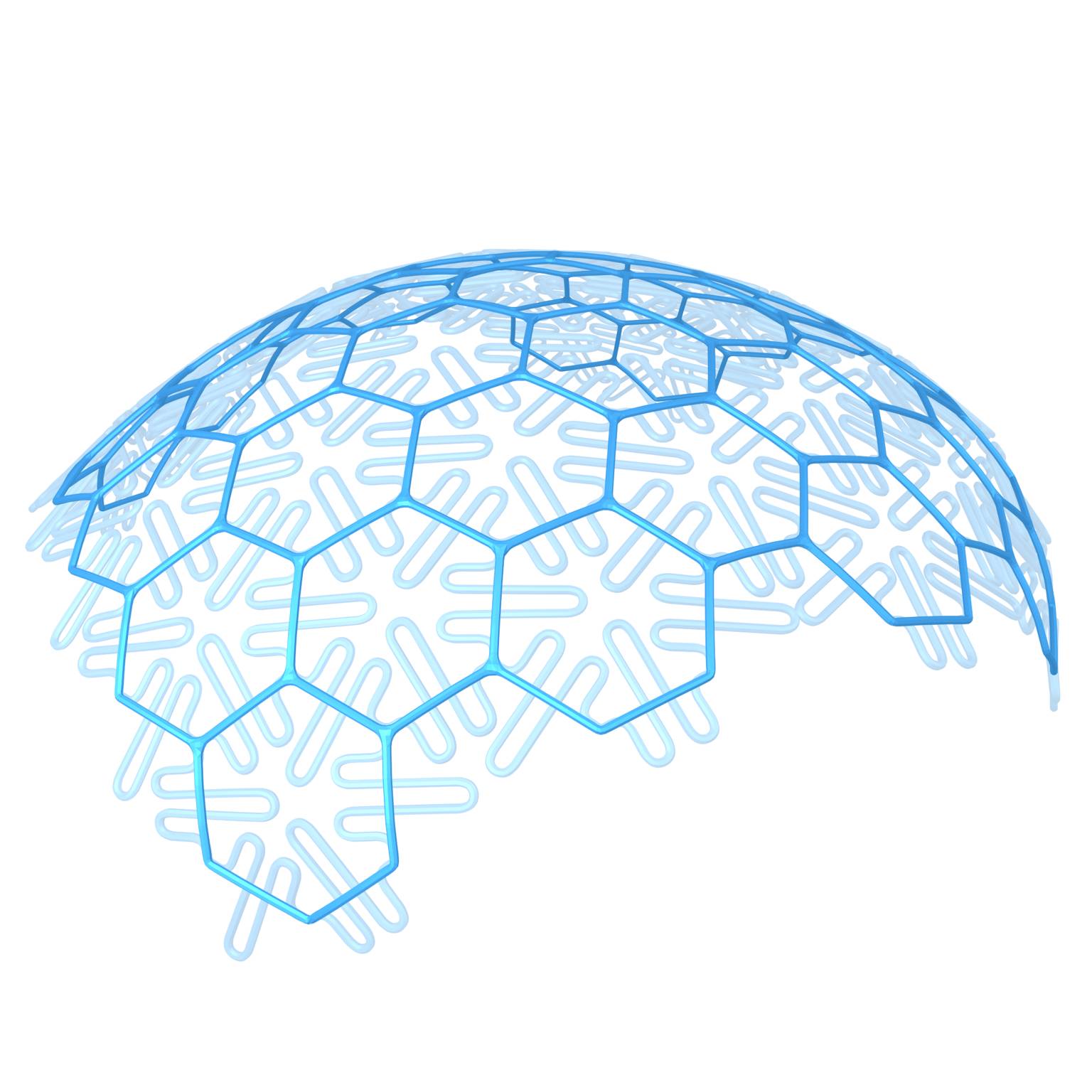}
		\includegraphics[width=\textwidth]{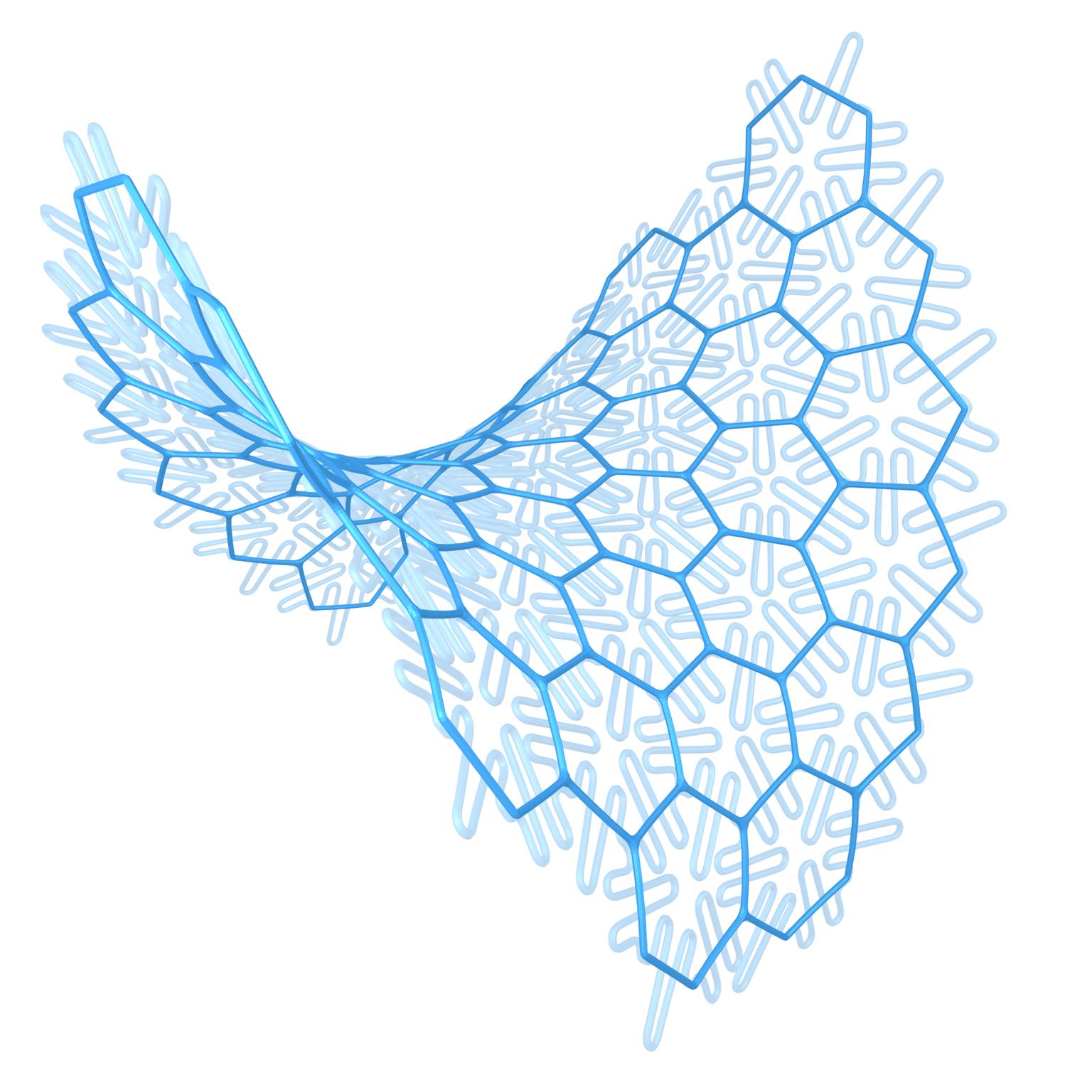}
		\includegraphics[width=\textwidth]{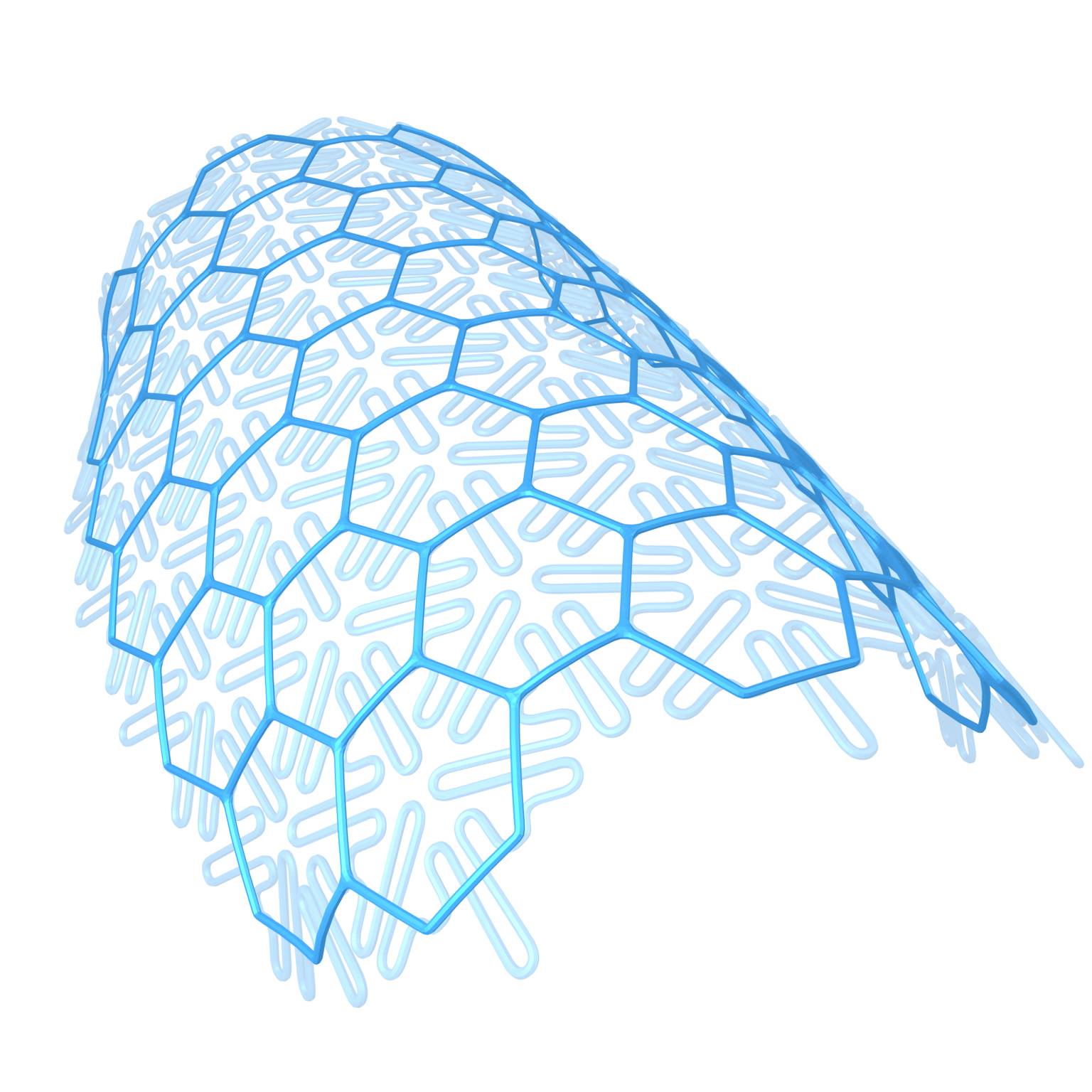}
		\includegraphics[width=\textwidth]{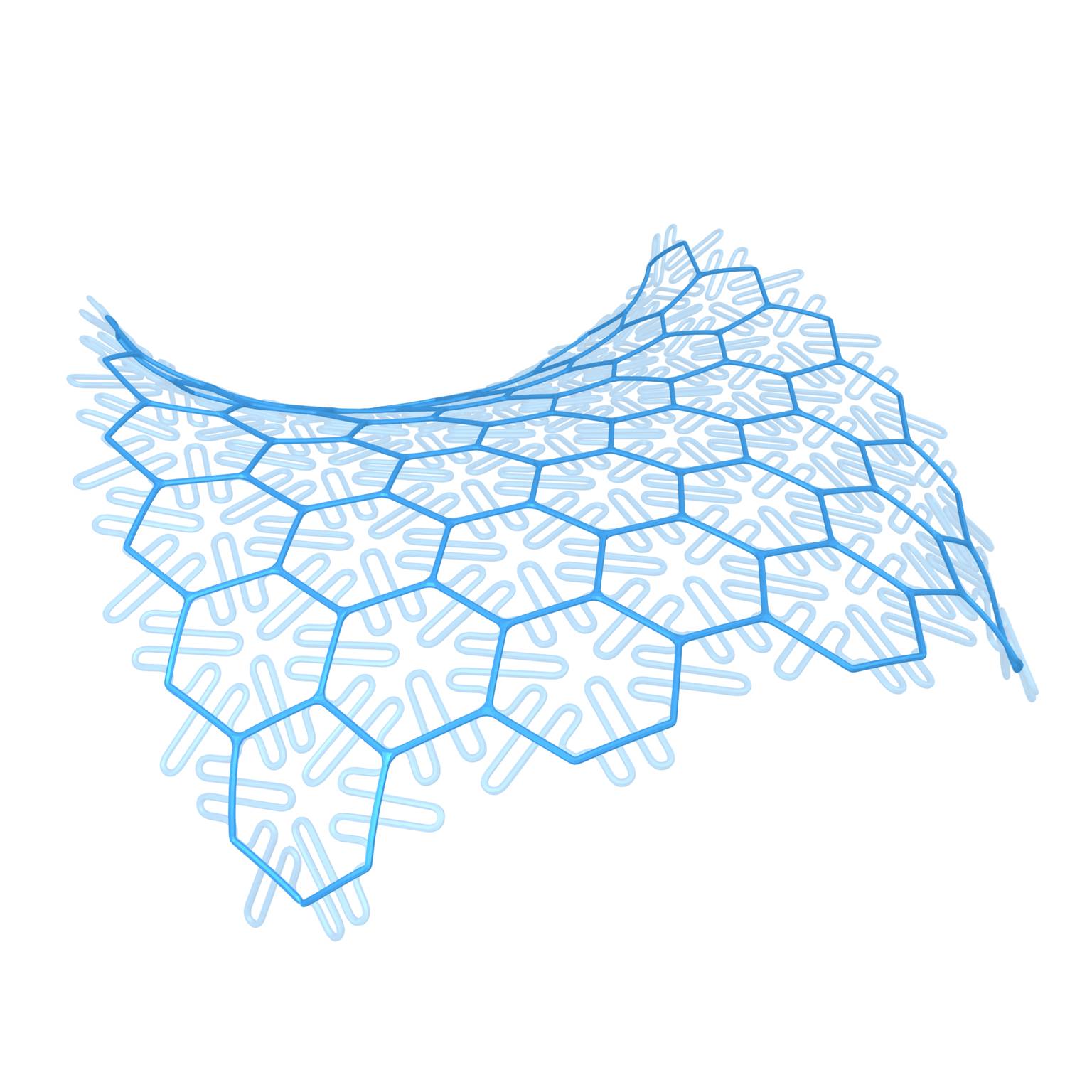}
		\caption{Overlay of final result}
	\end{subfigure}
	\caption{Results of the material parameter optimization. (a)  {Deformed zigzag pattern serving as the training shape}. (b) Simple pattern with initial material parameter values. (c) Simple pattern after a few iterations of optimization. (d) Overlay of optimized simple pattern with training shape.}
	\label{fig:stiffness_results}
	\end{centering}
\end{figure*}

The top of Figure \ref{fig:hexes} shows an example of how a single  {hexagon cell containing zigzag-springs} deforms when moving the two anchor points marked by circles along the x-axis. For the simplified hexagon cell (bottom), this in-plane bending deformation corresponds to both stretching and in-plane bending of the edges. By finding appropriate material stiffness parameters, the coordinates of the corner vertices become identical for  {both the original cell with zigzag-springs and the simplified cell.}

Figure \ref{fig:stiffness_results} shows larger patterns made of $7x6$ hexagon cells simulated using our approach.  {The zigzag patterns} on the left contain $12643$ vertices and $12684$ edge segments, while the simplified patterns on the right only contain $563$ vertices and $604$ edge segments, significantly reducing the computational cost of the simulation while yielding a good approximation of the resulting deformed patterns.
	
As an experiment, we use the proposed method to find the optimal parameters for the patterns shown in Figure \ref{fig:local_stiffness}. In all examples we use a structure of $7 \times 6$ hexagon cells with a circumradius of $7 mm$ and origin coordinates $c_0 = (0,0)$. The cross-section of  {the original pattern with zigzag springs} is $0.6 mm$ wide and $3 mm$ thick. For this experiment, we set the stiffness parameters with regard to stretching, bending, and twisting to $10^{10}$, $10^6$, and $10^6$ respectively. Together with the width and thickness of the cross-section, they form the material parameters $k_T$. 

For the simple pattern, we subdivide each hexagon edge into $4$ segments of equal length. With this choice, the x-coordinates of the vertices between the edge segments coincide with the peaks of the zigzag springs in the original pattern, where they are most flexible. Increasing the number of edge segments can increase the accuracy of the simulation, especially when using patterns with larger cells or zigzag springs with more windings, but also increases the computational cost. We evaluate the influence of this parameter in Table \ref{tab:numsegs}. 

The material parameters $k_S$ which we want to optimize are initialized as $10^6$ for the stretching, bending, and twisting stiffness and $1mm$ for both the thickness and width of the cross-section. We choose these parameter values as our initial guess since the edges of the simplified pattern are straight and discretized at a lower resolution than those of  {the original zigzag pattern}. Therefore, using the same material parameter values as in the original pattern for the initialization would make the pattern significantly less flexible, leading to worse convergence in our experiments. Please also note that we use only one set of material parameters for the entire pattern since the  {zigzag pattern} used in our experiments uses the same type of zigzag spring for every hexagon edge. However, it would also be possible to optimize for per-rod or even per-segment parameters if necessary. 

As explained in Section \ref{sub:material_parameters}, we would like to use shapes with different surface properties as examples to learn from in order to properly cover the range of possible deformations of the pattern. We therefore define $4$ sets of anchors to deform the patterns into the shapes shown on the left of Figure \ref{fig:local_stiffness}, which we will refer to as Dome, Saddle, Tunnel, and Manta from top to bottom. These shapes fulfill our aim of using surfaces of mainly positive, negative, zero, and varying Gaussian curvature respectively. The structure of the pattern and the vertices constrained by the anchors are the same for each training shape, only the prescribed positions and material directions at  {the constrained vertices} differ for each set of anchors. Please refer to Figure \ref{fig:grids} for a comparison to the fabricated examples.
We first simulate the  {original zigzag pattern} for each set of anchors to obtain the training shapes $T_j$ with $j=1,\dots,4$.

To evaluate the objective function, the simple pattern is simulated once for each set of anchors to obtain $S_j$ for the current iteration of the optimization. Using all four shapes as targets, the optimization process takes $291$ seconds on average and the resulting material parameters $k_S$ are $0.034 \cdot 10^6$, $0.79 \cdot 10^6$ and $1.48 \cdot 10^6$ for the stretching, bending, and twisting stiffness respectively, as well as cross-sectional thickness $1.54mm$ and width $0.61mm$. Table \ref{tab:objfun} shows the average distance between corresponding corner vertices of  {the original zigzag pattern and its simplified version} in millimeters (the size of the full pattern is $87mm$ by $83mm$). As an additional experiment, we perform the optimization using only a single training shape each. The computation times range from $66$ seconds for the Dome shape to $107$ seconds for the Manta shape. As can be seen in Table \ref{tab:objfun}, the quality of the optimization result greatly depends on the used training shape. Using only the Dome or Saddle shapes as targets results in significantly worse results overall, while the Tunnel or Manta shapes yield a better approximation. This suggests that the deformations of the Tunnel and Manta shapes contain more information about the deformation behavior of the pattern than the Dome or Saddle shapes. To account for the possibility of overfitting, we also evaluate our results against 3 larger sets of shapes (cf. Section \ref{sub:results_shape} and Table \ref{tab:numshapes}).

To compare the computational cost of the original and simplified patterns, we simulate the deformation of both patterns for each of the four training shapes and record the time until an equilibrium state is reached. For the  {original pattern with zigzag springs}, simulation takes $16.04$, $21.81$, $21.35$, and $30.68$ seconds for the Dome, Saddle, Tunnel, and Manta shapes respectively. For the simplified patterns, we obtain simulation times of $0.85$, $0.87$, $0.61$, and $1.07$ seconds respectively, which corresponds to a reduction of computation time ranging from $94.70$ to $97.13$ percent.

The approximation error can be reduced further by discretizing the hexagon edges of the simplified pattern at a higher resolution, with the drawback of increasing the computational cost of simulating the deformation of the pattern. Table \ref{tab:numsegs} shows the approximation error and computational cost using 2, 4, 8, or 16 segments per hexagon edge. As can be seen, the gain in accuracy is diminishing compared to the computational cost the more the resolution is increased. On the other hand, reducing the resolution even further does not result in a significant gain in speed, affirming our choice of using 4 segments per edge.

\subsection{Analysis of Training Set Influence} \label{sub:results_training}

\begin{table}
\caption{Influence of discretization resolution for the simplified pattern. We perform the material parameter optimization using 2, 4, 8 and or 16 segments for each hexagon edge. The columns show the average distance between vertices for the four target shapes, the number of optimization iterations, optimization time and time for simulating the deformation of the resulting pattern.}  \label{tab:numsegs}
\begin{tabular}{ c | c c c c}
 nSegs & Avg. Dist. & Iterations & Opt. Time & Sim. Time  \\
 \hline
 $2$ & $0.2966mm$ & $20$ & $195s$ & $0.75s$\\  
 $4$ & $0.1505mm$ & $18$ & $297s$ & $0.85s$\\ 
 $8$ & $0.0889mm$ & $21$ & $603s$ & $1.77s$ \\
 $16$ & $0.0620mm$ & $16$ & $1235s$ & $4.44s$ \\
\end{tabular}
\end{table}

\begin{table}
  \begin{center}
  \caption{Range of offsets used to generate the sets of training shapes. Distances are given in millimeters.}
    \begin{tabular}{c | c c c c}
  \label{tab:offsets}
    Shape & $\Bar{\rho}_p$ & $\Bar{\theta}_p$ & $\Bar{r}_p$ & $\Bar{\theta}_n$ \\
    \hline
    Elliptic & $[-\frac{\pi}{16},\frac{\pi}{16}]$ & $[-\frac{\pi}{8},\frac{\pi}{8}]$ & $[0,10]$ & $[-\frac{\pi}{4},\frac{\pi}{4}]$  \\
    Hyperbolic & $[-\frac{\pi}{32},\frac{\pi}{32}]$ & $[-\frac{\pi}{4},\frac{\pi}{4}]$ & $[2.5,7.5]$ & $[-\frac{\pi}{8},\frac{\pi}{8}]$  \\
    Cylindric & $[-\frac{\pi}{8},0]$ & $[-\frac{\pi}{8},\frac{\pi}{8}]$ & $[-5,5]$ & $[-\frac{\pi}{4},0]$
    \end{tabular}
    \end{center}
\end{table}

We furthermore evaluate our choice of training shapes for the optimization of the material parameters. For this purpose, we perform the optimization process using larger sets of training shapes, 
where we create 3 sets of shapes roughly corresponding to elliptic, hyperbolic, and cylindric surfaces by sampling possible positions and material directions for the anchor constraints.

\begin{figure} 
\centering
		\includegraphics[width=0.35\textwidth,trim=0cm 0cm 0cm 0cm, clip]{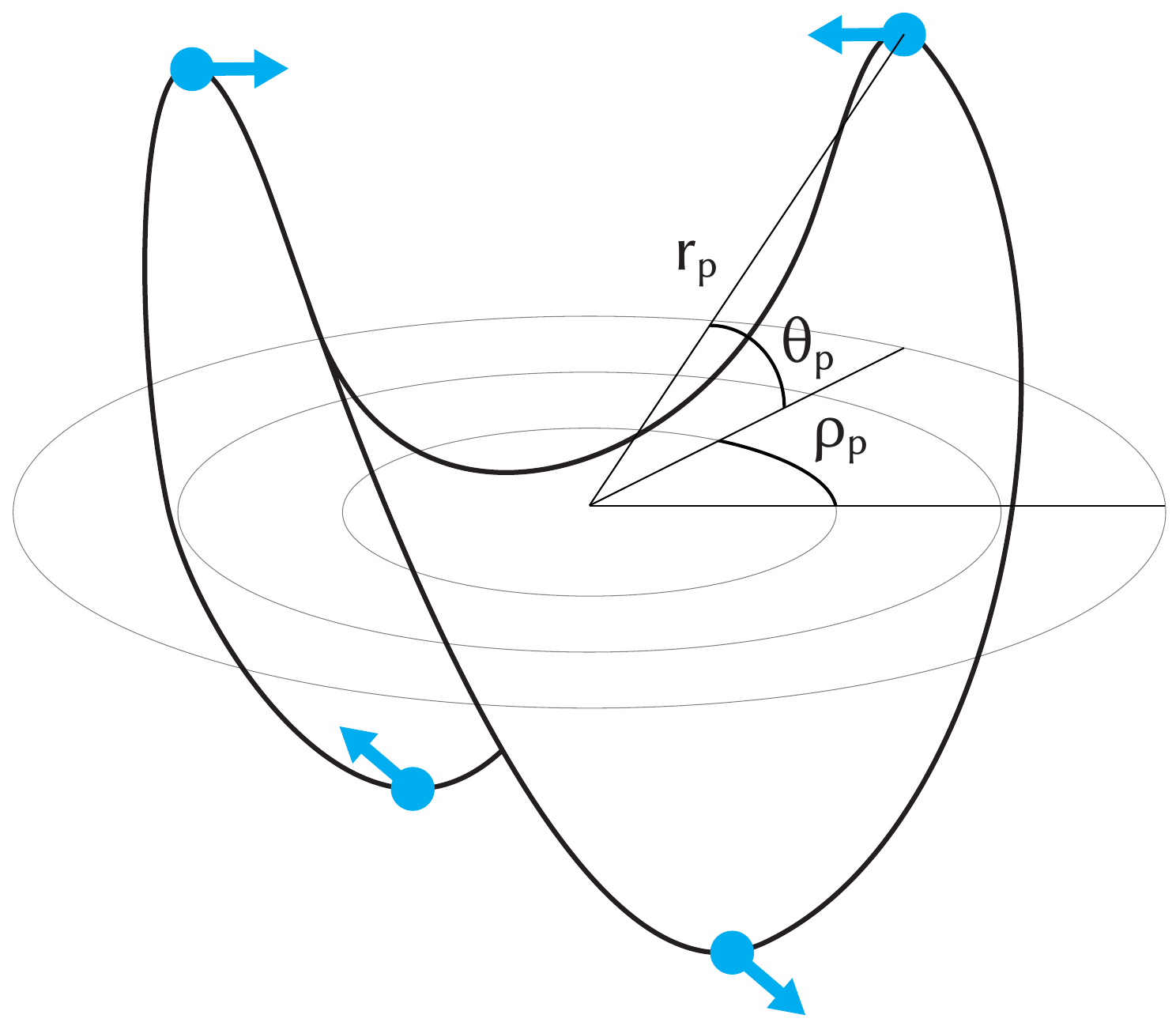}
\caption{Parameterization of the training samples with spherical polar coordinates $\rho_p$, $\theta_p$ and $r_p$. This allows to sample offsets $(\Bar{\rho}_p, \Bar{\theta}_p, \Bar{r}_p, \Bar{\theta}_n)$ efficiently and to apply them to the anchor coordinates (refer to Section~\ref{sub:results_training}). }
	\label{fig:polar}
\end{figure}

\begin{figure}
\centering

		\includegraphics[width=0.17\textwidth,trim=0cm 3cm 0cm 0cm, clip]{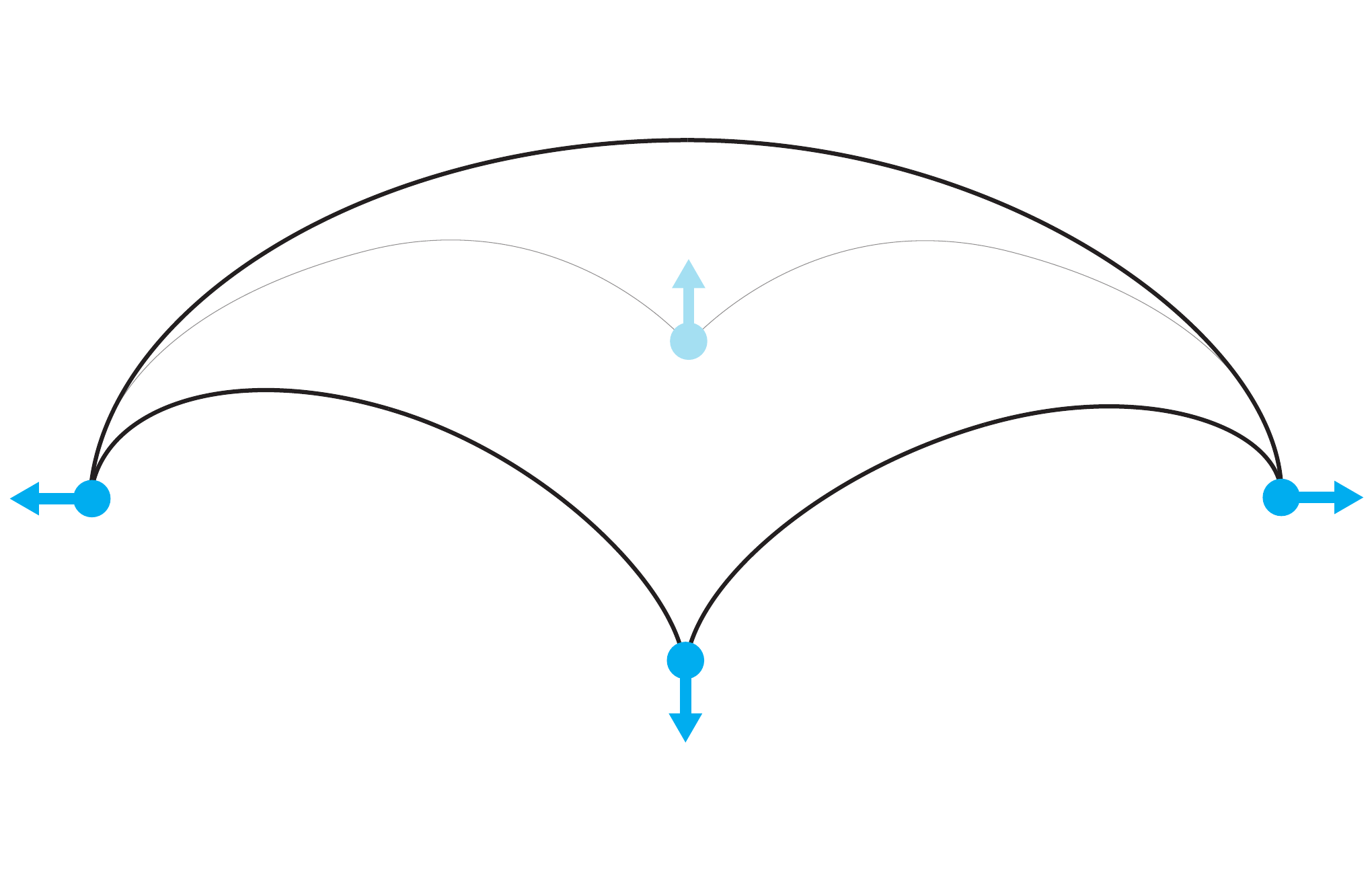}
		\hfill
		\includegraphics[width=0.12\textwidth,trim=5cm 0cm 5cm 0cm, clip]{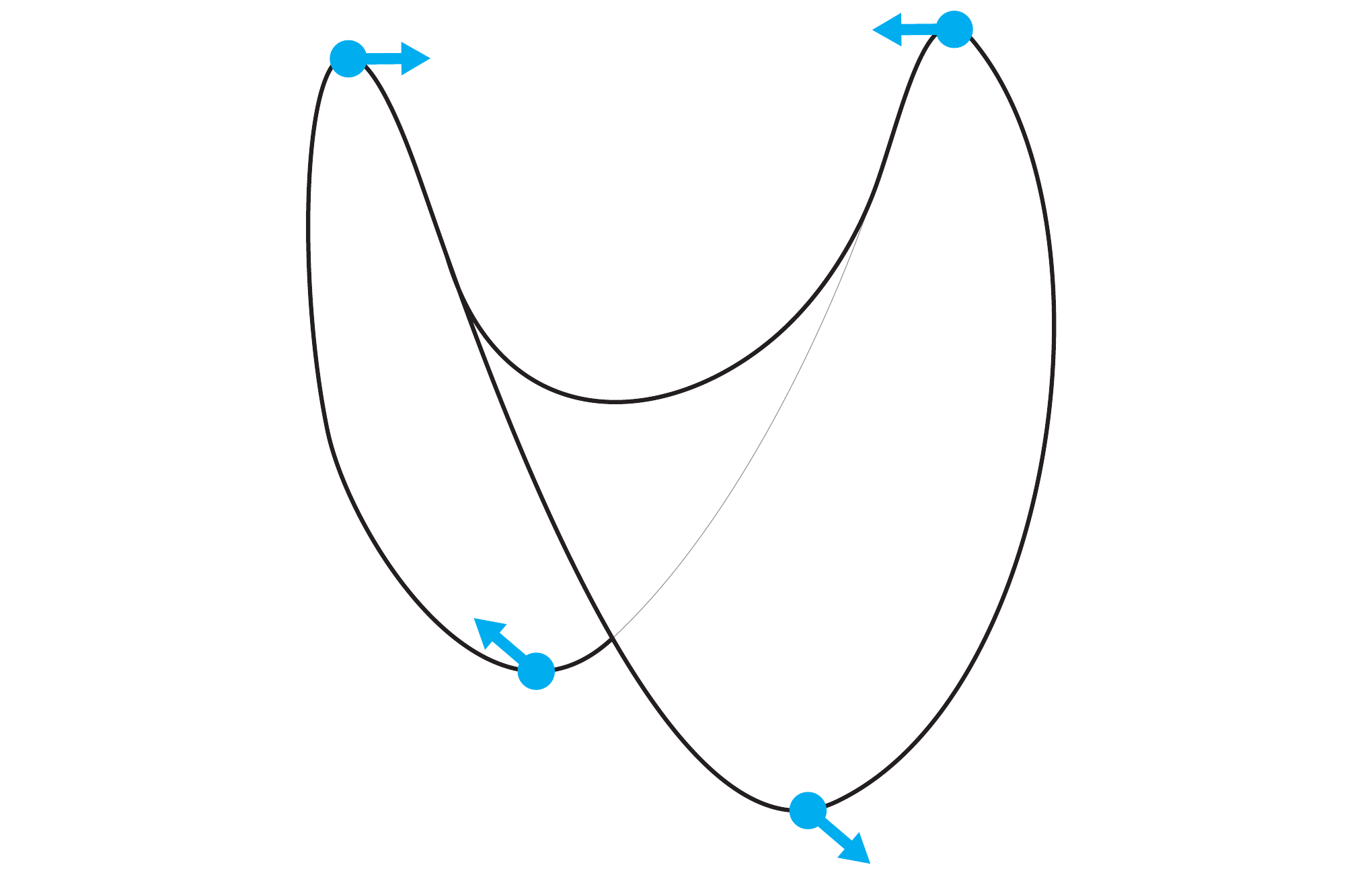}
		\hfill
		\includegraphics[width=0.17\textwidth,trim=0cm 3cm 0cm 0cm, clip]{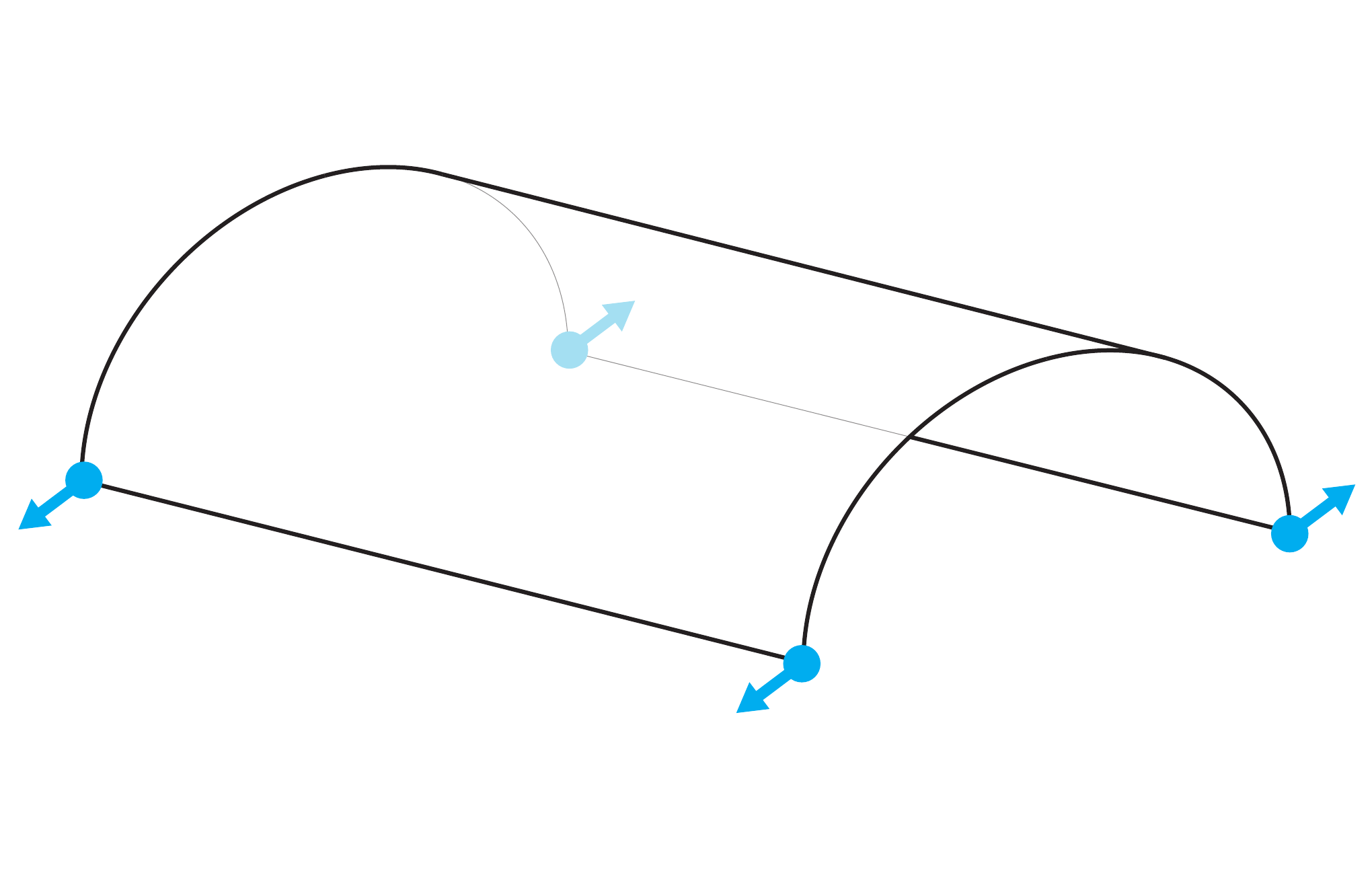}
	\caption{Outlines and initial anchor positions and material directions for the three sets of shapes used to evaluate our choice of training shapes. From left to right: Elliptic Hyperbolic, Cylindric.}
	\label{fig:anchors}
\end{figure}

For generating these sets of constraints it is useful to view the anchor positions in a spherical coordinate system with the origin at the center of the pattern, since changes to the spherical coordinates $\rho_p$, $\theta_p$ and $r_p$ are similar to in-plane bending, out-of-plane bending and stretching of the pattern respectively. 
The anchor material directions can also be converted to spherical coordinates $\rho_n$, $\theta_n$ and $r_n = 1$. 
We initialize these parameters based on the surface type (please refer to Figures \ref{fig:polar} and \ref{fig:anchors} for a visual depiction which shows the anchor positions and material directions in blue). 
To create the shape variation within a set, we sample offsets $(\Bar{\rho}_p, \Bar{\theta}_p, \Bar{r}_p, \Bar{\theta}_n)$ and apply them to the anchor coordinates. We define the range of these offset parameters empirically to avoid implausible deformations since our simulation does not account for the possibility of material failure (please refer to Table \ref{tab:offsets} for the range of parameters for each set).
We then uniformly sample this parameter space to create the shape variations, taking 3, 4, 3, and 3 samples each and forming every possible combination to create 108 shapes per set. Additionally, we create a mixed set by randomly choosing 36 shapes from each of the 3 sets.

The material parameter optimization process is performed once for each set of training shapes and the results are evaluated on every set. The computation times for the optimization are $12931s$, $4649s$, $24670s$, and $14401s$ for the Elliptic, Hyperbolic, Cylindric, and Mixed sets respectively.

We furthermore compare the accuracy of the optimized material parameters to the results obtained in our initial experiment using 4 training shapes for the optimization. The results in Table \ref{tab:numshapes} show that the increased number of training shapes does not lead to an overall improvement in accuracy, especially considering the increased computational cost of the optimization process. Therefore, we conclude that a small set of training shapes with varying surface properties is sufficient to capture the deformation behavior of  {the original pattern with zigzag springs}.

\begin{table}[t]
\caption{Evaluation of the optimized stiffness parameters. Each row shows which training shape was used in the optimization. The values show the average distance between corresponding corner vertices of the  {zigzag pattern and its simplified version} in millimeters (the size of the full pattern is 87mm by 83mm).  \label{tab:objfun}}
\begin{center}
\begin{tabular}{ c | c c c c }
 Training & \multicolumn{4}{c}{Evaluated Shape} \\
 Shape & Dome & Saddle & Tunnel & Manta  \\
 \hline
 All & $0.0940$ & $0.1271$ & $0.1525$ & $0.2284$ \\
 Dome & $0.0823$ & $0.1696$ & $0.1826$ & $0.3901$ \\  
 Saddle & $2.2855$ & $0.2737$ & $3.0581$ & $1.5823$ \\ 
 Tunnel & $0.2273$ & $0.1996$  & $0.1234$ & $0.2401$ \\
 Manta & $0.1523$ & $0.1787$  & $0.2036$ & $0.2062$

\end{tabular}
\end{center}
\end{table}

\begin{table}
  \begin{center}
  \caption{Evaluation of our choice of training shapes (bottom row) for the material parameter optimization. Performing the optimization process using larger training sets (108 shapes each) shows that the increased number of shapes does not result in a better fit of the material parameters. The values show the average distance between corresponding corner vertices of the  {original pattern with zigzag springs and the simplified pattern} in millimeters.}
    \begin{tabular}{c | c c c}
  \label{tab:numshapes}
    Training & \multicolumn{3}{c}{Evaluated Set} \\
    Set & Elliptic & Hyperbolic & Cylindric  \\
    \hline
    Elliptic & 0.3321 & 0.1560 & 0.2559 \\
    Hyperbolic & 2.9984 & 1.2974 & 1.6355 \\
    Cylindric & 0.3165 & 0.1719 & 0.2376 \\
    Mixed & 0.3494 & 0.1575 & 0.3241 \\
    4 Shapes & 0.2978 & 0.1451 & 0.2704 \\
    \end{tabular}
    \end{center}
\end{table}

\subsection{Shape Optimization Results} \label{sub:results_shape}

\begin{figure}[t]
    \includegraphics[width=0.48\textwidth]{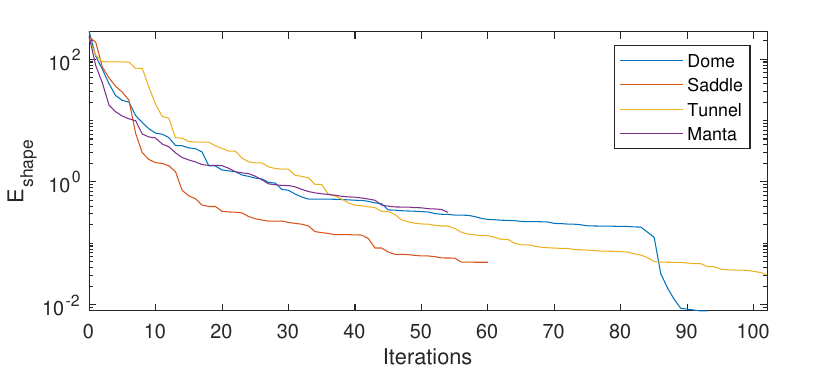}
	\caption{Plot of the objective value $E_{shape}$ during shape optimization. The x-axis shows the number of iterations, while the y-axis shows the objective value on a logarithmic scale.}
	\label{fig:plot_eshape}
\end{figure}

\begin{figure*}[t] 
\begin{centering}
    \begin{subfigure}{0.23\textwidth}
		\includegraphics[width=\textwidth,trim=0 10cm 0 10cm, clip]{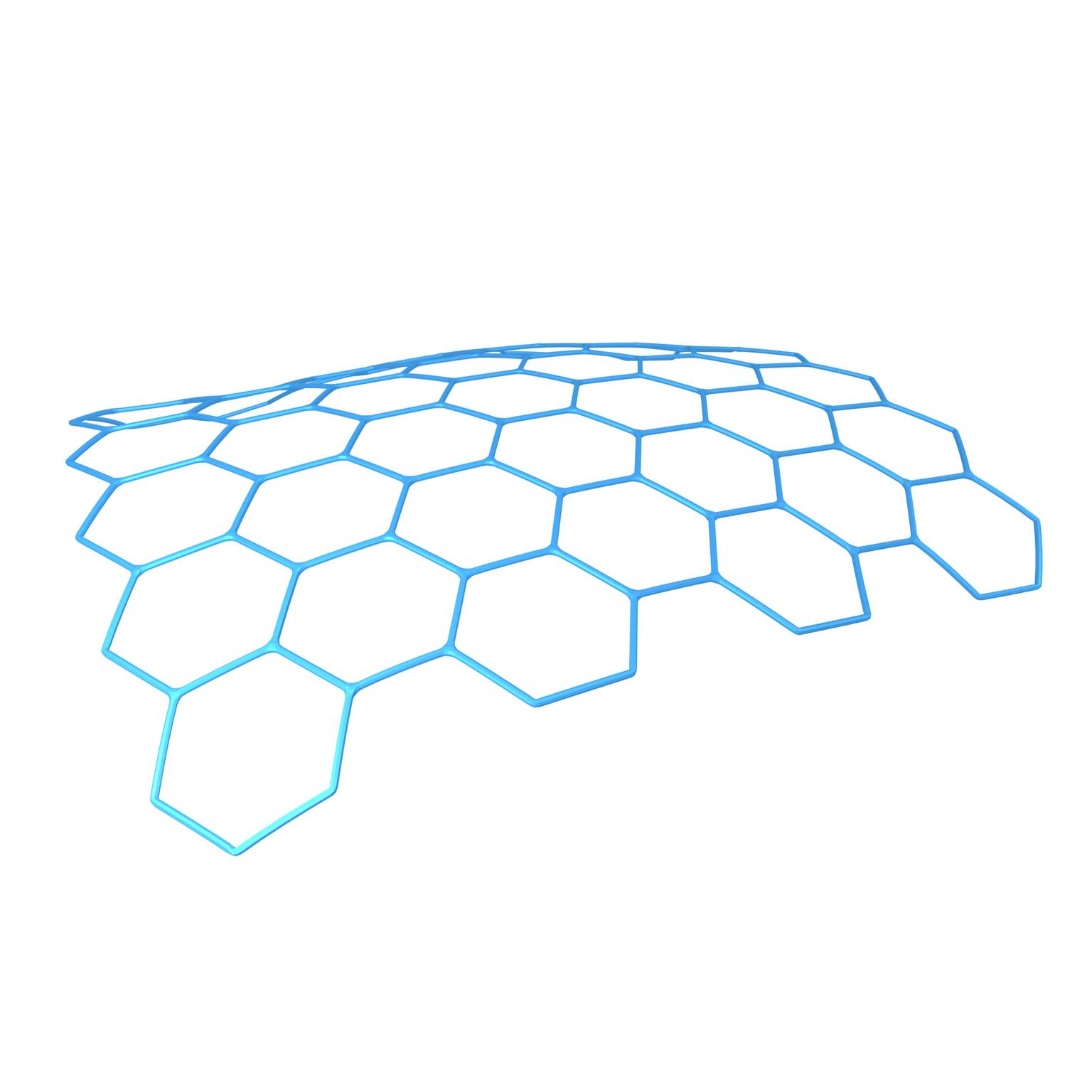}
		\includegraphics[width=\textwidth,trim=0 8cm 0 10cm, clip]{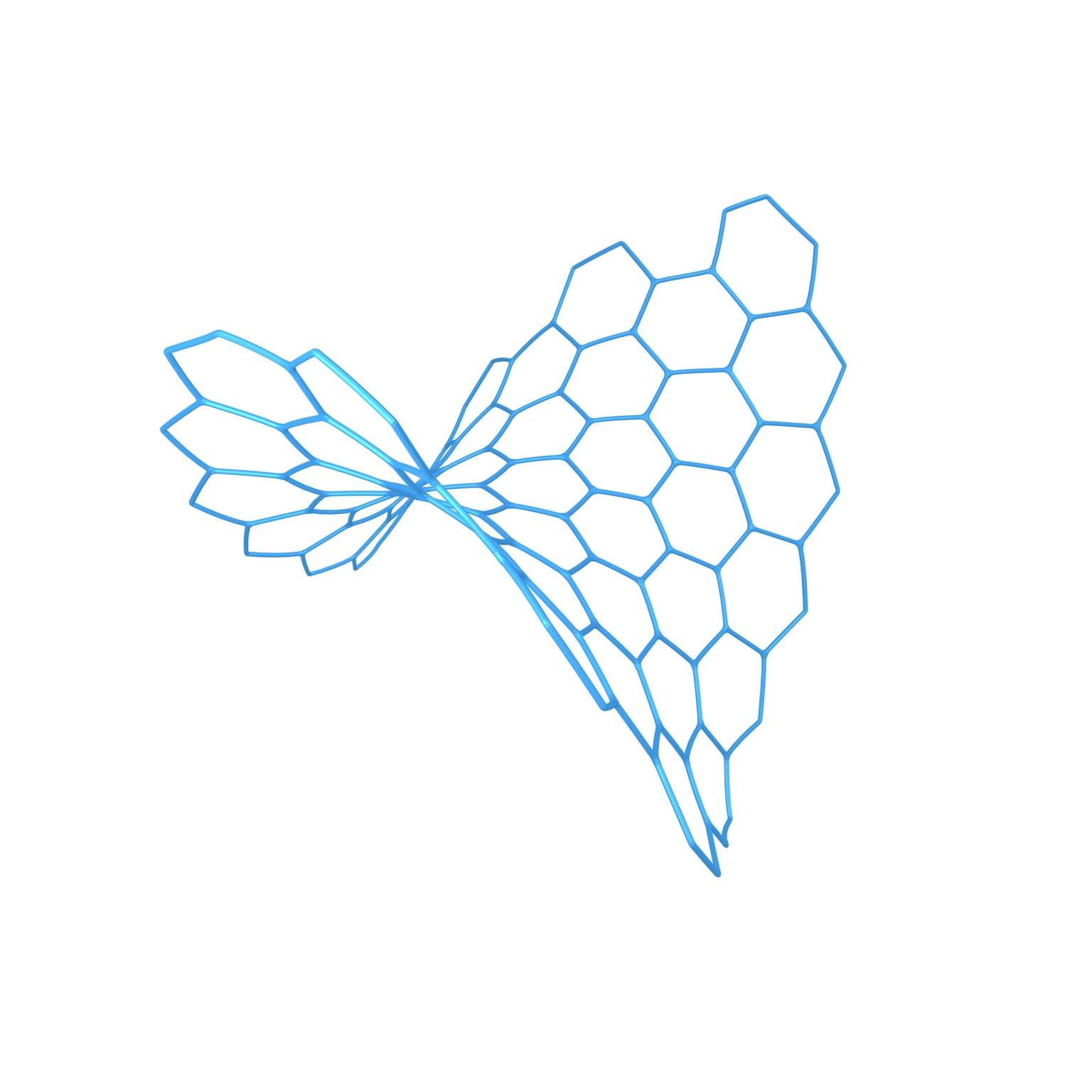}
		\includegraphics[width=\textwidth,trim=0 12cm 0 8cm, clip]{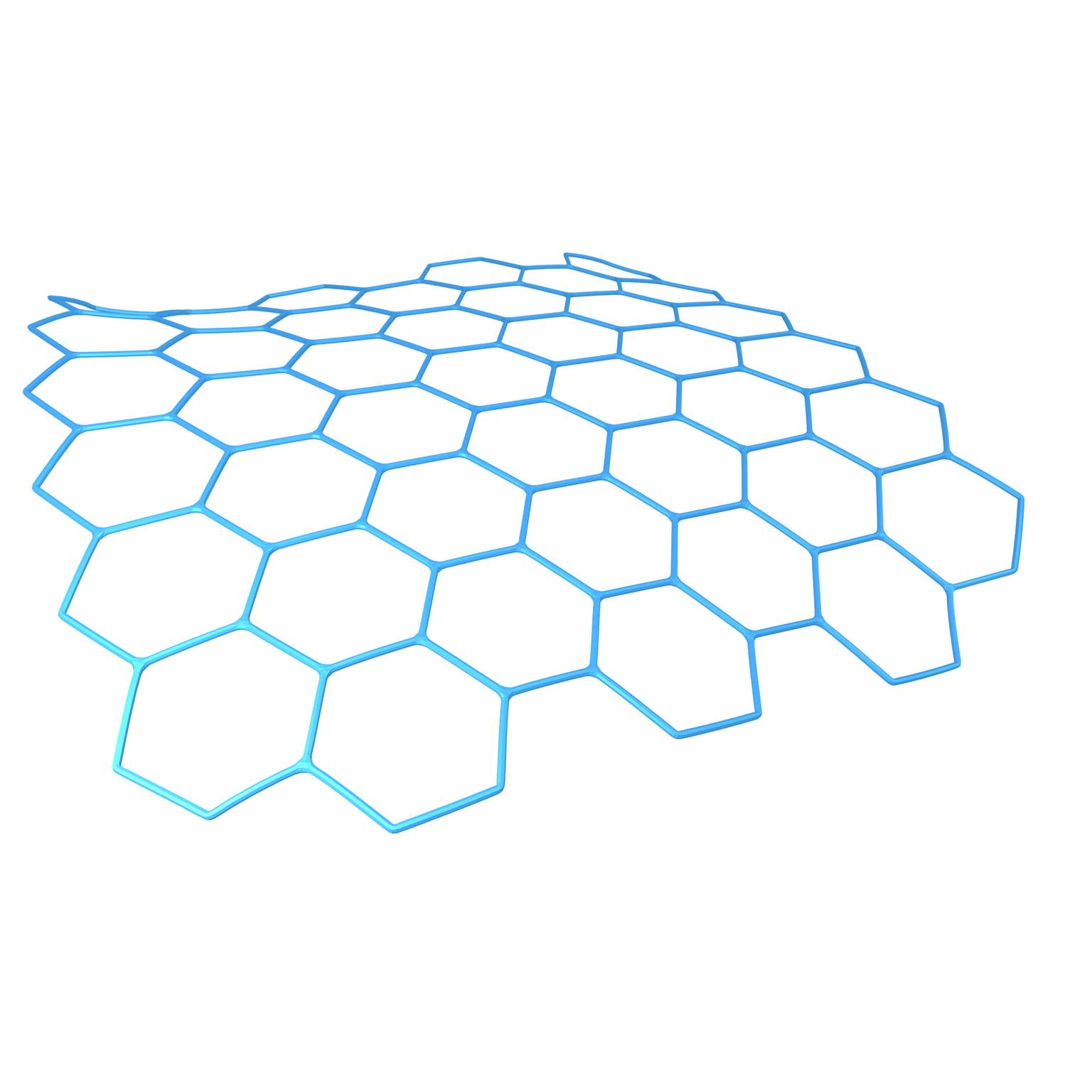}
		\includegraphics[width=\textwidth,trim=0 10cm 0 12cm, clip]{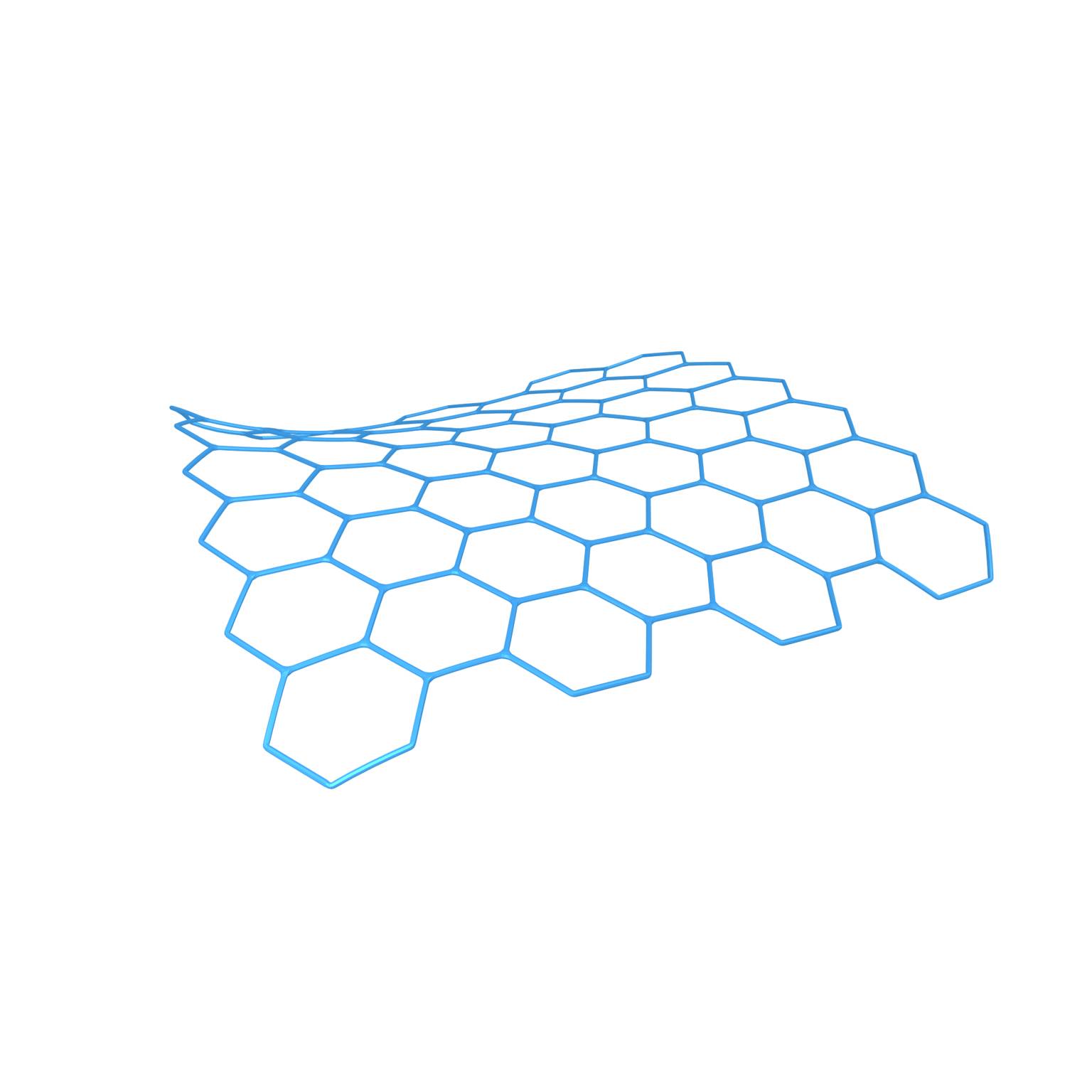}
		\caption{5 Iterations}
	\end{subfigure}
	\begin{subfigure}{0.23\textwidth}
		\includegraphics[width=\textwidth,trim=0 10cm 0 10cm, clip]{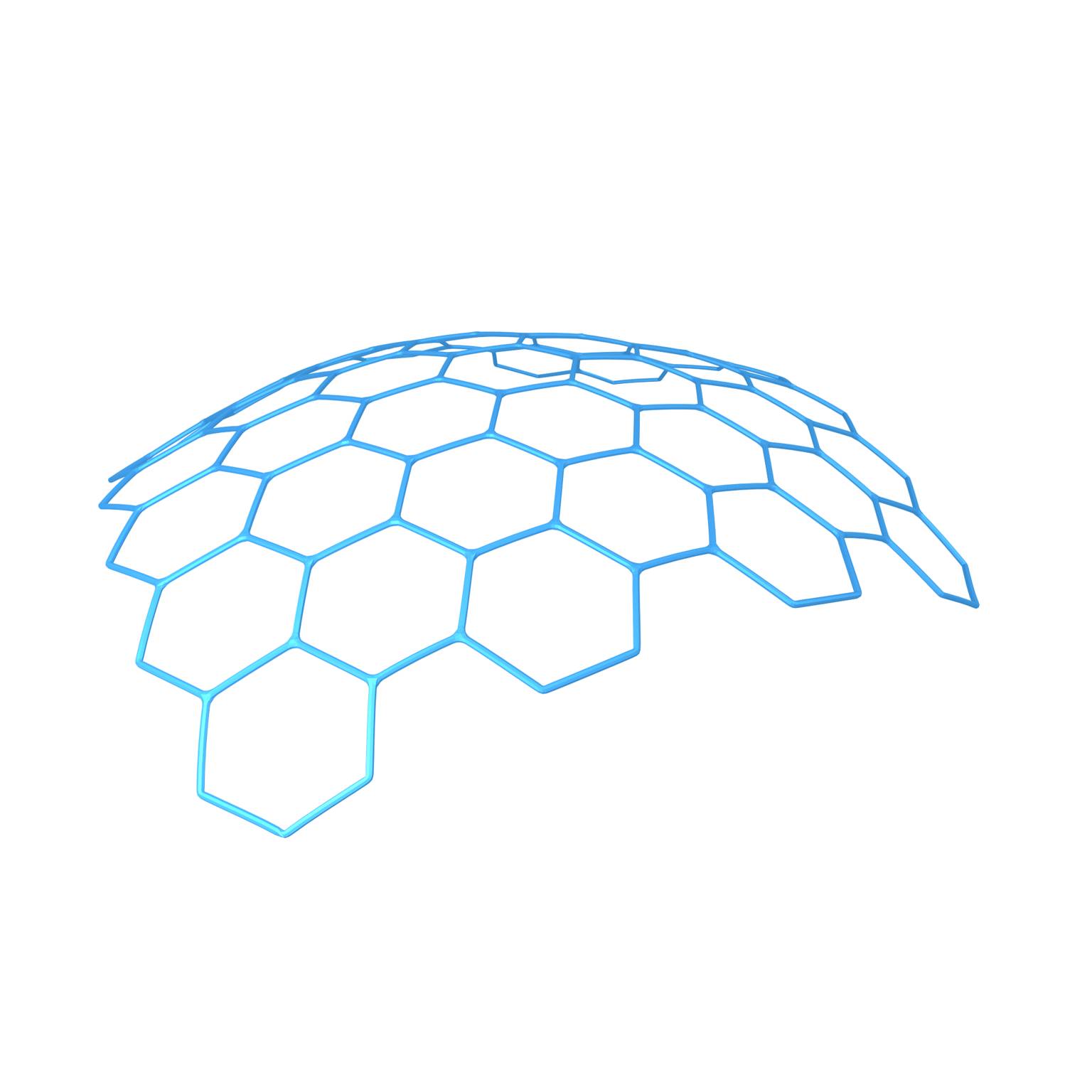}
		\includegraphics[width=\textwidth,trim=0 8cm 0 10cm, clip]{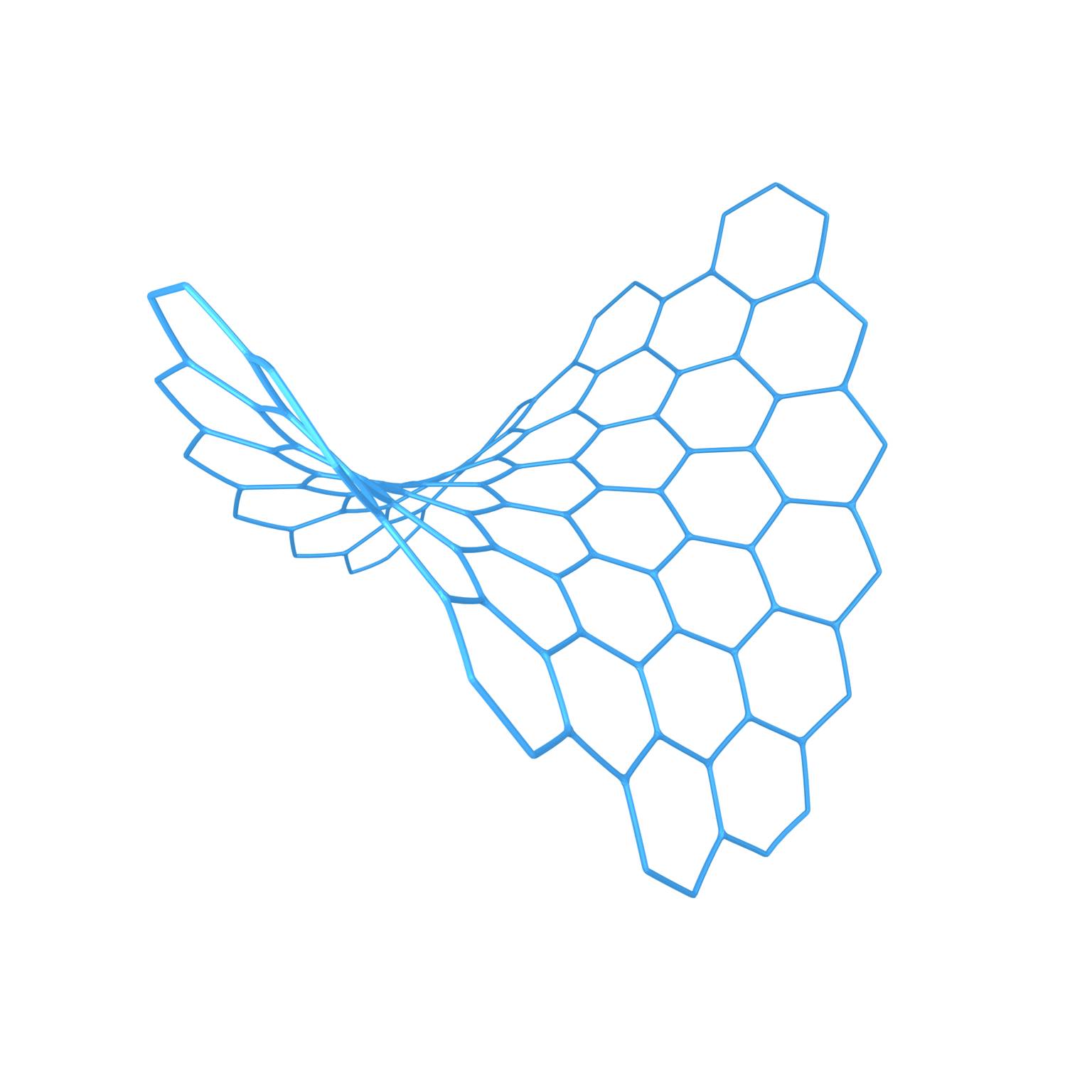}
		\includegraphics[width=\textwidth,trim=0 12cm 0 8cm, clip]{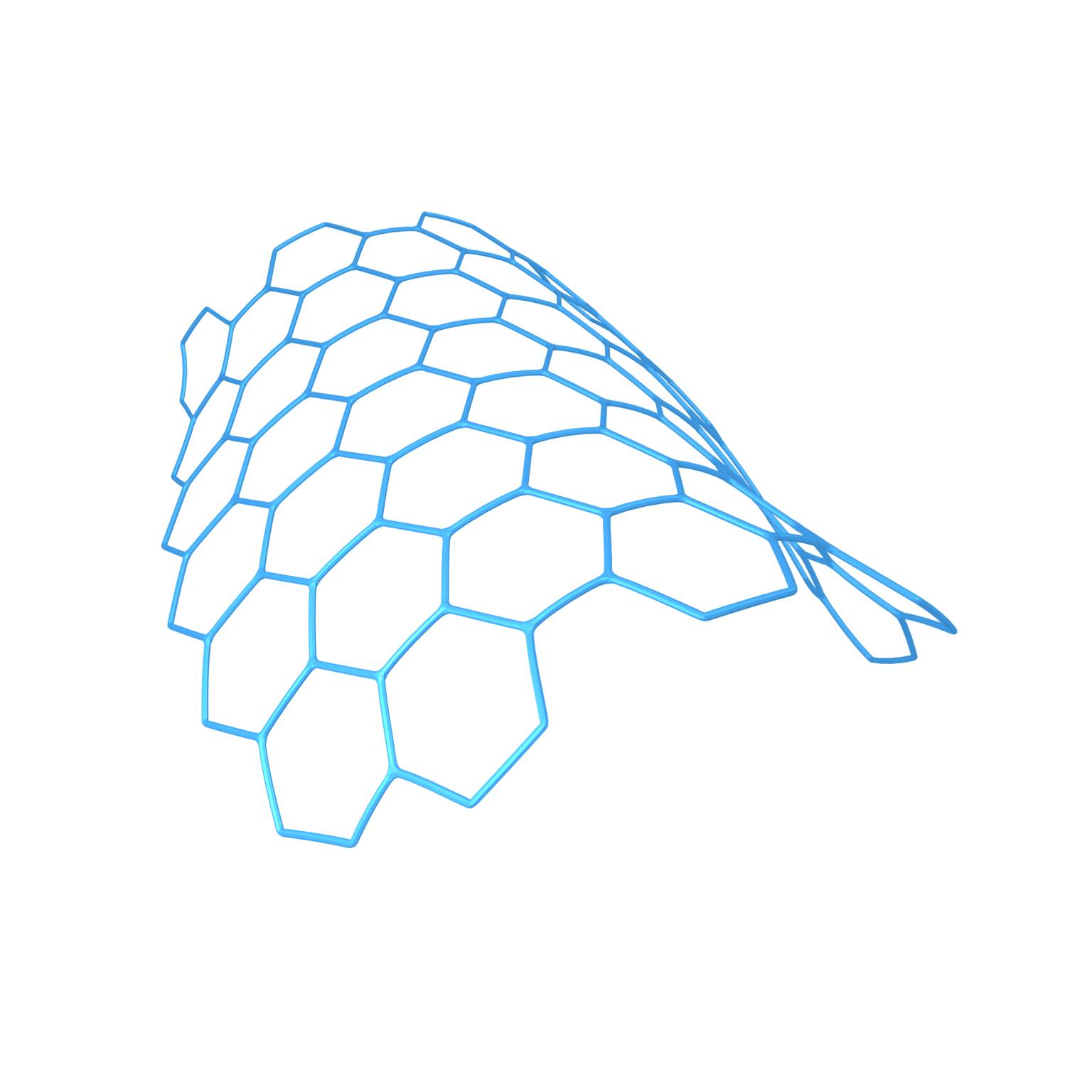}
		\includegraphics[width=\textwidth,trim=0 10cm 0 12cm, clip]{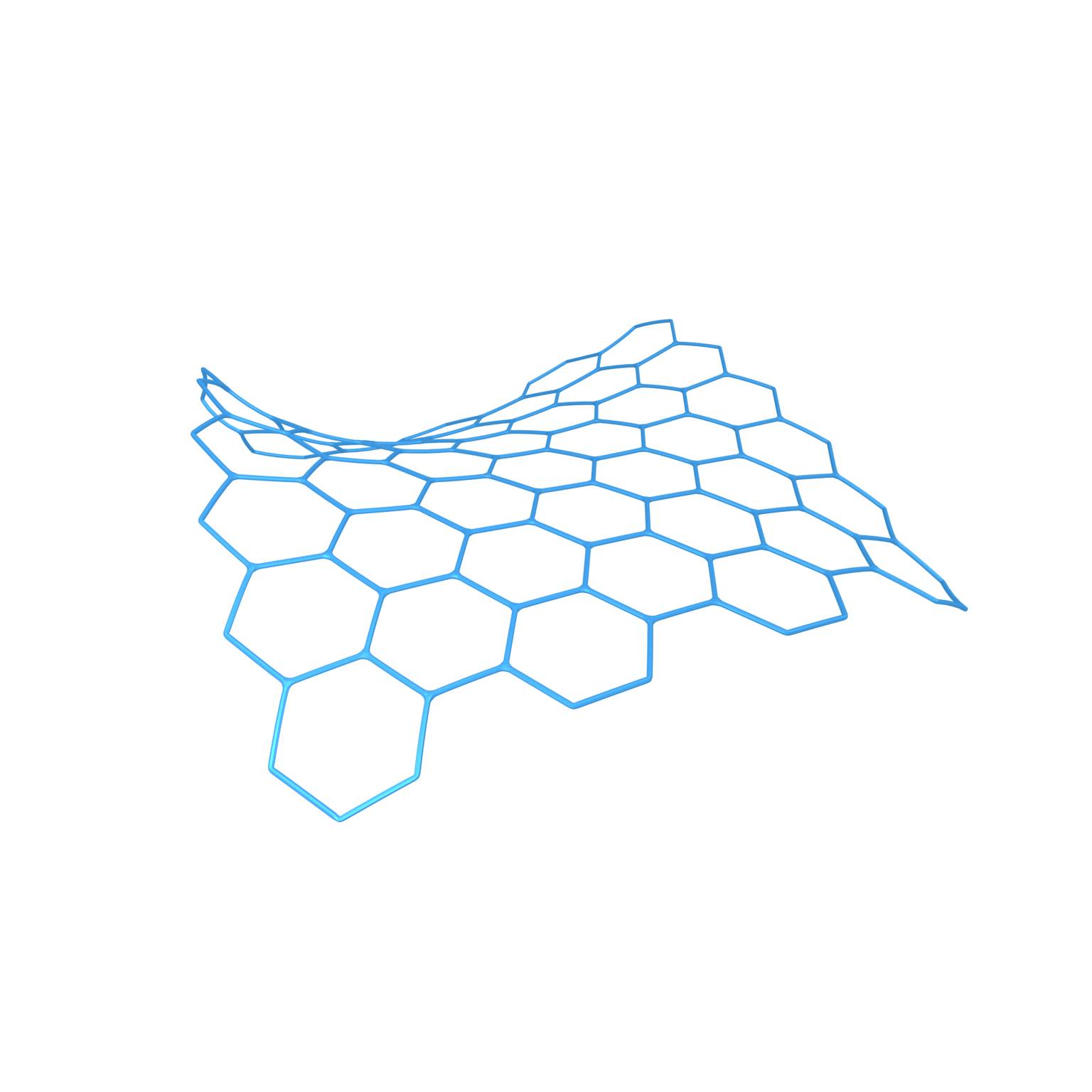}
		\caption{10 Iterations}
	\end{subfigure}
	\begin{subfigure}{0.23\textwidth}
		\includegraphics[width=\textwidth,trim=0 10cm 0 10cm, clip]{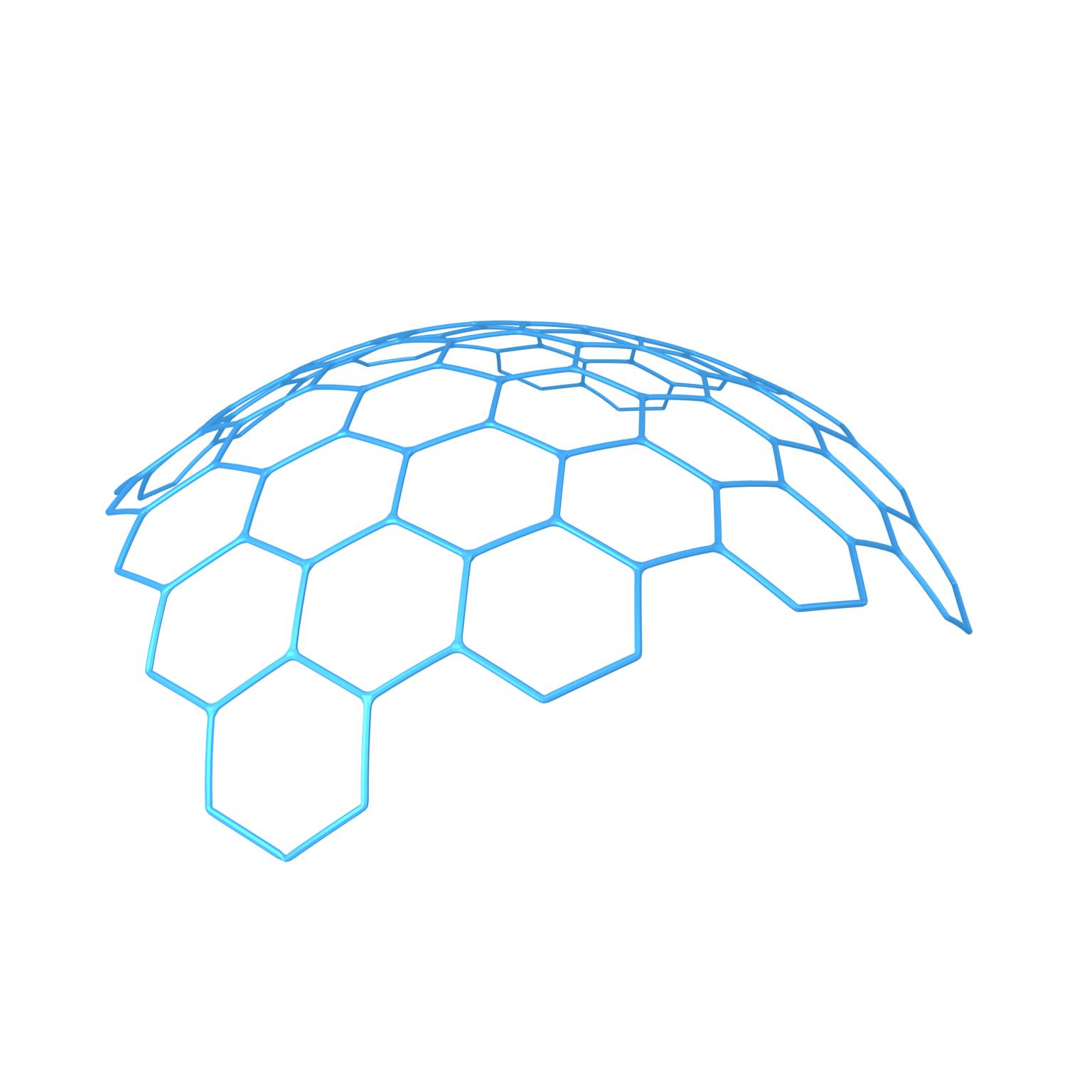}
		\includegraphics[width=\textwidth,trim=0 8cm 0 10cm, clip]{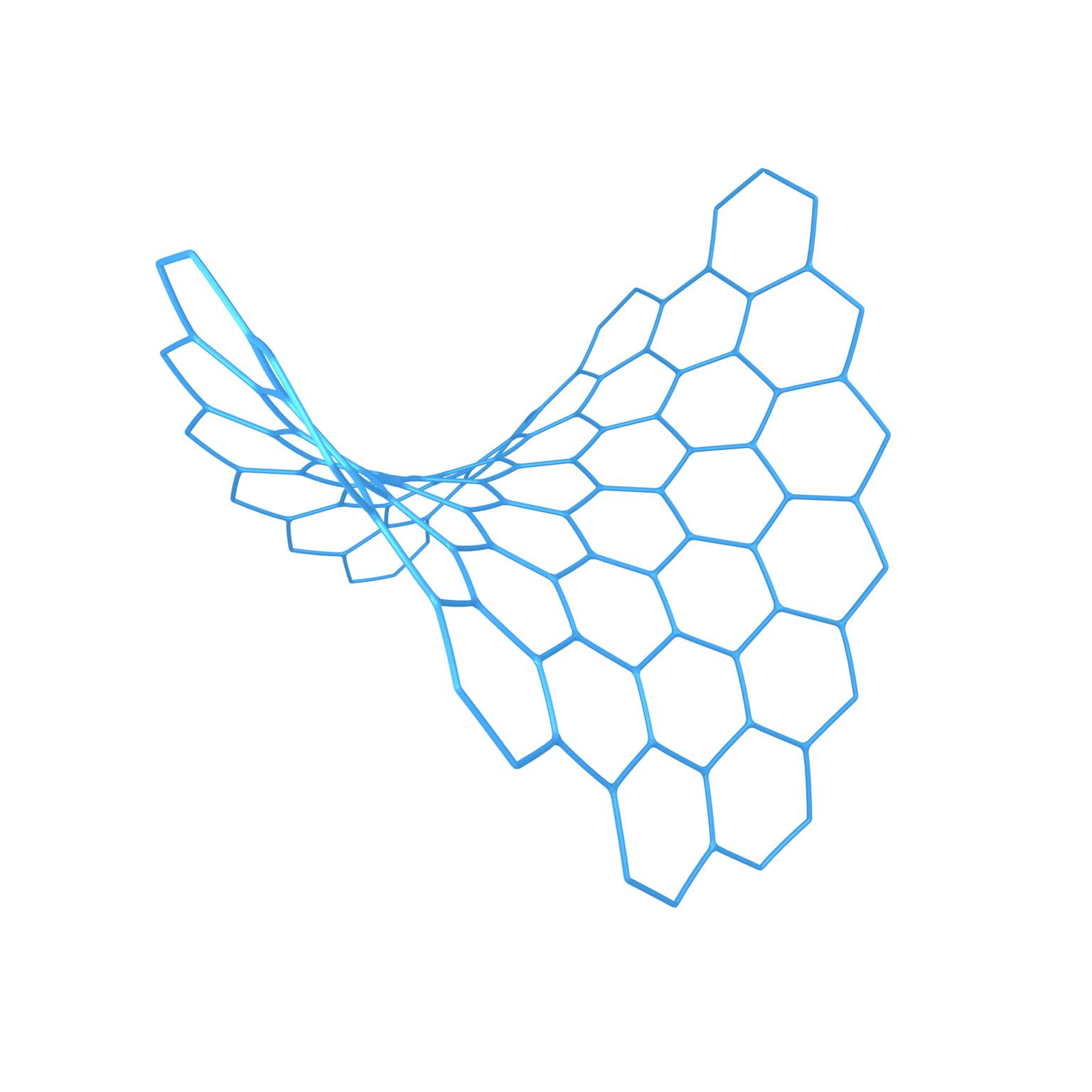}
		\includegraphics[width=\textwidth,trim=0 12cm 0 8cm, clip]{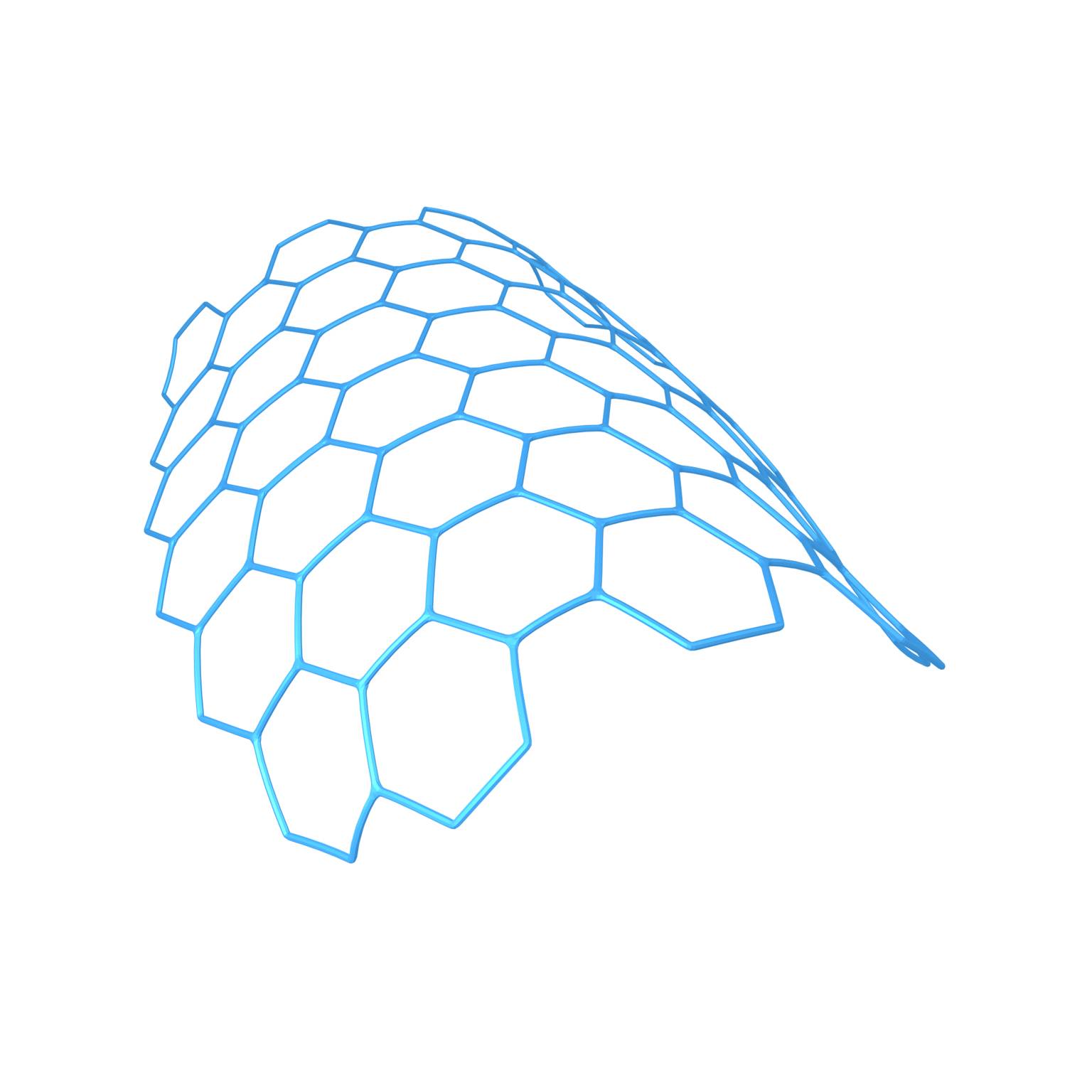}
		\includegraphics[width=\textwidth,trim=0 10cm 0 12cm, clip]{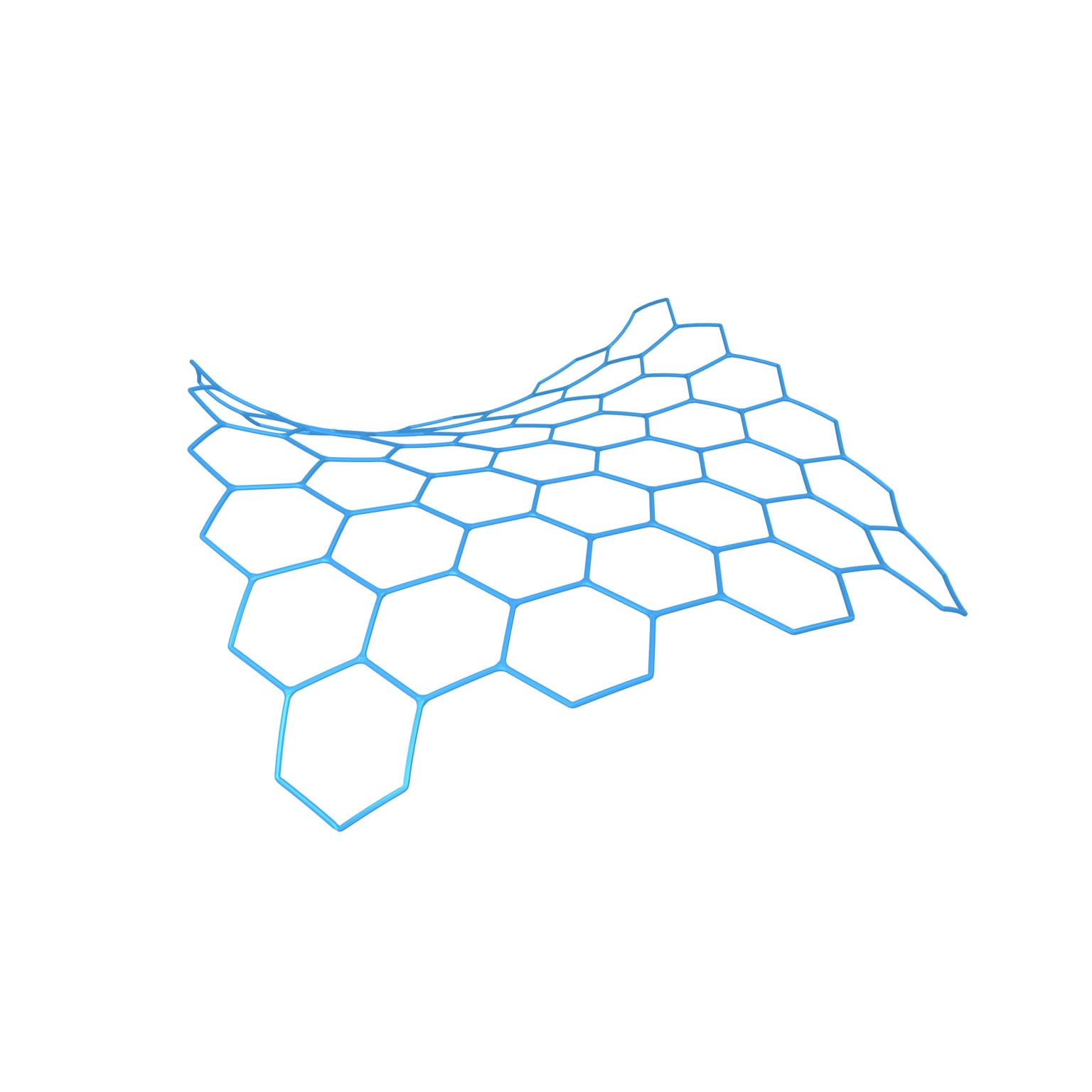}
		\caption{20 Iterations}
	\end{subfigure}
	\begin{subfigure}{0.23\textwidth}
		\includegraphics[width=\textwidth,trim=0 10cm 0 10cm, clip]{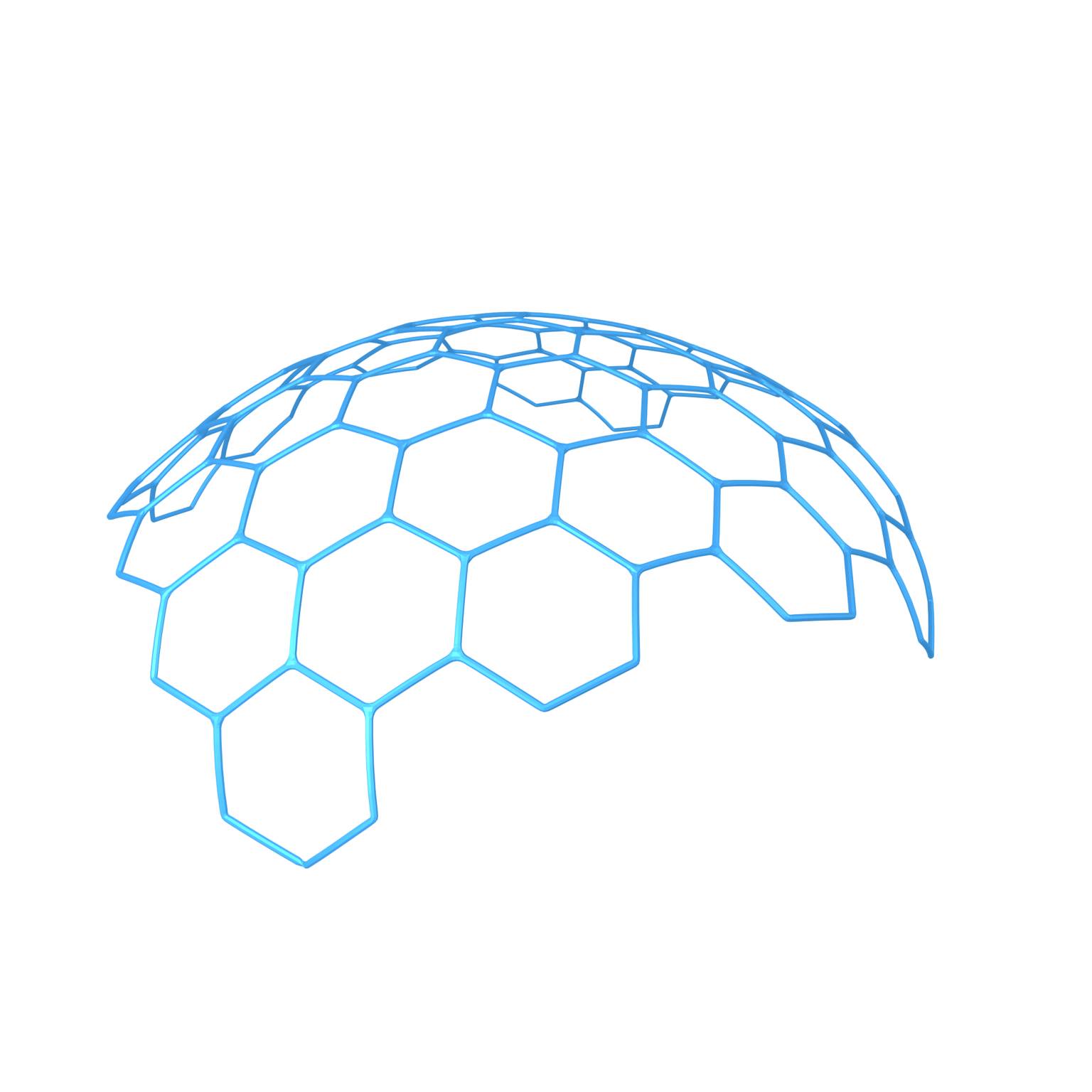}
		\includegraphics[width=\textwidth,trim=0 8cm 0 10cm, clip]{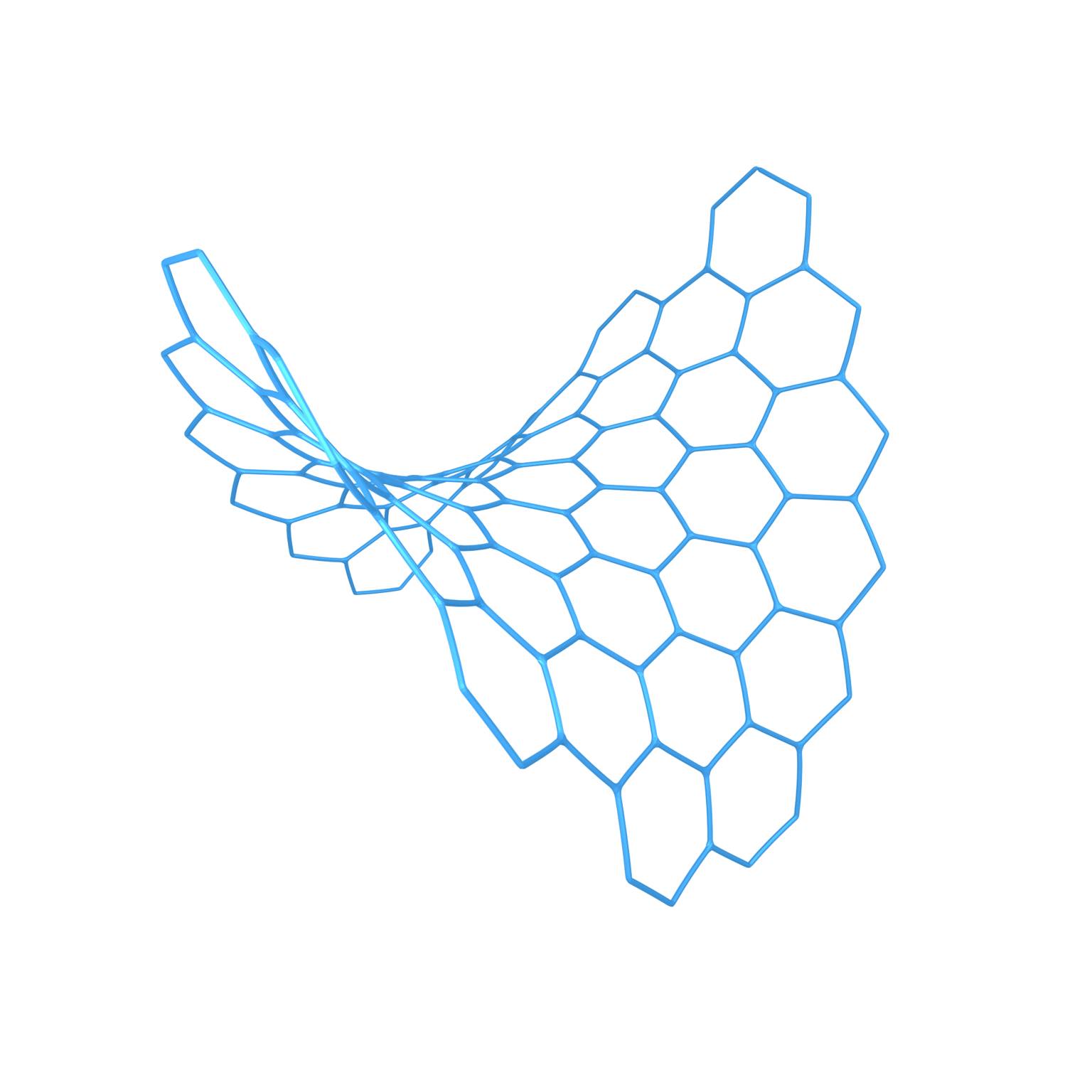}
		\includegraphics[width=\textwidth,trim=0 12cm 0 8cm, clip]{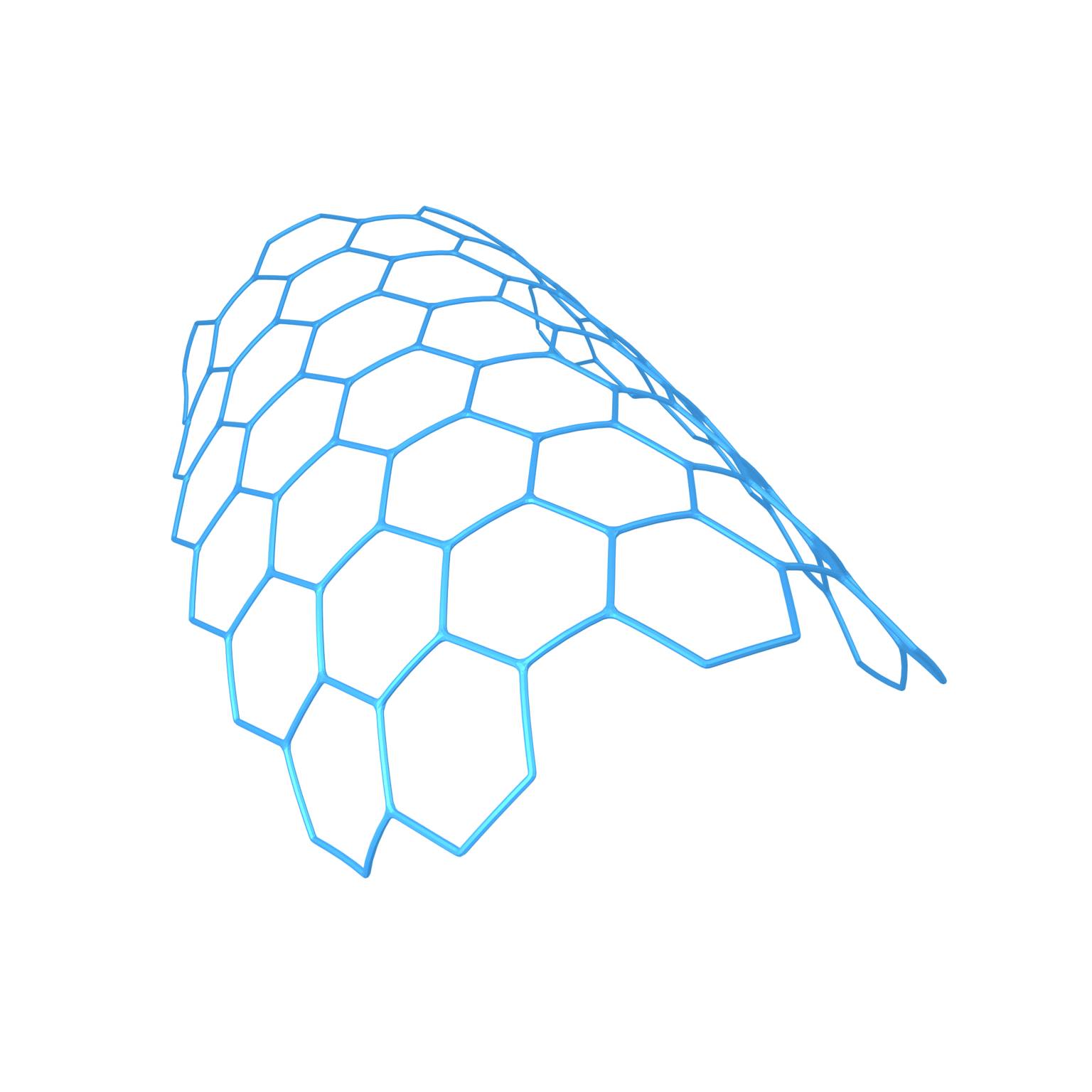}
		\includegraphics[width=\textwidth,trim=0 10cm 0 12cm, clip]{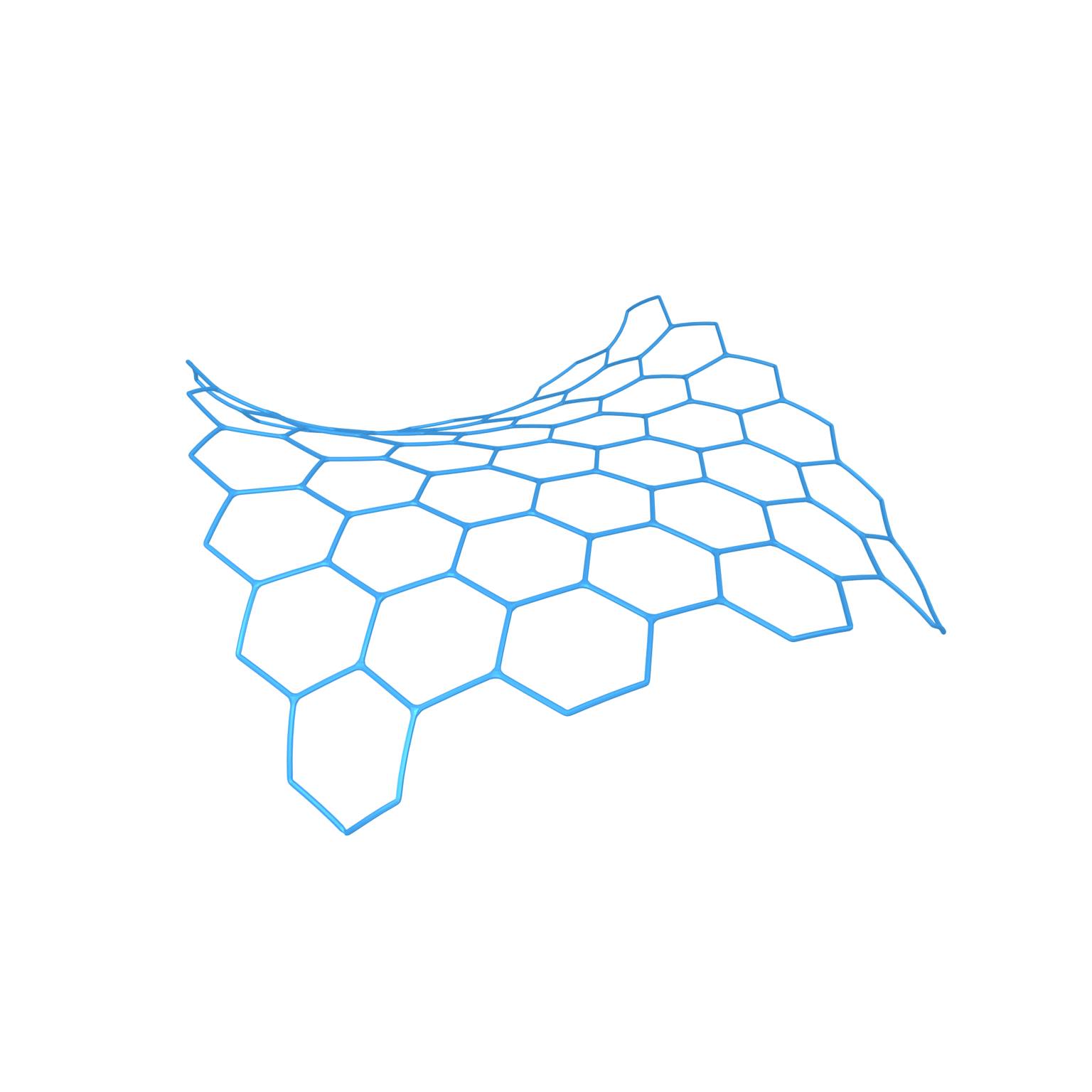}
		\caption{Result}
	\end{subfigure}
	\caption{Shape optimization results. Starting with a flat pattern, the position and material direction of the anchor constraints at the four corners of the pattern are optimized to deform the pattern into the target shape.}
	\label{fig:shape_optim}
\end{centering}
\end{figure*}

To evaluate our shape optimization method, we follow the example described in Section \ref{sec:optimization}. Our input is a flat pattern and four anchor constraints whose initial values correspond to the location and material direction of the four corners of the pattern. The material parameters $k_S$ of the pattern are set to the optimized values we obtained in the pattern simplification step described in Section \ref{sub:results_stiffness}. Our goal is to deform the pattern into the target shape by finding the optimal values for the anchor constraints. As targets we use the four shapes shown on the right of Figure \ref{fig:stiffness_results} that are the result of our pattern simplification process. Once again we use the SQP algorithm to solve the optimization problem. Figure \ref{fig:shape_optim} shows the intermediate and final results for the shape optimization, while Table \ref{tab:shape_optim} shows the quantitative evaluation in terms of the objective function value (see Equation \eqref{eq:shape_energy}), average distance between vertices in millimeters, number of optimization iterations, and computation time. Finally, Figure \ref{fig:plot_eshape} shows a logarithmic plot of how the shape energy term $E_{shape}$ changes during the course of the optimization.

\begin{figure} 
\begin{centering}
    \begin{subfigure}[b]{0.23\textwidth}
		\includegraphics[width=\textwidth]{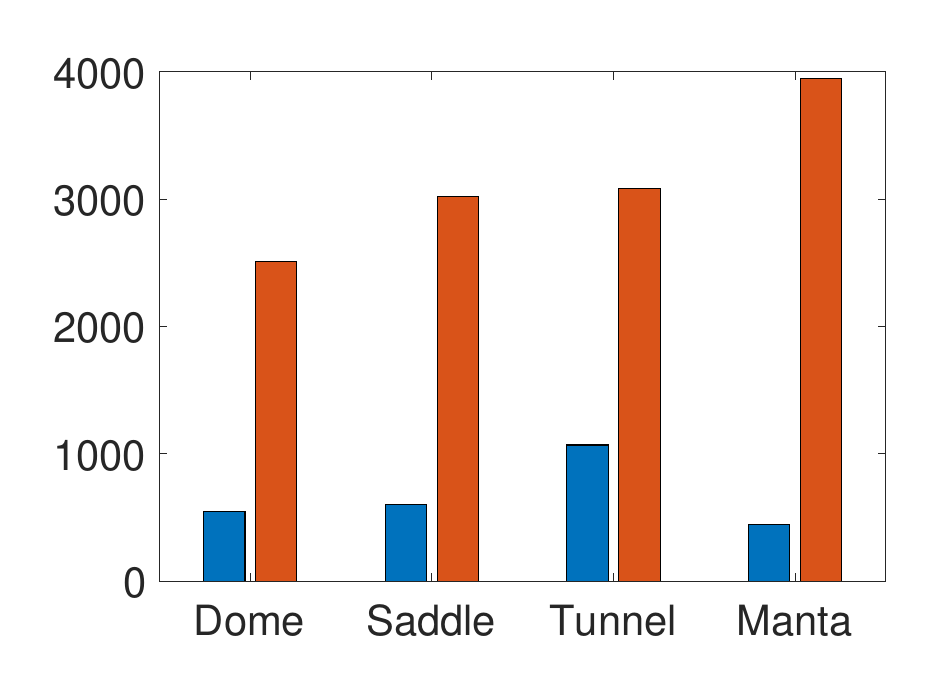}
		\caption{Computation time (seconds)}
	\end{subfigure}
	\begin{subfigure}[b]{0.23\textwidth}
		\includegraphics[width=\textwidth]{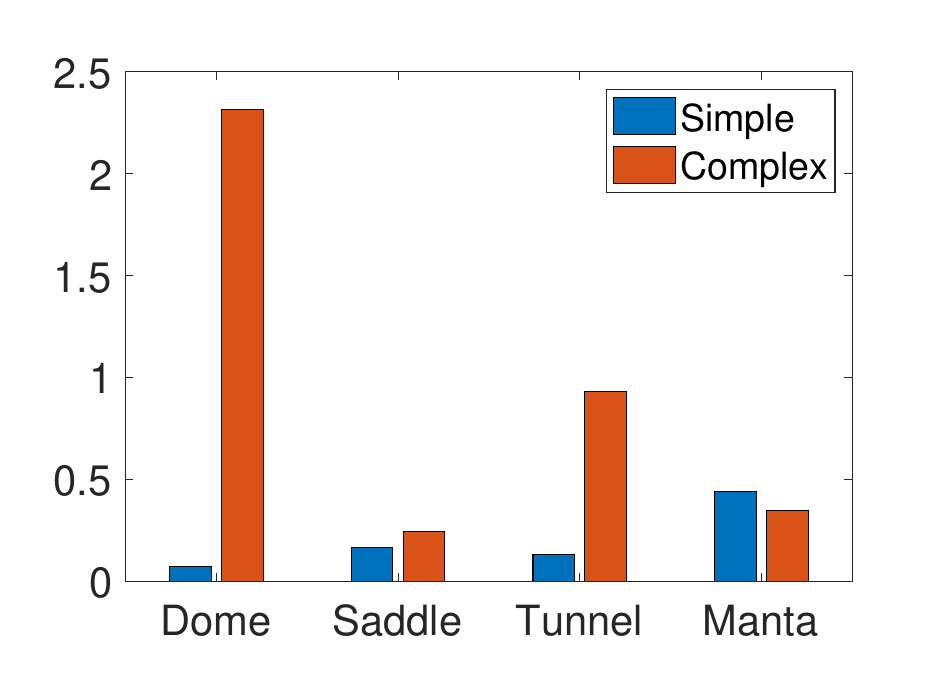}
		\caption{Average distance (millimeters)}
	\end{subfigure}
	\caption{Comparison of computation time and average distance between vertices when using the  {the original zigzag pattern or its simplified version} for the task of shape design.}
	\label{fig:compare_plot}
\end{centering}
\end{figure}

To compare the computational cost of shape optimization using a  {pattern with zigzag springs} against using a simplified pattern, we also perform the experiment using  {the zigzag pattern} for both the input and target shapes. While the computation time of the optimization for the simplified pattern ranges from $445$ to $1070$ seconds, optimization with  {the zigzag pattern} takes $2511$ to $3948$ seconds, as shown in Table \ref{tab:shape_optim}. Furthermore, the results we obtained from optimization using  {the zigzag pattern} have lower accuracy on average, as the optimization is more likely to converge early at a local optimum. A direct comparison of computation time and accuracy in terms of the average distance between vertices can be seen in Figure \ref{fig:compare_plot}.

\begin{table}
\caption{Evaluation of the shape optimization experiment. The columns show the objective function value (see Equation \eqref{eq:shape_energy}), average distance between vertices in millimeters, number of optimization iterations and computation time for each of the four target shapes.}  \label{tab:shape_optim}
\begin{tabular}{ c | c c c c}
 \multicolumn{5}{c}{Simplified Pattern} \\
 Shape & $E_{shape}$ & Avg. Distance & Iterations & Time  \\
 \hline
 Dome & $0.0079$ & $0.0724mm$ & $93$ & $547s$ \\  
 Saddle & $0.0489$ & $0.1678mm$ & $60$ & $599s$\\ 
 Tunnel & $0.0305$ & $0.1317mm$  & $102$ & $1070s$ \\
 Manta & $0.3185$ & $0.4385mm$  & $54$ & $445s$ \\
 \multicolumn{5}{c}{} \\
 \multicolumn{5}{c}{Complex Pattern} \\
 Shape & $E_{shape}$ & Avg. Distance & Iterations & Time  \\
 \hline
 Dome & $7.7539$ & $2.3148mm$ & $19$ & $2511s$ \\  
 Saddle & $0.1116$ & $0.2464mm$ & $46$ & $3023s$\\ 
 Tunnel & $1.1352$ & $0.9315mm$  & $61$ & $3085s$ \\
 Manta & $0.1875$ & $0.3452mm$  & $58$ & $3948s$ \\
\end{tabular}
\end{table}

\subsection{Implementation}\label{sec:implementation}

All our algorithms for pattern generation, pattern simplification, and shape optimization are implemented in MATLAB, using the SQP algorithm to solve the optimization problems defined in Equations \eqref{eq:simpl} and \eqref{eq:shape_optim}. The DER simulation is implemented in C++, building upon the framework of Vekhter et al.~\cite{vekhter}. For fabrication, we additionally make use of the 3d-modeling software Rhinoceros with parametric modeling plugin Grasshopper to create a 3d-printable mesh. The pattern is then fabricated using an Ultimaker 3 3d-printer.

\section{Future Work}\label{sec:future}

\begin{figure}[t]
		\centering
		\includegraphics[width=0.23\textwidth]{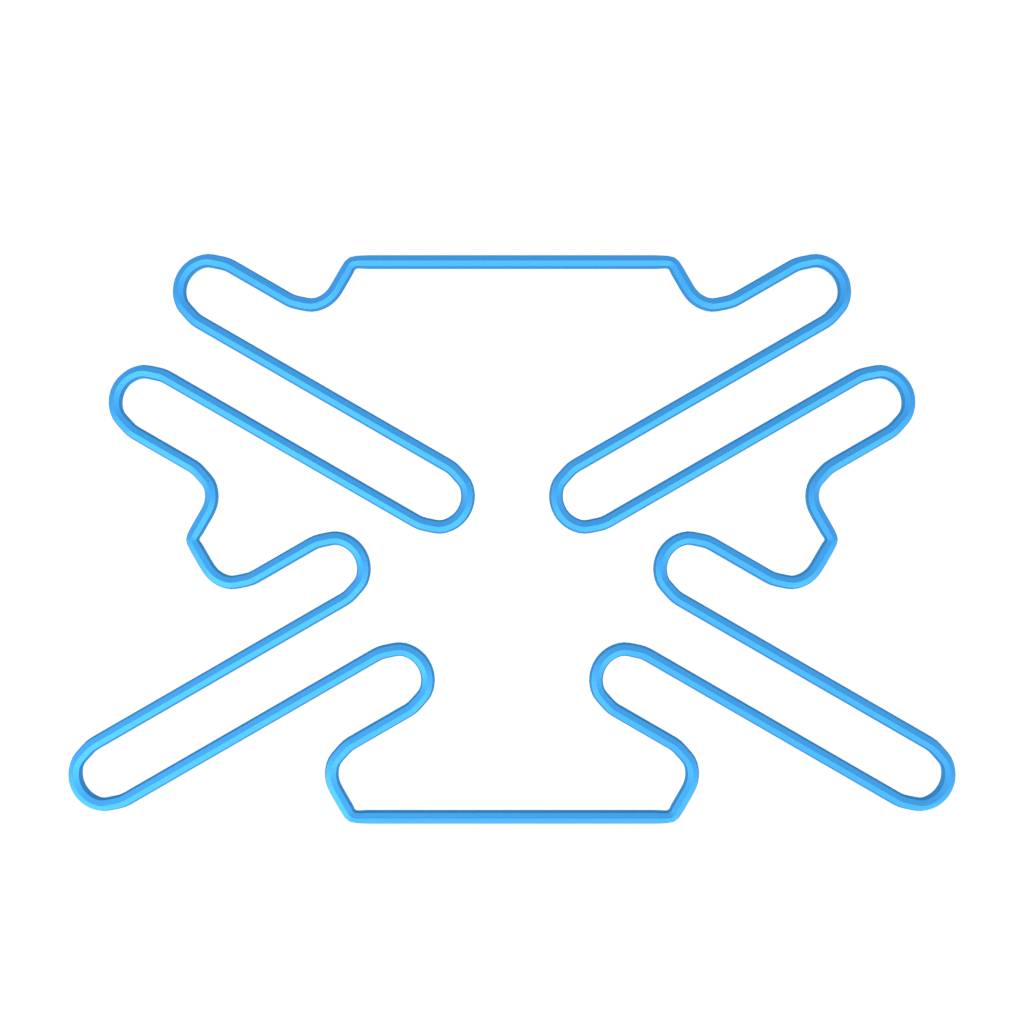}
		\includegraphics[width=0.23\textwidth]{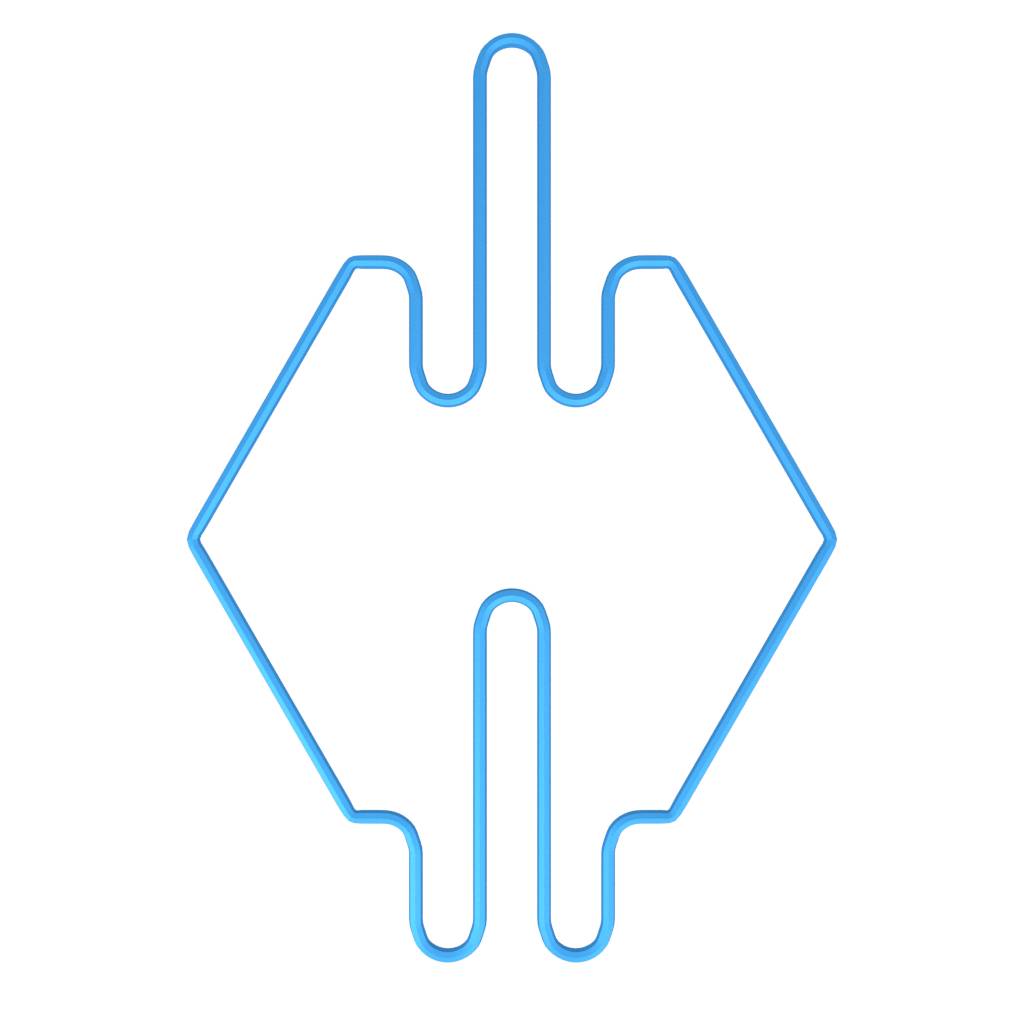}
		\includegraphics[width=0.23\textwidth]{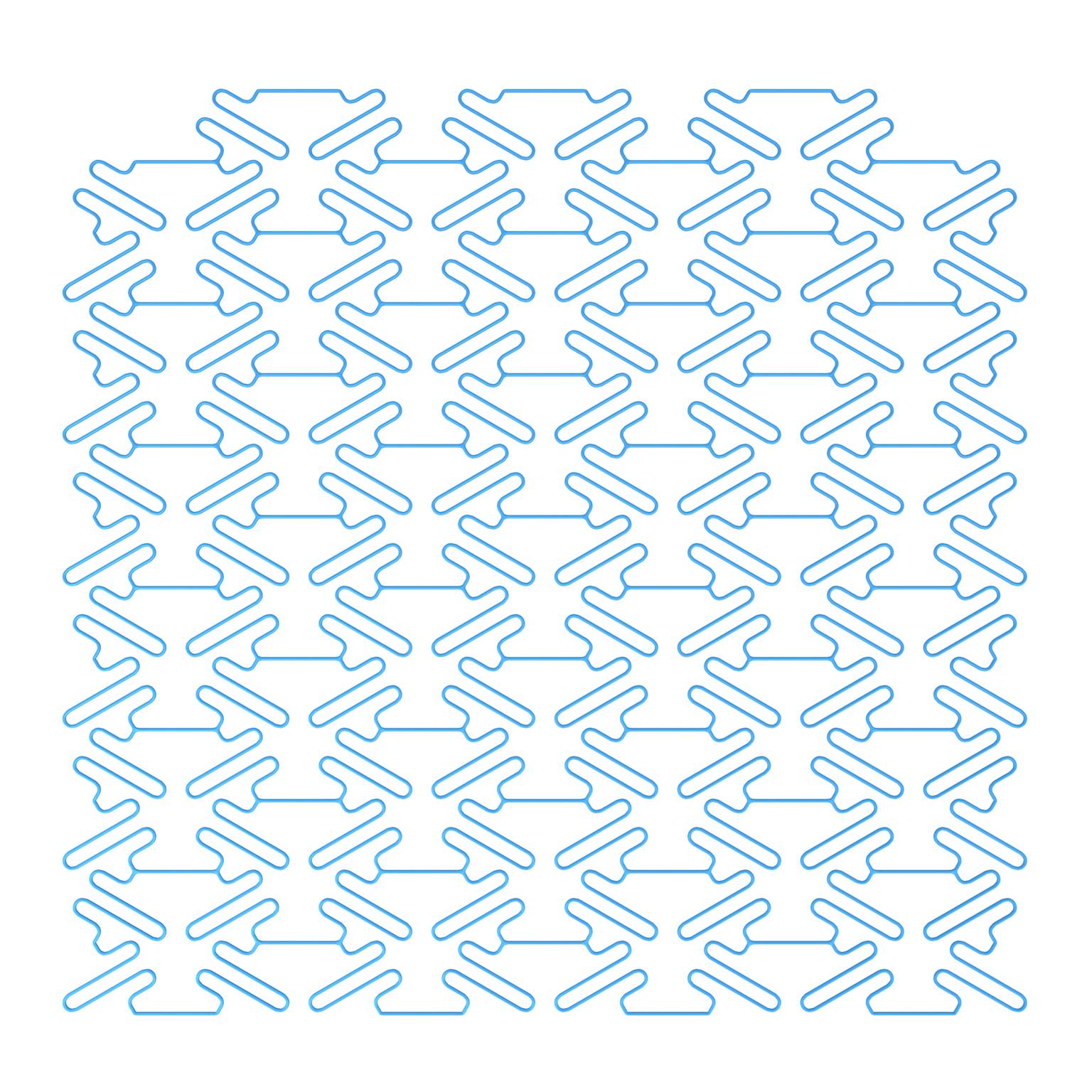}
		\includegraphics[width=0.23\textwidth]{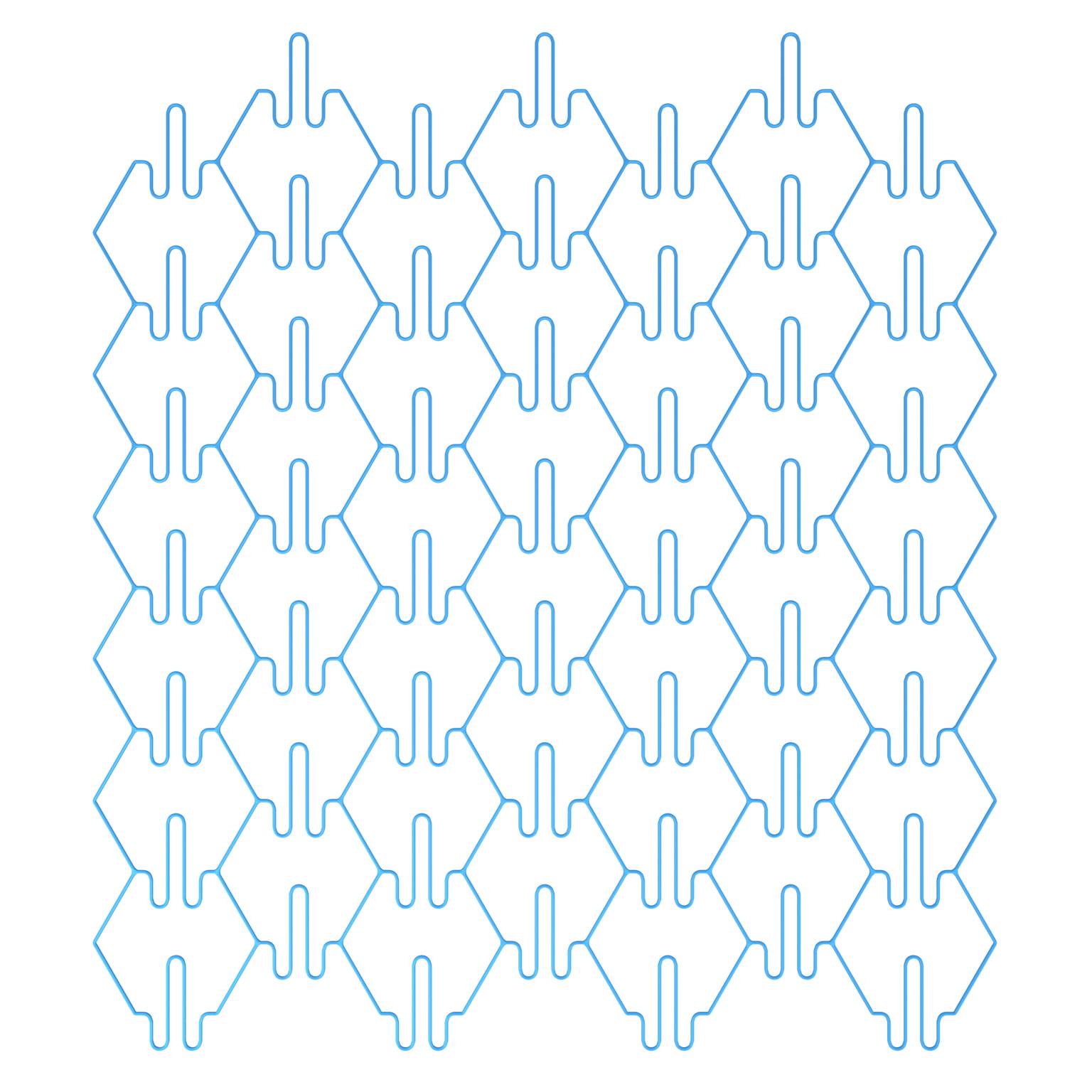}
		\caption{Different cell types created by changing the input parameters to our pattern generator. }
		\label{fig:local_stiffness}
\end{figure}

In the future, we would like to extend our method to compute the gradient of the material parameter optimization objective function analytically by means of sensitivity analysis. The gain in speed would allow us to consider per-edge material parameters instead of global parameters to account for local variations in deformation behavior (see Figure \ref{fig:local_stiffness} for an example).

We would furthermore like to examine the relationship between the free parameters of our  {zigzag spring meso-structure} and the material parameters of the simplified grid. Creating a mapping between these two sets of parameters would make it possible to reverse the grid simplification process, allowing us to perform shape optimization with the simplified pattern using the local material parameters as design variables, then replacing the hexagon edges with the appropriate zigzag springs that yield the desired deformation behavior.

We are also interested in further examining the space of possible pattern deformations. Defining anchor constraints that result in an interesting and physically plausible deformation is not a trivial task. Extending the simulation to account for the possibility of material failure, for example, due to excessive stretching of the pattern, would make it possible to check a deformation for physical plausibility. However, it would be even more interesting to examine how reasonable bounds for the anchor constraints can be defined based on the geometry of the pattern.

 {Additionally, the general applicability of our pattern simplification approach is a topic for future investigation. We have demonstrated our approach using zigzag springs as our choice of meso-structure, which have the advantage of providing a direct correspondence between the original and simplified pattern since the zigzag springs are simply replaced by straight edges in the simplification. In theory, our approach can be used with other types of meso-structures that can be approximated by straight edges during the simplification process, but we cannot guarantee that the accuracy of the resulting reduced model is comparable to our experiments, since our approach may not yield a good approximation for all types of meso-structures. Examining the suitability of our approach for different types of meso-structures from a theoretical point of view could be an interesting topic for future research.}

Finally, we have only considered the use of regular tilings for the generation of our patterns so far. However, it may not be possible to approximate very complex shapes using such tilings. In the future, it would be interesting to consider the use of non-regular tilings and how such tilings can be generated for the best possible approximation of the target shape. This would allow us to increase the dimensions of the design space and hence approximate more complex shapes.

\section{Conclusions}\label{sec:conclusions}

We presented a framework for the generation and optimization of reduced-order flexible patterns. Our pattern generator is capable of  {creating complex patterns containing zigzag-spring meso-structures} based on hexagonal or rectangular tilings whose deformation behavior can then be simulated. Since the simulation of such complex patterns has a high computational cost, we simplify the pattern by encoding the complex deformation behavior of the structure into the meta-material parameters of a more simple pattern. We furthermore showed that a small but varied set of example deformations of  {the zigzag pattern} is sufficient to properly fit the material parameters of the simplified pattern and that increasing the number of samples does not yield an improved approximation. The simple pattern can then be used for shape design, which we demonstrated by finding the optimal boundary constraints that deform the pattern into the desired target shape.

\section*{Acknowledgments}
\noindent
This research was funded by the Vienna Science and Technology Fund (\href{https://www.wwtf.at/programmes/information_communication/ICT15-082/index.php}{WWTF ICT15-082}).

\bibliographystyle{alpha} 
\bibliography{sample-base,shapes}

\newcommand{\etalchar}[1]{$^{#1}$}
\begin{thebibliography}{GSFD{\etalchar{+}}14}

\bibitem[Ash06]{ashby2006}
Michael~F Ashby.
\newblock The properties of foams and lattices.
\newblock {\em Philosophical Transactions of the Royal Society A: Mathematical,
  Physical and Engineering Sciences}, 364(1838):15--30, 2006.

\bibitem[BAV{\etalchar{+}}10]{bergou2010}
Mikl{\'o}s Bergou, Basile Audoly, Etienne Vouga, Max Wardetzky, and Eitan
  Grinspun.
\newblock Discrete viscous threads.
\newblock {\em ACM Transactions on Graphics (TOG)}, 29(4):1--10, 2010.

\bibitem[BBO{\etalchar{+}}09]{Bickel2009}
Bernd Bickel, Moritz B{\"{a}}cher, Miguel~A. Otaduy, Wojciech Matusik,
  Hanspeter Pfister, and Markus Gross.
\newblock {Capture and modeling of non-linear heterogeneous soft tissue}.
\newblock {\em ACM Transactions on Graphics}, 28(3):1, jul 2009.

\bibitem[BBO{\etalchar{+}}10]{Bickel2010}
Bernd Bickel, Moritz B{\"{a}}cher, Miguel~A. Otaduy, Hyunho~Richard Lee,
  Hanspeter Pfister, Markus Gross, and Wojciech Matusik.
\newblock {Design and fabrication of materials with desired deformation
  behavior}.
\newblock {\em ACM Transactions on Graphics}, 29(4):1, jul 2010.

\bibitem[BWR{\etalchar{+}}08]{bergou2008}
Mikl\'{o}s Bergou, Max Wardetzky, Stephen Robinson, Basile Audoly, and Eitan
  Grinspun.
\newblock Discrete elastic rods.
\newblock {\em ACM Transactions on Graphics (TOG)}, 27(3):1–12, August 2008.

\bibitem[CWJ{\etalchar{+}}15]{clausen2015}
Anders Clausen, Fengwen Wang, Jakob~S Jensen, Ole Sigmund, and Jennifer~A
  Lewis.
\newblock Topology optimized architectures with programmable poisson’s ratio
  over large deformations.
\newblock {\em Adv. Mater}, 27(37):5523--5527, 2015.

\bibitem[DLL{\etalchar{+}}15]{dumas2015}
J{\'e}r{\'e}mie Dumas, An~Lu, Sylvain Lefebvre, Jun Wu, and Christian Dick.
\newblock By-example synthesis of structurally sound patterns.
\newblock {\em ACM Transactions on Graphics (TOG)}, 34(4):1--12, 2015.

\bibitem[EP15]{Eidini2015}
Maryam Eidini and Glaucio~H. Paulino.
\newblock {Unraveling metamaterial properties in zigzag-base folded sheets}.
\newblock {\em Science Advances}, 1(8):e1500224, sep 2015.

\bibitem[GASR82]{Gibson1982}
L~J Gibson, M~F Ashby, G~S Schajer, and C~I Robertson.
\newblock {The Mechanics of Two-Dimensional Cellular Materials}.
\newblock {\em Proceedings of the Royal Society A: Mathematical, Physical and
  Engineering Sciences}, 382(1782):25--42, jul 1982.

\bibitem[GMB17]{Guseinov2017}
Ruslan Guseinov, Eder Miguel, and Bernd Bickel.
\newblock {CurveUps}.
\newblock {\em ACM Transactions on Graphics}, 36(4):1--12, jul 2017.

\bibitem[GMP{\etalchar{+}}20]{Guseinov2020}
Ruslan Guseinov, Connor McMahan, Jes{\'{u}}s P{\'{e}}rez, Chiara Daraio, and
  Bernd Bickel.
\newblock {Programming temporal morphing of self-actuated shells}.
\newblock {\em Nature Communications}, 11(1):1--7, dec 2020.

\bibitem[GSFD{\etalchar{+}}14]{Garg2014}
Akash Garg, Andrew~O. Sageman-Furnas, Bailin Deng, Yonghao Yue, Eitan Grinspun,
  Mark Pauly, and Max Wardetzky.
\newblock {Wire mesh design}.
\newblock {\em ACM Transactions on Graphics}, 33(4):1--12, jul 2014.

\bibitem[HG92]{Haftka1992}
Raphael~T. Haftka and Zafer G{\"{u}}rdal.
\newblock {\em {Elements of Structural Optimization}}.
\newblock Springer, 3rd rev. a edition, 1992.

\bibitem[IFW{\etalchar{+}}16]{ion2016}
Alexandra Ion, Johannes Frohnhofen, Ludwig Wall, Robert Kovacs, Mirela Alistar,
  Jack Lindsay, Pedro Lopes, Hsiang-Ting Chen, and Patrick Baudisch.
\newblock Metamaterial mechanisms.
\newblock In {\em Proceedings of the 29th Annual Symposium on User Interface
  Software and Technology}, pages 529--539, 2016.

\bibitem[JCC{\etalchar{+}}17]{jenett2017}
Benjamin Jenett, Sam Calisch, Daniel Cellucci, Nick Cramer, Neil Gershenfeld,
  Sean Swei, and Kenneth~C Cheung.
\newblock Digital morphing wing: active wing shaping concept using composite
  lattice-based cellular structures.
\newblock {\em Soft robotics}, 4(1):33--48, 2017.

\bibitem[JTV{\etalchar{+}}15]{jiang}
Caigui Jiang, Chengcheng Tang, Amir Vaxman, Peter Wonka, and Helmut Pottmann.
\newblock Polyhedral patterns.
\newblock {\em ACM Transactions on Graphics (TOG)}, 34(6):1--12, 2015.

\bibitem[LGS05]{Li2005}
K.~Li, X.-L. Gao, and G.~Subhash.
\newblock {Effects of cell shape and cell wall thickness variations on the
  elastic properties of two-dimensional cellular solids}.
\newblock {\em International Journal of Solids and Structures},
  42(5-6):1777--1795, mar 2005.

\bibitem[LM20a]{Leimer2020b}
Kurt Leimer and Przemyslaw Musialski.
\newblock Reduced-order simulation of flexible meta-materials.
\newblock In {\em Symposium on Computational Fabrication}, SCF '20, New York,
  NY, USA, 2020. Association for Computing Machinery.

\bibitem[LM20b]{Leimer2020a}
Kurt Leimer and Przemyslaw Musialski.
\newblock {Simulation of Flexible Patterns by Structural Simplification}.
\newblock In {\em ACM SIGGRAPH 2020 Posters on - SIGGRAPH '20}. ACM Press,
  2020.

\bibitem[MDG14]{meza2014}
Lucas~R Meza, Satyajit Das, and Julia~R Greer.
\newblock Strong, lightweight, and recoverable three-dimensional ceramic
  nanolattices.
\newblock {\em Science}, 345(6202):1322--1326, 2014.

\bibitem[ME96]{Masters1996}
I.G. Masters and K.E. Evans.
\newblock {Models for the elastic deformation of honeycombs}.
\newblock {\em Composite Structures}, 35(4):403--422, aug 1996.

\bibitem[MPI{\etalchar{+}}18]{malomo}
Luigi Malomo, Jes{\'u}s P{\'e}rez, Emmanuel Iarussi, Nico Pietroni, Eder
  Miguel, Paolo Cignoni, and Bernd Bickel.
\newblock Flexmaps: computational design of flat flexible shells for shaping 3d
  objects.
\newblock {\em ACM Transactions on Graphics (TOG)}, 37(6):1--14, 2018.

\bibitem[MSDL17]{Martinez2017}
Jon{\`{a}}s Mart{\'{i}}nez, Haichuan Song, J{\'{e}}r{\'{e}}mie Dumas, and
  Sylvain Lefebvre.
\newblock {Orthotropic {\textless}i{\textgreater}k{\textless}/i{\textgreater}
  -nearest foams for additive manufacturing}.
\newblock {\em ACM Transactions on Graphics}, 36(4):1--12, jul 2017.

\bibitem[MSS{\etalchar{+}}19]{Martinez2019}
Jon{\`{a}}s Mart{\'{i}}nez, M{\'{e}}lina Skouras, Christian Schumacher, Samuel
  Hornus, Sylvain Lefebvre, and Bernhard Thomaszewski.
\newblock {Star-shaped metrics for mechanical metamaterial design}.
\newblock {\em ACM Transactions on Graphics}, 38(4):1--13, jul 2019.

\bibitem[NMK{\etalchar{+}}06]{Nealen2006a}
Andrew Nealen, Matthias M{\"{u}}ller, Richard Keiser, Eddy Boxerman, and Mark
  Carlson.
\newblock {Physically Based Deformable Models in Computer Graphics}.
\newblock {\em Computer Graphics Forum}, 25(4):809--836, dec 2006.

\bibitem[NW06]{nocedal2006numerical}
Jorge Nocedal and Stephen Wright.
\newblock {\em {Numerical Optimization}}.
\newblock Springer Series in Operations Research and Financial Engineering.
  Springer New York, 2006.

\bibitem[PJH{\etalchar{+}}15]{Pottmann2015}
Helmut Pottmann, Caigui Jiang, Mathias H{\"{o}}binger, Jun Wang, Philippe
  Bompas, and Johannes Wallner.
\newblock {Cell packing structures}.
\newblock {\em Computer Aided Design}, 60:70--83, mar 2015.

\bibitem[PKLI{\etalchar{+}}19]{Panetta2019}
Julian Panetta, MINA Konakovi{\'c}-Lukovi{\'c}, Florin Isvoranu, Etienne
  Bouleau, and Mark Pauly.
\newblock X-shells: A new class of deployable beam structures.
\newblock {\em ACM Transactions on Graphics (TOG)}, 38(4):1--15, 2019.

\bibitem[PLBM20]{pillwein2020}
Stefan Pillwein, Kurt Leimer, Michael Birsak, and Przemyslaw Musialski.
\newblock {On Elastic Geodesic Grids and Their Planar to Spatial Deployment}.
\newblock {\em ACM Transactions on Graphics}, 39(4):12, jul 2020.

\bibitem[PM21]{Pillwein2021a}
Stefan Pillwein and Przemyslaw Musialski.
\newblock {Generalized Deployable Elastic Geodesic Grids}.
\newblock {\em ACM Transactions on Graphics (Proc. SIGGRAPH Asia 2021)},
  40(6):in press, dec 2021.

\bibitem[POT17]{Perez2017a}
Jes{\'{u}}s P{\'{e}}rez, Miguel~A. Otaduy, and Bernhard Thomaszewski.
\newblock {Computational design and automated fabrication of kirchhoff-plateau
  surfaces}.
\newblock {\em ACM Transactions on Graphics}, 36(4):1--12, jul 2017.

\bibitem[PTC{\etalchar{+}}15]{Perez2015}
Jes{\'{u}}s P{\'{e}}rez, Bernhard Thomaszewski, Stelian Coros, Bernd Bickel,
  Jos{\'{e}}~A. Canabal, Robert Sumner, and Miguel~A. Otaduy.
\newblock {Design and fabrication of flexible rod meshes}.
\newblock {\em ACM Transactions on Graphics}, 34(4):138:1--138:12, jul 2015.

\bibitem[PZM{\etalchar{+}}15]{Panetta2015}
Julian Panetta, Qingnan Zhou, Luigi Malomo, Nico Pietroni, Paolo Cignoni, and
  Denis Zorin.
\newblock {Elastic textures for additive fabrication}.
\newblock {\em ACM Transactions on Graphics}, 34(4):135:1--135:12, jul 2015.

\bibitem[SC16]{schaedler2016}
Tobias~A Schaedler and William~B Carter.
\newblock Architected cellular materials.
\newblock {\em Annual Review of Materials Research}, 46:187--210, 2016.

\bibitem[SMGT18]{schumacher}
Christian Schumacher, Steve Marschner, Markus Gross, and Bernhard Thomaszewski.
\newblock Mechanical characterization of structured sheet materials.
\newblock {\em ACM Transactions on Graphics (TOG)}, 37(4):1--15, 2018.

\bibitem[TPBF87]{Terzopoulos1987}
Demetri Terzopoulos, John Platt, Alan Barr, and Kurt Fleischer.
\newblock {Elastically deformable models}.
\newblock {\em ACM SIGGRAPH Computer Graphics}, 21(4):205--214, aug 1987.

\bibitem[TSME11]{Taylor2011}
C.M. Taylor, C.W. Smith, W.~Miller, and K.E. Evans.
\newblock {The effects of hierarchy on the in-plane elastic properties of
  honeycombs}.
\newblock {\em International Journal of Solids and Structures},
  48(9):1330--1339, may 2011.

\bibitem[VZF{\etalchar{+}}19]{vekhter}
Josh Vekhter, Jiacheng Zhuo, Luisa F~Gil Fandino, Qixing Huang, and Etienne
  Vouga.
\newblock Weaving geodesic foliations.
\newblock {\em ACM Transactions on Graphics (TOG)}, 38(4):1--22, 2019.

\bibitem[WFE03]{Wadley2003}
H.N.G. Wadley, N.A. Fleck, and A.G. Evans.
\newblock {Fabrication and structural performance of periodic cellular metal
  sandwich structures}.
\newblock {\em Composites Science and Technology}, 63(16):2331--2343, dec 2003.

\bibitem[ZCT16]{Zehnder2016}
Jonas Zehnder, Stelian Coros, and Bernhard Thomaszewski.
\newblock Designing structurally-sound ornamental curve networks.
\newblock {\em ACM Transactions on Graphics (TOG)}, 35(4):1--10, 2016.

\bibitem[ZLW{\etalchar{+}}14]{zheng2014}
Xiaoyu Zheng, Howon Lee, Todd~H Weisgraber, Maxim Shusteff, Joshua DeOtte,
  Eric~B Duoss, Joshua~D Kuntz, Monika~M Biener, Qi~Ge, Julie~A Jackson, et~al.
\newblock Ultralight, ultrastiff mechanical metamaterials.
\newblock {\em Science}, 344(6190):1373--1377, 2014.

\end{thebibliography}

\appendix
\section{Anchor Constraint Derivatives} \label{app:derivatives}

Since any anchor constraint is acting on a given edge segment on a specific rod, we will omit the indices for the rod and segment for ease of notation unless necessary. Following from the anchor constraint energy defined in Equation \eqref{eq:anchor}, the corresponding entries of the Jacobian for the position term are given by
\begin{equation*}
    \frac{\partial r_{pos}}{\partial p_i} = (1-\beta) I,
\end{equation*}
\begin{equation*}
    \frac{\partial r_{pos}}{\partial p_{i+1}} = \beta I,
\end{equation*}
\begin{equation*}
    \frac{\partial r_{pos}}{\partial p_a} = -I ,
\end{equation*}
with $I$ denoting the $3$-by-$3$ identity matrix. The entries of the Jacobian corresponding to the anchor direction terms are
\begin{equation*}
    \frac{\partial r_{dir}}{\partial e} = -\frac{(z^T t) m}{\norm{e}} ,
\end{equation*}
\begin{equation*}
    \frac{\partial r_{dir}}{\partial \theta} = -z^T (-d_1 \sin(\theta) + d_2 \cos(\theta)) ,
\end{equation*}
\begin{equation*}
    \frac{\partial r_{dir}}{\partial m_a} = \frac{z}{\norm{m_a}} ,
\end{equation*}
with
\begin{equation*}
    z = -\frac{m \times m_a}{\norm{m \times m_a}} \times m ,
\end{equation*}
where $e = p_{i+1} - p_i$ denotes the $i$-th edge segment vector and $t = e / \norm{e}$ is its tangent. $d_1$ and $d_2$ denote the directors of the edge segment that, together with the angle $\theta$, define the material direction $m = d_1 \cos(\theta) + d_2 \sin(\theta)$.

Finally, we need the entries for the Hessian. The second derivatives of the anchor position terms vanish. For the entries of the direction terms, we first need some preliminary computations:
\begin{equation*}
    \frac{\partial t}{\partial e} = \frac{I}{\norm{e}} - \frac{e e^T}{\norm{e}^3} ,
\end{equation*}
\begin{equation*}
    \frac{\partial m}{\partial e} = -\frac{t m^T}{\norm{e}} ,
\end{equation*}
\begin{equation*}
    \frac{\partial m}{\partial \theta} = -d_1 \sin(\theta) + d_2 \cos(\theta)  ,
\end{equation*}
\begin{equation*}
    \frac{\partial \norm{m \times m_a}}{\partial e} = - \frac{m \times m_a}{\norm{m \times m_a}}^T \left[ m_a \right]_{\times} \frac{\partial m}{\partial e},
\end{equation*}
\begin{equation*}
    \frac{\partial \norm{m \times m_a}}{\partial m_a} = \frac{m \times m_a}{\norm{m \times m_a}}^T \left[ m \right]_{\times},
\end{equation*}
\begin{equation*}
 \begin{aligned}
    \frac{\partial z}{\partial e} =  \frac{\frac{\partial m}{\partial e} (m^T m_a) + m (m_a^T \frac{\partial m}{\partial e}) 
    - 2 m_a (m \frac{\partial m}{\partial e})}{\norm{m \times m_a}} 
    \\
    - \frac{m \times (m \times m_a) \frac{\partial \norm{m \times m_a}}{\partial e}^T}{\norm{m \times m_a}\norm{m \times m_a}},
     \end{aligned}
\end{equation*}
\begin{equation*}
    \begin{aligned}
        \frac{\partial z}{\partial \theta} = \left( - \frac{\frac{\partial m}{\partial \theta} \times m_a}{\norm{m \times m_a}} 
        + \frac{(m \times m_a) \frac{m \times m_a}{\norm{m \times m_a}}^T \frac{\partial m}{\partial \theta} \times m_a}{\norm{m \times m_a}\norm{m \times m_a}} \right) \times m \\
        - \frac{m \times m_a}{\norm{m \times m_a}} \times \frac{\partial m}{\partial \theta},
    \end{aligned}
\end{equation*}
\begin{equation*}
    \frac{\partial z}{\partial m_a} = \frac{m m^T - I}{\norm{m \times m_a}} 
    - \frac{m \times (m \times m_a) \frac{\partial \norm{m \times m_a}}{\partial m_a}^T}{\norm{m \times m_a}\norm{m \times m_a}} ,
\end{equation*}
with $\left[ m_a \right]_{\times}$ denoting the skew-symmetric matrix of the vector $m_a$. The entries of the Hessian are then given by
\begin{equation*}
    \frac{\partial^2 r_{dir}}{\partial e \partial e} = -\frac{m (t^T \frac{\partial z}{\partial e} + z^T \frac{\partial t}{\partial e}) + (z^T t) \frac{\partial m_i}{\partial e}}{\norm{e}}
    - \frac{(z^T t) (m t^T)}{\norm{e}\norm{e}},
\end{equation*}
\begin{equation*}
    \frac{\partial^2 r_{dir}}{\partial e \partial \theta} = - \frac{(\frac{\partial z}{\partial \theta}^T t) m
    - (z^T t) \frac{\partial m}{\partial \theta}}{\norm{e}} ,
\end{equation*}
\begin{equation*}
    \frac{\partial^2 r_{dir}}{\partial \theta \partial \theta} = \frac{\partial z}{\partial \theta}^T \frac{\partial m}{\partial \theta}
    + z^T (-d_1 \cos(\theta) - d_2 \sin(\theta)) ,
\end{equation*}
\begin{equation*}
    \frac{\partial^2 r_{dir}}{\partial e \partial m_a} = - \frac{m (t^T \frac{\partial z}{\partial m_a})}{\norm{e}} ,
\end{equation*}
\begin{equation*}
    \frac{\partial^2 r_{dir}}{\partial \theta \partial m_a} = \frac{\partial m}{\partial \theta}^T \frac{\partial z}{\partial m_a} \,.
\end{equation*}
\end{document}